\documentclass[%
 reprint,
 superscriptaddress,
 amsmath,amssymb,
 aps,
prb,
]{revtex4-2}

\usepackage{bm}
\usepackage{xcolor}
\usepackage{blindtext}
\usepackage{graphicx}
\usepackage{hyperref}
\usepackage{braket}
\usepackage{nicefrac}
\usepackage{float}
\usepackage{booktabs}
\usepackage{soul}

\hypersetup{
    colorlinks,
    linkcolor={blue},
    citecolor={blue},
    urlcolor={blue}
}

\graphicspath{ {./figures/} }
\newcommand{\mc}[1]{\mathcal{#1}}
\newcommand{\ve}[1]{\boldsymbol{#1}}

\newcommand{\e}{\mathrm{e}}
\newcommand{\du}{\mathrm{d}} % d upright

\newcommand{\FigRGvelocities}{\ref{fig:anisotropy}}

\begin{document}

\title{Nematic quantum criticality in Dirac systems}


\author{Jonas Schwab}
%\thanks{}
\affiliation{Institut f\"ur Theoretische Physik und Astrophysik and W\"urzburg-Dresden Cluster of Excellence ct.qmat, Universit\"at W\"urzburg, 97074 W\"urzburg, Germany}

\author{Lukas Janssen}
\affiliation{Institut f\"ur Theoretische Physik and W\"urzburg-Dresden Cluster of Excellence ct.qmat, Technische Universit\"at Dresden, 01062 Dresden, Germany}

\author{Kai Sun}
\affiliation{Physics Department, University of Michigan, Ann Arbor, Michigan 48109, USA}

\author{Zi Yang Meng}
\affiliation{Department of Physics and HKU-UCAS Joint Institute of Theoretical and Computational Physics, The University of Hong Kong, Pokfulam Road, Hong Kong SAR, China}
\affiliation{Beijing National Laboratory for Condensed Matter Physics,
	Institute of Physics, Chinese Academy of Sciences, Beijing 100190, China}

\author{Igor F.  Herbut}
\affiliation{Department of Physics, Simon Fraser University, British Columbia, Canada V5A 1S6}

\author{Matthias Vojta}
\affiliation{Institut f\"ur Theoretische Physik and W\"urzburg-Dresden Cluster of Excellence ct.qmat, Technische Universit\"at Dresden, 01062 Dresden, Germany}

\author{Fakher F. Assaad}
\affiliation{Institut f\"ur Theoretische Physik und Astrophysik and W\"urzburg-Dresden Cluster of Excellence ct.qmat, Universit\"at W\"urzburg, 97074 W\"urzburg, Germany}


\date{\today}

% !TEX root = master_main.tex

\begin{abstract}

We investigate nematic quantum phase transitions in two different Dirac fermion models. The models feature twofold and fourfold, respectively, lattice rotational symmetries that are spontaneously broken in the ordered phase. Using negative-sign-free quantum Monte Carlo simulations and an $\epsilon$-expansion renormalization group analysis, we show that both models exhibit continuous phase transitions. In contrast to generic Gross-Neveu dynamical mass generation,  the  quantum critical regime is   characterized by  large velocity anisotropies, with fixed-point values being approached very slowly.  Both  experimental and  numerical   investigations  will  not   be representative  of the infrared fixed point,  but of a quasiuniversal regime  where  the  drift of the  exponents tracks the  velocity  anisotropy.  
\end{abstract}

%\keywords{}

\maketitle

	In a strongly correlated electron system, global symmetries, such as spin rotation,  point group, or translational symmetries, can be spontaneously broken as a function of some external tuning parameter.  This   challenging problem has been studied extensively numerically and experimentally over the  last years and impacts our understanding of quantum criticality \cite{Sachdev_book}   in cuprates  \cite{Sachdev03}  and heavy fermions \cite{Lohneysen_rev}.  The problem greatly simplifies when  the Fermi surface reduces to isolated Fermi points in $2+1$  dimensions  and   the  critical point  features  emergent Lorentz symmetry.  In this context, spin, time reversal, 
 and translational symmetry breaking  generically    correspond to the dynamical generation of mass terms \cite{Ryu09},  and the   semimetal-to-insulator transition belongs to one of the various Gross-Neveu universality classes  \cite{Gross74,Herbut06,Herbut09a,Janssen14a,Zerf17,Janssen18,Ray21a}. 

	Across nematic transitions, rotational symmetry is spontaneously broken \cite{Oganesyan01,Schattner15}.   For continuum  Dirac fermions with Hamiltonian $H(\ve{k}) =  v \left( k_x \sigma_x  + k_y \sigma_y \right) $ in momentum space, where $\ve{\sigma}$  are Pauli spin matrices and $v$ is the Fermi velocity,  nematic transitions correspond to  the dynamical  generation of nonmass terms, such as $m \sigma_x$.
They shift the position of the Dirac cone and as such break rotational, and therewith also Lorentz, symmetries.
 Such nematic transitions have been studied theoretically in the past in the context of $d$-wave superconductors \cite{Vojta00,Vojta00a,Huh08,Kim08,Wang13} and bilayer graphene~\cite{Ray21b}. Fundamental questions pertaining to the very nature of the transition  remain open: While initial renormalization group (RG) calculations based on the $\epsilon$ expansion suggested a first-order transition~\cite{Vojta00, Vojta00a},  a continuous transition has been found in large-$N$ analyses~\cite{Kim08, Huh08}. In this Letter, we use quantum Monte Carlo (QMC) simulations and a revised $\epsilon$-expansion analysis to study these transitions. We introduce two different models of Dirac fermions with twofold and fourfold, respectively, lattice rotational symmetries and demonstrate numerically and analytically that both models feature a continuous nematic transition, realizing a new family of quantum universality classes in Dirac systems \emph{without} emergent Lorentz invariance. 

\begin{figure}[b]
\includegraphics[width=\linewidth]{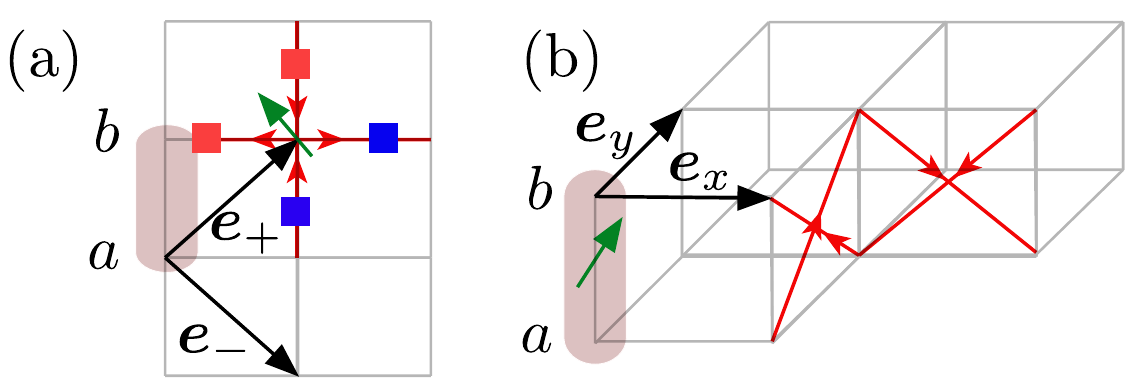}
\caption{
\label{fig:models}
Sketch of (a) $C_{2v}$  and (b) $C_{4v}$ models, defined on $\pi$-flux single-layer and bilayer square lattices, with lattice vectors $\ve{e}_{\nicefrac{+}{-}}$ and $\ve{e}_{\nicefrac{x}{y}}$, respectively.  Dark pink regions indicate unit cells, containing two orbitals ($a$ and $b$) and one Ising spin (green arrow) in both cases. 
Fermions hop  along the red lines and acquire  a  phase  factor  $e^{i \pi/4}$  when following the  direction of the arrow. 
Red and blue squares in (a) indicate the sign structure in the Yukawa coupling of the $C_{2v}$ model.
}
\end{figure}
	
%%%%%%%%%%%%%%%%%%%%%%%%%%%%%%%%%%%%%%%%%%%%%%%%%%%%%%%
\paragraph*{Models.}
%%%%%%%%%%%%%%%%%%%%%%%%%%%%%%%%%%%%%%%%%%%%%%%%%%%%%%%
Inspired from Refs.~\cite{Schattner15,XYXu2017,YYHe2018}, we  design two models
of (2+1)-dimensional  Dirac fermions, $\mathcal{H}_0$, coupled to a transverse-field Ising model (TFIM), 
\begin{align}
  \mathcal{H}_\text{Ising} = -J \sum_{\langle \ve{R},\ve{R}' \rangle} \hat{s}_{\ve{R}}^z \hat{s}_{\ve{R}'}^z -h \sum_{\ve{R}} \hat{s}_{\ve{R}}^x,
\end{align}
where  $\ve{R}$   denotes a unit cell  and  $\langle \ve{R},\ve{R}' \rangle  $  runs  over  adjacent unit  cells.  
A Yukawa coupling,  $\mathcal{H}_\text{Yuk}$,  between the Ising field and  nematic fermion bilinear yields  the desired models,  $\mathcal{H}= \mathcal{H}_0 + \mathcal{H}_{\text{Ising} } +\mathcal{H}_{\text{Yuk}}$,  that  correspond to one of many possible lattice  regularizations  of  continuum field theories of  Eqs.~\eqref{eq:field-theory1} and \eqref{eq:field-theory2}.

In the $C_{2v}$ model, depicted in Fig.~\ref{fig:models}(a), we employ a $\pi$-flux Hamiltonian on the square lattice as
\begin{subequations}
\label{eq:model1}
\begin{align}
  \mathcal{H}_0^{C_{2v}} & = -t\sum_{\ve{R}} \sum_{\sigma=1}^{N_\sigma}
                        \hat{a}_{\ve{R},\sigma}^\dag 
                        \Big( \hat{b}_{\ve{R},\sigma}           \e^{-i\tfrac{\pi}{4}}
                             + \hat{b}_{\ve{R}+\ve{e}_-,\sigma}      \e^{ i\tfrac{\pi}{4}} \nonumber \\
        &\quad    + \hat{b}_{\ve{R}+\ve{e}_--\ve{e}_+,\sigma} \e^{-i\tfrac{\pi}{4}}
                             + \hat{b}_{\ve{R}-\ve{e}_+,\sigma}      \e^{ i\tfrac{\pi}{4}}
                               \Big) + \mathrm{H.c.},
\end{align}
where $\hat a$ and $\hat b$ with spin index $\sigma$ are fermion annihilation operators on the two sublattices, $t$ is the hopping parameter, and $N_\sigma=2$ is the number of spin degrees of freedom. $\mathcal{H}_0$ features two inequivalent Dirac points per spin component in the Brillouin zone (BZ). The Ising spins $\hat{s}_{\ve{R}}$ couple, with the sign structure  indicated in Fig.~\ref{fig:models}(a), to the nearest-neighbor fermion hopping terms,
\begin{align}
  \mathcal{H}_\text{Yuk}^{C_{2v}} & = -\xi \sum_{\ve{R}} \sum_{\sigma=1}^{N_\sigma}
                 \hat{s}_{\ve{R}}^z \hat{a}^\dag_{\ve{R},\sigma}
                 \Big( \hat{b}_{\ve{R},\sigma}           \e^{-i\tfrac{\pi}{4}}
                      - \hat{b}_{\ve{R}+\ve{e}_-,\sigma}      \e^{ i\tfrac{\pi}{4}} 
                      \nonumber\\
   &\quad         - \hat{b}_{\ve{R}+\ve{e}_--\ve{e}_+,\sigma} \e^{-i\tfrac{\pi}{4}}
                      + \hat{b}_{\ve{R}-\ve{e}_+,\sigma}      \e^{ i\tfrac{\pi}{4}}
                               \Big) + \mathrm{H.c.},
\end{align}
\end{subequations}
where $\xi$ denotes the coupling strength.
The model has a $C_{2v}$ point group symmetry, composed of reflections, $\hat{T}_{\pm}$ on the  $\ve{e}_\pm = \ve{e}_x \pm \ve{e}_y$  axis.
$\hat{T}_{\pm}$  pins the  Dirac  cones to the $\ve{K}_{\pm} = (\pi/2, \pm \pi/2) $ points in the BZ. 
Aside from the above reflections,  $\pi$ rotations about the $z$ axis are obtained as $\hat{T}_+ \hat{T}_- $. 
Further, the model exhibits an explicit $\text{SU}(N_\sigma)$ spin symmetry that is enlarged to $O(2N_\sigma)$ \cite{suppl}.

\begin{figure}
\includegraphics[width=\linewidth]{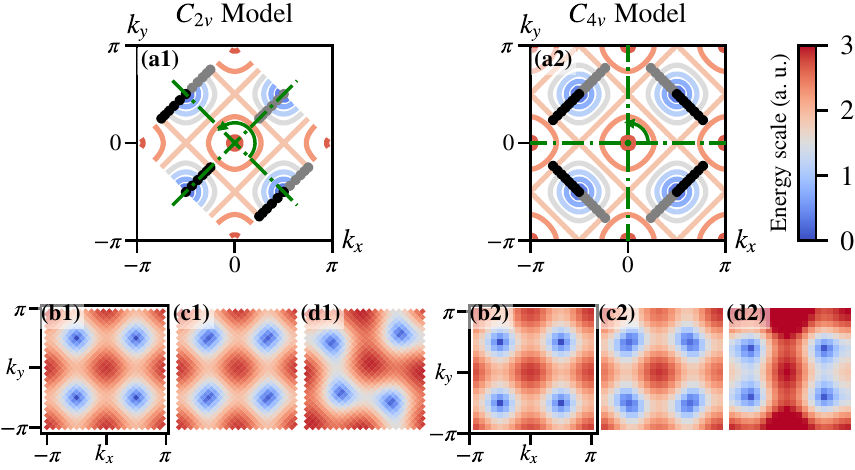}
\caption{
\label{fig:dispersion}
%Fermion dispersion across nematic transition in $C_{2v}$  (left) and $C_{4v}$ (right)  models.
%
{(a)}~Contour plot of the fermion dispersion in the disordered phase from mean-field theory. % The plotted area indicates the Brillouin zone. 
The green lines and arrows indicate the point group symmetries. Black (gray) dots sketch the meandering of the Dirac cones in the nematic phase for $\langle \hat{s}_{\ve{R}}^z \rangle > 0$ ($<0$).
{(b)-(d)}~Fermion dispersion from QMC  at  $L = 20$ for (b) $h = 5.0 > h_c$ featuring isotropic Fermi velocities  
(c) $h \simeq h_c$, $h \approx 3.27$ (left) and $h \approx 3.65$ (right), and (d)  at $h= 1.0  < h_c$  featuring broken point-group symmetries.
Color  scale applies to all plots.
}
\end{figure}

The $C_{4v}$ model   corresponds  to  a bilayer $\pi$-flux model, in which the Ising spins are located on the rungs, Fig.~\ref{fig:models}(b). 
The fermion hopping Hamiltonian is
\begin{subequations}
\label{eq:model2}
\begin{align}
   \mathcal{H}_0^{C_{4v}} & = -t\sum_{\ve{R}} \sum_{\sigma=1}^{N_\sigma}
                        \hat{a}_{\ve{R},\sigma}^\dag 
                        \Big( \hat{b}_{\ve{R}+\ve{e}_x,\sigma}           \e^{i\tfrac{\pi}{4}}
                             + \hat{b}_{\ve{R}-\ve{e}_x,\sigma}      \e^{ i\tfrac{\pi}{4}} 
  \nonumber \\
        &\quad    + \hat{b}_{\ve{R}+\ve{e}_y,\sigma} \e^{-i\tfrac{\pi}{4}}
                             + \hat{b}_{\ve{R}-\ve{e}_y,\sigma}      \e^{ -i\tfrac{\pi}{4}}
                               \Big) + \mathrm{H.c.},
\end{align}
featuring  four Dirac cones per spin component. The Yukawa coupling reads
\begin{align}
  \mathcal{H}_\text{Yuk}^{C_{4v}} & = -\xi \sum_{\ve{R}} \sum_{\sigma=1}^{N_\sigma}
                 i \hat{s}_{\ve{R}}^z \hat{a}^\dag_{\ve{R},\sigma} \hat{b}_{\ve{R},\sigma} + \mathrm{H.c.},
\end{align}
\end{subequations}
amounting to a coupling of the Ising spins to the interlayer fermion current.
The $C_{4v}$ Hamiltonian  commutes with $\hat{T}_{\pi/2}$, corresponding to $\pi/2$ rotation about the $z$ axis.
The model is invariant under  reflections $\hat T_x$ and $\hat T_y$ along the $x$ and $y$ axes, respectively.
Reflections along $\ve{e}_{\pm} = \ve{e}_x \pm \ve{e}_y$, denoted by $\hat{T}_{\pm}$,  can be derived from $\hat{T}_{\pi/2}$, $\hat{T}_x$, and $\hat{T}_{y}$, and therefore also leave the model invariant.
The model hence has a $C_{4v}$ symmetry. 
Particle-hole symmetry,
imposes  $A(\ve{k},\omega)  =  A(-\ve{k} + \ve{Q},-\omega) $, where  $\ve{Q}=(\pi,\pi)$  such that alongside with 
the   $C_{4v}$ symmetry  the Dirac  cones are pinned to the $\pm\ve{K}_{\pm}$ points in the  BZ \cite{suppl}. 

%%%%%%%%%%%%%%%%%%%%%%%%%%%%%%%%%%%%%%%%%%%%%%%%%%%%%%%
\paragraph*{Lattice mean-field theory.}
%%%%%%%%%%%%%%%%%%%%%%%%%%%%%%%%%%%%%%%%%%%%%%%%%%%%%%%
%
The key point of  both models  is that the   point  group and particle-hole symmetries are  tied to the  flipping of  the  Ising spin  degree of freedom.  In  the large-$h$  limit,   the  ground state  has  the full symmetry of the model Hamiltonian  and at the mean-field  level   we can set $ \langle \hat{s}_{\ve{R}}^z \rangle = 0$. In this limit,  the   Dirac  cones are pinned  by   symmetry. 
In the opposite  small-$h$  limit,  the Ising spins order,  $ \langle \hat{s}_{\ve{R}}^z \rangle \neq 0$.  Thereby, the $C_{2v}$ ($C_{4v}$)  symmetry  is reduced to  $\hat{T}_+$ ($C_{2v}$).  At  the  mean-field  level, this induces 
a meandering of  the  Dirac points   in the  BZ, see Fig.~\ref{fig:dispersion}(a),   and  an anisotropy  in the Fermi velocities.  
A  detailed  account of the mean-field  calculations is   presented in the Supplemental Material (SM)~\cite{suppl}, and at this level of  approximation  the  transition  turns  out to be continuous, in agreement with the large-$N$ analysis~\cite{Kim08}.

%%%%%%%%%%%%%%%%%%%%%%%%%%%%%%%%%%%%%%%%%%%%%%%%%%%%%%%
\paragraph*{Continuum field theory.}
%%%%%%%%%%%%%%%%%%%%%%%%%%%%%%%%%%%%%%%%%%%%%%%%%%%%%%%
%
In order to investigate whether the  above  remains true upon the inclusion of order-parameter fluctuations, we derive corresponding continuum field theories, which are amenable to RG analyses.
To leading order in the gradient expansion around the nodal points, we obtain the Euclidean action $S = \int \du^2x \du \tau (\mathcal L_\Psi + \mathcal L_\phi)$ with
\begin{align}
\label{eq:field-theory1}
	\mathcal L_{\Psi}^{C_{2v}} & = \Psi_{\sigma}^\dagger \left(\partial_\tau + \gamma_0\gamma_1 v_\parallel \partial_+ + \gamma_0\gamma_2 v_\perp \partial_- + g \phi \gamma_2 \right) \Psi_\sigma 
\end{align}
for the four-component Dirac spinors $\Psi_{\sigma} \equiv (\hat a_{\sigma}^{+},\hat b_{\sigma}^{+},\hat a_{\sigma}^{-},\hat b_{\sigma}^{-})^\top$ in the $C_{2v}$ model, where $\hat a_{\sigma}^{\pm}$ and $\hat b_{\sigma}^{\pm}$ corresponds to hole excitations near $\ve{K}_\pm$ on the $A$ and $B$ sublattices, respectively, and
\begin{align}
\label{eq:field-theory2}
	\mathcal L_\Psi^{C_{4v}} & = \Psi_{\sigma}^\dagger \bigl[\partial_\tau + \tilde{\gamma}_0 (\tilde{\gamma}_1v_\parallel  \oplus \tilde{\gamma}_2v_\perp) \partial_+
	\nonumber \\&\quad 
	 + \tilde{\gamma}_0(\tilde{\gamma}_2v_\perp \oplus \tilde{\gamma}_1 v_\parallel) \partial_- + g \phi (\tilde{\gamma}_2 \oplus \tilde{\gamma}_2)\bigr] \Psi_{\sigma} 
\end{align}
for the eight-component Dirac spinors $\Psi_{\sigma} \equiv (\hat a_{\sigma}^{++},\hat b_{\sigma}^{++},\hat a_{\sigma}^{-+},\hat b_{\sigma}^{-+},\hat a_{\sigma}^{+-},\hat b_{\sigma}^{+-},\hat a_{\sigma}^{--},\hat b_{\sigma}^{--})^\top$ in the $C_{4v}$ model, where $\hat a_{\sigma}^{+\pm}$ and $\hat b_{\sigma}^{+\pm}$ ($\hat a_{\sigma}^{-\pm}$ and $\hat b_{\sigma}^{-\pm}$) correspond to hole excitations near $\ve{K}_\pm$ ($-\ve{K}_\pm$).
In the above Lagrangians, we have assumed the summation convention over repeated indices, and $\oplus$ denotes the matrix direct sum. The Fermi velocities $v_\parallel$ and $v_\perp$ correspond to the directions parallel and perpendicular to the shift of the Dirac cones in the ordered phase, with $v_\parallel = v_\perp \sim t$ at the UV cutoff scale $\Lambda$.
$\partial_\pm$ denotes the spatial derivative in the direction along $\ve{K}_\pm$.
The two sets of Dirac matrices $\gamma_\mu, \tilde{\gamma}_\mu$ realize four-dimensional representations of the Clifford algebra $\{\gamma_\mu,\gamma_\nu\}=\{\tilde{\gamma}_\mu,\tilde{\gamma}_\nu\}=2\delta_{\mu\nu}$, $\mu,\nu=0,1,2$.
The fermions couple via $g \sim \xi$ to the Ising order-parameter field $\phi$, the dynamics of which is governed by the usual $\phi^4$ Lagrangian,
$\mathcal L_\phi = \frac{1}{2} \phi(r - \partial_\tau^2 - c_+^2 \partial_+^2 - c_-^2 \partial_-^2)\phi + \lambda \phi^4$,
with the tuning parameter $r$, the boson velocities $c_\pm$, and the bosonic self-interaction $\lambda$.

\begin{figure}
\includegraphics[width=\linewidth]{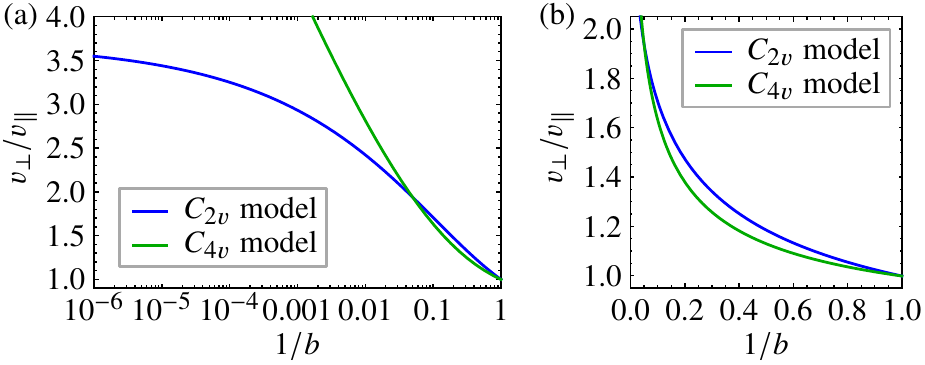}
\caption{Ratio of Fermi velocities $v_\perp/v_\parallel$ as function of RG scale $1/b$ for both models. We assume ultraviolet initial values of $v_\parallel(b=1) = v_\perp(b=1) = 0.25$, and  set $g^2/(v_\parallel v_\perp)(b=1)$ to the value at the respective stable fixed point. (a) Semilogarithmic, (b) linear plots.  Starting at a temperature scale representative of the ultraviolet initial parameters,  one has to cool the system by 2 orders of magnitude to  \textit{start}   observing the differences between both models. 
}
\label{fig:rg_anisotropy_main}
\end{figure}

%%%%%%%%%%%%%%%%%%%%%%%%%%%%%%%%%%%%%%%%%%%%%%%%%%%%%%%
\paragraph*{$\epsilon$ expansion.}
%%%%%%%%%%%%%%%%%%%%%%%%%%%%%%%%%%%%%%%%%%%%%%%%%%%%%%%
The presence of a unique upper critical spatial dimension of three allows an $\epsilon = 3-d$ expansion, with $\epsilon = 1$ corresponding to the physical case.
Because of the lack of Lorentz and continuous spatial rotational symmetries in the low-energy models, it is useful to employ a regularization in the frequency only, which allows us to rescale the different momentum components independently, and evaluate the loop integrals analytically~\cite{suppl}.
Two central properties of nematic quantum phase transitions in Dirac systems are revealed by the one-loop RG analysis:
First, both models admit a stable fixed point featuring anisotropic power laws of the fermion and order parameter correlation functions.
 In the $C_{2v}$ model,  both components of the Fermi velocity remain finite at the stable fixed point with $0 < v_\parallel^* < v_\perp^*$. 
At the critical point, a unique timescale $\tau$ emerges for both fields $\Psi$ and $\phi$~\cite{Meng12,Janssen15}, which scales with the two characteristic length scales $\ell_+$ and $\ell_-$ as $\tau \sim \ell_+^{z_+} \sim \ell_-^{z_-}$, with associated dynamical critical exponents $z_\pm = [1-\frac12 \eta_\phi + \frac12 \eta_\pm]^{-1}$ as $(z_+, z_-) = (1 + 0.3695\epsilon, 1 + 0.1086\epsilon) + \mathcal O(\epsilon^2)$, reflecting the absence of Lorentz and rotational symmetries at criticality.
By contrast, in the $C_{4v}$ model, the fixed point is characterized by a maximal velocity anisotropy with $(v_\parallel^*,v_\perp^*) = (0,1)$ in units of fixed boson velocities $c \equiv c_+ = c_- = 1$.
This result is consistent with the large-$N$ RG analysis in fixed $d=2$~\cite{Huh08}.
The fact that $v_\parallel^*$ vanishes leads to the interesting behavior that the fixed-point couplings $g^2_*$ and $\lambda_*$ are bound to vanish in this case as well. This happens in a way that the ratio $(g^2/v_\parallel)_*$ remains finite, such that the boson anomalous dimensions become $\eta_\phi = \eta_+ = \eta_- = \epsilon$. Importantly, as the fixed-point couplings $g^2_*$ and $\lambda^*$ vanish, we expect the one-loop result for the critical exponents to hold at \emph{all} loop orders in the $C_{4v}$ model.
For the correlation-length exponent, we find $1/\nu = 2 - \epsilon$.
The remaining exponents can then be computed by assuming the usual hyperscaling relations~\cite{HerbutBook}. The susceptibility exponent, for instance, becomes $\gamma = 1$, independent of $\epsilon$. This result is again consistent with the large-$N$ calculation and has previously already been argued to hold exactly~\cite{Huh08}.
We note that the values of the exponents in the $C_{4v}$ model are independent of the number of spinor components, in contrast to the situation in the $C_{2v}$ model, as well as to the usual Gross-Neveu universality classes~\cite{Herbut09a,Janssen14a,Zerf17,Janssen18,YYHe2018,YZLiu2020}.
The unique dynamical critical exponent in the $C_{4v}$ model becomes $z = 1$.
%\st{, reflecting the fact that temporal and spatial correlations of the order parameter exhibit the same power-law decays at criticality}. \fa{Possible space saving} 
 We emphasize, however, that the critical point still does \emph{not} feature emergent Lorentz symmetry~\cite{Roy16} due to the anisotropic fermion spectral function.
The second important property revealed by the RG analysis is that the stable fixed points in both models are approached only extremely slowly as function of RG scale,  Fig.~\ref{fig:rg_anisotropy_main}. This is universally true for the $C_{4v}$ model, in which case $v_\parallel$ corresponds to a marginally irrelevant parameter, hence scaling only logarithmically to zero while other irrelevant operators  rapidly   die  out.  This defines a  quasiuniversal  flow \cite{Nahum15, Nahum19}   in 
which  only  the  velocity anisotropy and not the initial ultraviolet values of other parameters   determine  the  slow  drift of  the   exponents.  The  RG  
suggests that  this  regime emerges  at  scales    $1/b \lesssim 0.05$  (see Ref.~\cite{suppl}), such  that it will  dominate    numerical  as  well as  experimental
realizations of  this  critical phenomena.
 For a reasonable set of ultraviolet starting values and $\epsilon = 1$, we find that the effective correlation-length exponent $1/\nu_\text{eff}$ (anomalous dimension $\eta_\phi^\text{eff}$) approaches one from above (below), with sizable deviations at intermediate RG scales, see Ref.~\cite{suppl} for details. Moreover, we also observe that the initial flows at high energy in the two models resemble each other, despite the fact that they substantially deviate from each other at low energy. This suggests that the flow is generically slow in the $C_{2v}$ model as well.

%%%%%%%%%%%%%%%%%%%%%%%%%%%%%%%%%%%%%%%%%%%%%%%%%%%%%%%
\paragraph*{QMC setup.}
%%%%%%%%%%%%%%%%%%%%%%%%%%%%%%%%%%%%%%%%%%%%%%%%%%%%%%%
For the numerical simulations, we used the ALF program package \cite{ALF_v2} that  provides a general implementation of the  finite-temperature auxiliary field QMC  algorithm \cite{Blankenbecler81,White89,Assaad08_rev}.   To formulate  the path integral,   we  use a Trotter decomposition with time step $\Delta_\tau t=0.1$ and choose a basis where $\hat{s}_{\ve{R}}^{z} | s_{\ve{R}} \rangle =  s_{\ve{R}}  | s_{\ve{R}} \rangle $. The configuration space is that of a $(2+1)$-dimensional Ising model and we use a single-spin-flip update to sample it. As shown in the Supplemental Material~\cite{suppl}  both models are negative-sign-problem free for all values of $N_{\sigma}$  \cite{Li16}. For our simulations, we have used an inverse temperature $\beta = 4 L$  for $L \times L$ lattices, and have checked that this choice of $\beta$ reflects ground-state properties.
For the results shown in the main text, we have fixed the parameters as $J = t = 1$  and  $N_\sigma = 2$. 
In  the  $C_{2v}$ model,   we choose  $\xi = 0.25$,  as larger values of $\xi$ lead to spurious size effects  that could falsely be interpreted as first-order transitions, see Ref.~\cite{YYHe2018} and the Supplemental Material~\cite{suppl} for a detailed discussion.
In the  $C_{4v}$ model, we set $\xi = 1 $.   As shown in the Supplemental Material~\cite{suppl}, other values of $\xi$ and $N_\sigma$ do not alter the continuous nature of the transition.  

%%%%%%%%%%%%%%%%%%%%%%%%%%%%%%%%%%%%%%%%%%%%%%%%%%%%%%%
\paragraph*{QMC results.}
%%%%%%%%%%%%%%%%%%%%%%%%%%%%%%%%%%%%%%%%%%%%%%%%%%%%%%%
We  compute the spin structure factor,
$S({\bf k}) = \sum_{\bf R} e^{i {\bf k \cdot R}} \Braket{\hat{s}^z_{\bf 0} \hat{s}^z_{\bf R}}$,
the spin susceptibility,
$\chi({\bf k}) =  
\sum_{\bf r} e^{i {\bf k \cdot R}} 
\int_{0}^{\beta}\!{\rm d}\tau\,  
\Braket{\hat{s}^z_{\bf R}(\tau) \hat{s}^z_{{\bf 0}}(0)}$
and moments of the total spin $\hat{s}^z = \sum_{\bf R} \hat{s}^z_{\bf R}$ to derive RG-invariant quantities such as the correlation ratio \cite{Kaul15},
\begin{equation}
R_O = 1 - \frac{O(\ve{k}_{\text{min}})}{O(\ve{k}=\ve{0})}
\quad \text{with} \quad O = S, \chi,
\end{equation}
and  the Binder ratio,  $B = \left( 3 - \frac{\langle (\hat{s}^{z})^4\rangle}{\langle (\hat{s}^{z})^2\rangle^2} \right) / 2$. Here, $\ve{k}_{\text{min}}$ corresponds to the longest wavelength on a given finite-size lattice. From the single-particle Green's function, we can extract quantities such as the fermion dispersion relation and Fermi velocities. 

\begin{figure}
\includegraphics[width=\linewidth]{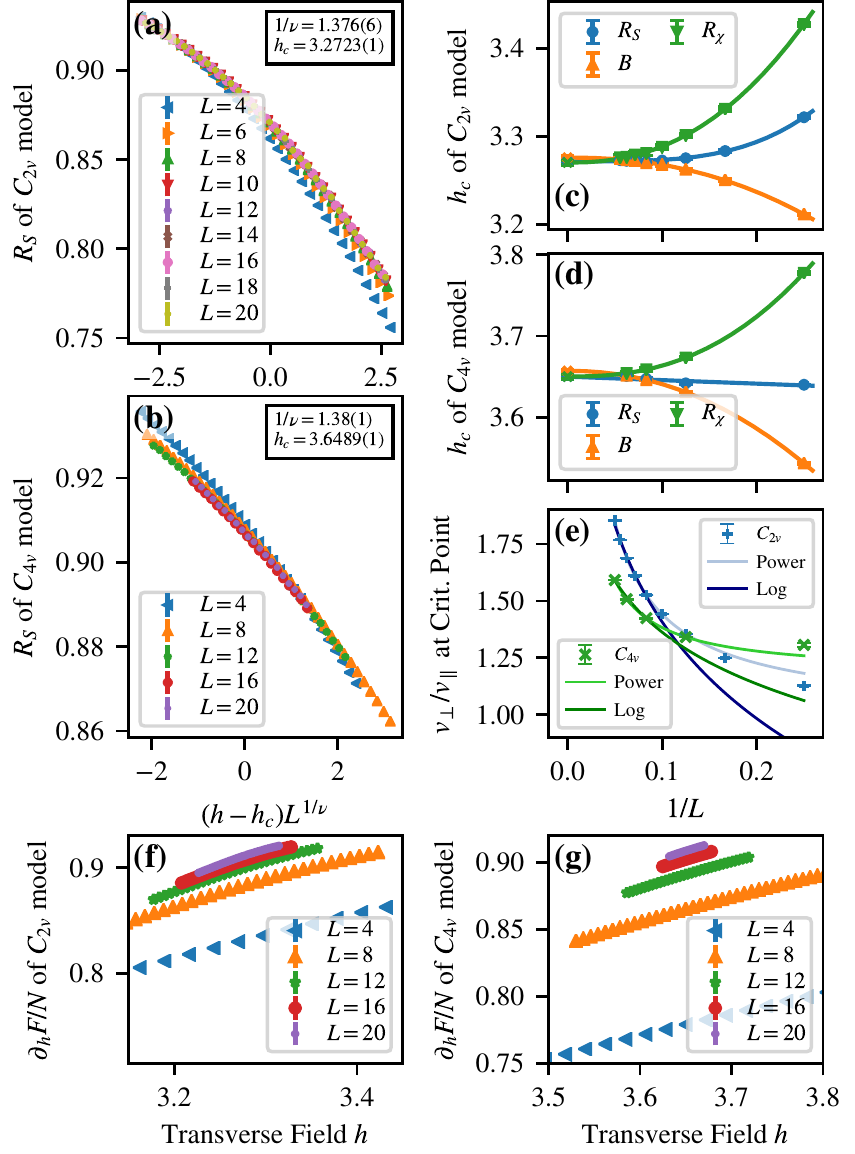}
\caption{
\label{fig:R_hc_an_X}
%Demonstration of continuous nature of transition.
%
(a)~%Correlation ratio of spin structure factor 
$R_S$ as function of $(h-h_\mathrm{c}) L^{1/\nu}$ for the $C_{2v}$ model, revealing data collapse for $L \gtrsim 12$, assuming $1/\nu = 1.376$.
(b)~Same as (a), but for $C_{4v}$ model, assuming $1/\nu = 1.38$.
(c)~Crossing points of different RG-invariant quantities as function of $1/L$ with $\Delta L = 2$ in $C_{2v}$ model, indicating a unique critical point $h_\mathrm{c} = 3.27$ for $L \to \infty$.
(d)~Same as (c), but for $C_{4v}$ model and $\Delta L = 4$, extrapolating to $h_\mathrm{c} = 3.65$
(e)~Ratio of Fermi velocities $v_\perp/v_\parallel$ as function of $1/L$ at $h_\mathrm{c}$, revealing that the velocity anisotropy increases with increasing system size. The solid lines show power law fits for $L \geq 8$ and logarithmic fits for $L \geq 12$.
%Fits for $C_{2v}$ model:
%  $v_\perp/v_\parallel = 0.99(3) + 0.05(1) L^{0.92(5)}$, 
%  $v_\perp/v_\parallel = 0.00(8) + 0.61(3) \log(L)$.
%Fits for $C_{4v}$ model:
%  $v_\perp/v_\parallel = 1.179(1) + 0.0197(3) L^{1.016(5)}$,
%  $v_\perp/v_\parallel = 0.61(5) + 0.33(2) \log(L)$
%
(f),(g)~Derivative of free energy as function of $h$, exhibiting no discontinuities.
}
\end{figure}

At a quantum critical point, RG-invariant quantities follow the form $f[ L^z/\beta, (h-h_c)L^{1/\nu},  L^{-\Delta z  }, L^{-\omega} ]$ \cite{suppl}.   Here  we  have taken into account  the  possibility of two characteristic length  scales:    $\Delta z  = 1 - z_-/z_+$.  
Since  our  temperature is  representative of the ground state, we can neglect the dependence on $L^{z}/\beta$.  Up to corrections to scaling, $\omega$, and  the possibility of  $z_- \neq z_+$,  which   would result in another  correction to  scaling term,  the  data for different lattice sizes  cross at the critical field $h_\mathrm{c}$. Figs.~\ref{fig:R_hc_an_X}(c,d) show the crossing points between $L$ and $L+\Delta L$ lattices,  with $\Delta L=2$ $(4)$ for the $C_{2v}$ ($C_{4v}$) model.  As  apparent,  we  obtain  consistent  results for  $h_\mathrm{c}$   when  considering different RG-invariant  quantities.  We   estimate  the correlation-length exponents $1/\nu$ by data collapse for the two models in Figs.~\ref{fig:R_hc_an_X}(a,b). Considering values of  $L  \geq L_{\text{min}}  =12$  we obtain  $1/\nu = 1.376(6)$ [$1/\nu = 1.38(1)$] for the $C_{2v}$ ($C_{4v}$) model. These values are in the  ballpark of the  $\epsilon$-expansion  results  in the  quasiuniversal  regime  \cite{suppl}.
The data for various values of $L_{\text{min}}$ are given in the Supplemental Material~\cite{suppl}, and stand in agreement with the above values.     Although  seemingly    converged,  
the fact that the velocity anisotropy is expected to flow extremely slowly suggest that the exponents are subject to considerable size effects, see below.
Figures \ref{fig:R_hc_an_X}(f,g) show the derivative of the free energy with respect to the tuning parameter, $\partial F/\partial h$, confirming the absence of any discontinuity at $h_\mathrm{c}$.
The impact of critical fluctuations on the fermion spectrum is displayed in Figs.~\ref{fig:dispersion}(b,c).
In the disordered phase,  Fig.~\ref{fig:dispersion}(b), the dispersion relation exhibits rotational symmetry around the Dirac points. On the other hand, at criticality, Fig.~\ref{fig:dispersion}(c), the dispersion relation suggests  a velocity anisotropy, $v_{\parallel} < v_{\perp}$  at the Dirac point. Figure~\ref{fig:R_hc_an_X}(e) demonstrates that this anisotropy grows as a function of system size, in qualitative agreement with the RG predictions. 
  Although our system sizes  are  too small to detect convergence or   divergence of the velocity ratio, we find it reassuring that its dependence on system size qualitatively resembles the scale dependence predicted from the integrated RG flow, cf.\ Fig.~\ref{fig:R_hc_an_X}(e) with Fig.~\ref{fig:rg_anisotropy_main}.

%%%%%%%%%%%%%%%%%%%%%%%%%%%%%%%%%%%%%%%%%%%%%%%%%%%%%%%
\paragraph*{Summary.}     
%%%%%%%%%%%%%%%%%%%%%%%%%%%%%%%%%%%%%%%%%%%%%%%%%%%%%%%

Both the $\epsilon$-expansion analysis and the QMC simulations show that our two symmetry  distinct models of Dirac fermions support continuous nematic transitions. In both cases, the key  feature of the quantum critical point is a velocity anisotropy that is best seen in the  QMC data of Fig.~\ref{fig:dispersion}(c). For the $C_{4v}$ model, the $\epsilon$-expansion shows that it diverges logarithmically with system size, in agreement with previous large-$N$ results~\cite{Huh08}. 
This law is supported by finite-size analysis based on QMC data up to linear system size $L=20$, which is close to the upper bound allowed by current computational approaches.
Since the effective exponents flow with the velocity anisotropy, we foresee that lattice sizes beyond the reach of our numerical approach and  experiments at ultralow temperatures will be required  to obtain converged values. The  QMC  data  captures  a  quasiuniversal  regime \cite{Nahum15, Nahum19},  in  which   irrelevant  operators  aside from the   velocity anisotropy   die  out. In fact, the  RG prediction  for exponents in this intermediate-energy regime is  roughly consistent with the finite-size QMC measurements, Fig.~\FigRGvelocities(c) of Ref.~\cite{suppl}.  Furthermore,  for  a reasonable set of starting values, the integrated RG flows of the two models are initially very similar and deviate from each other only at very low energy scales. A similar behavior of the two models is also observed in the QMC data.

An advantage  of our  models is that the   Dirac  points  are   pinned by  symmetry, such that QMC  approaches  that take  momentum-space  patches  around  these points into account  \cite{Liu19} represent an attractive  direction for future work.  
  Our  models  equally  allow 
for   large-$N$ generalizations,    such  that QMC and  analytical  large-$N$ calculations can  be  compared as a function of increasing $N$.    Finally,  we  can make  contact to  nematic transitions in $(2+1)$-dimensional Fermi liquids \cite{Oganesyan01,Schattner15},  since   our models   do not suffer  from  the negative-sign problem under doping.

%%%%%%%%%%%%%%%%%%%%%%%%%%%%%%%%%%%%%%%%%%%%%%%%%%%%%%%
\begin{acknowledgments}
\paragraph*{Acknowledgements.}
This research has been funded by the Deutsche Forschungsgemeinschaft (DFG) through the W\"urzburg-Dresden Cluster of Excellence on Complexity and Topology in Quantum Matter \textit{ct.qmat} -- Project No.~390858490 (L.\,J., M.\,V., F.\,F.\,A.),  the SFB~1170 on Topological and Correlated Electronics at Surfaces and Interfaces -- Project No.~258499086 (J.\,S., F.\,F.\,A.), the SFB 1143 on Correlated Magnetism -- Project No.~247310070 (L.\,J., M.\,V.), the Emmy Noether program -- Project No.~411750675 (L.\,J.),  the National Science and Engineering Council (NSERC) of Canada (I.\,F.\,H.), and Grant No.~AS~120/14-1 (F.\,F.\,A.).
Z.\,Y.\,M. acknowledges the Research Grants Council of Hong Kong China (Grants No.~17303019, No.~17301420, No.~17301721 and No.~AoE/P-701/20) and the Strategic Priority Research Program of the Chinese Academy of Sciences (Grant No.~XDB33000000), the K.~C. Wong Education Foundation (Grant No.~GJTD-2020-01) and the Seed Funding Quantum-Inspired explainable-AI at the HKU-TCL Joint Research Centre for Artificial Intelligence.
We are grateful to the Gauss Centre for Supercomputing e.V.\ (www.gauss-centre.eu) for providing computing time on the GCS Supercomputer SUPERMUC-NG at Leibniz Supercomputing Centre (www.lrz.de).
\end{acknowledgments}
%%%%%%%%%%%%%%%%%%%%%%%%%%%%%%%%%%%%%%%%%%%%%%%%%%%%%%%

\bibliographystyle{longapsrev4-2}
%longapsrev4-2.bst hand-edited version by Lukas Janssen of apsrev4-2.bst

\clearpage
\onecolumngrid
\begin{center}
\Large{\bf Supplemental Material: Nematic quantum criticality in Dirac systems}
\end{center}
% !TEX root = master_suppl.tex

\tableofcontents

%%%%%%%%%%%%%%%%%%%%%%%%%%%%%%%%%%%%%%%%%%%%%%%%%%%%%%%
\section{Absence of negative sign problem }
%%%%%%%%%%%%%%%%%%%%%%%%%%%%%%%%%%%%%%%%%%%%%%%%%%%%%%%
Here we use the Majorana representation to demonstrate, using the results of Ref.~\cite{Li16}, the absence of negative sign problem for all values of $N_{\sigma}$. Both models have $SU(N_{\sigma})$ symmetry. Since the Ising spins couple symmetrically to the  fermion  spins, $SU(N_{\sigma})$ symmetry is present for all Ising spin configurations. Thereby, the fermion determinant of the $SU(N_{\sigma})$ model  corresponds to that of the $U(1)$ model 
($N_{\sigma} =1$) elevated to the power $N_{\sigma}$. It hence suffices to demonstrate the absence of negative sign problem at $N_{\sigma}  =1 $.   In this section, we will hence omit the spin index.  Additionally  we include a  chemical potential term $\mathcal{H}_{\mu}$, to show that there is also no sign problem under doping  for  even values  of $N_{\sigma}$.

%%%%%%%%%%%%%%%%%%%%%%%%%%%%%%%%%%%%%%%%%%%%%%%%%%%%%%%
\subsection{The \texorpdfstring{$C_{2v}$}{C2v} model} 
Consider the  canonical transformation, 
\begin{equation}
\begin{pmatrix}
   \hat{a}_{\ve{k}}  \\
   \hat{b}_{\ve{k}} 
\end{pmatrix} 
\rightarrow 
 \begin{pmatrix}
   1   & 0 \\
   0  & e^{-i 3 \pi / 4 }
\end{pmatrix} 
\begin{pmatrix}
   \hat{a}_{\ve{k} - \Delta \ve{K}} \\
   \hat{b}_{\ve{k} - \Delta \ve{K}} 
\end{pmatrix} 
\end{equation}
with
\begin{equation}
	\hat{b}_{\ve{R}}   = \frac{1}{\sqrt{N}} \sum_{\ve{k}  \in BZ }  e^{i \ve{k} \cdot \ve{R}} \hat{b}_{\ve{k}}
\end{equation}
and   $ \Delta \ve{K} =    \frac{1}{4}  \left( \ve{b}_- -  \ve{b}_+ \right)  $.  Here ,  $  \ve{e}_i \cdot \ve{b}_j  =  2 \pi  \delta_{i,j}  $.   This  canonical transformation renders the Hamiltonian real: the $\pi$-flux, is realized by changing  the sign of the  intra  unit-cell hopping  with respect to the other   hoppings.  More precisely     after the transformation, the   Fermionic part of the Hamiltonian takes the form:
\begin{subequations}
  \label{eq:ham_c2v_real}
\begin{align}
  &\mathcal{H}_0^{C_{2v}} = -t\sum_{\ve{R}} 
                        \hat{a}_{\ve{R}}^\dag
                        \Big(  - \hat{b}_{\ve{R}}
                             + \hat{b}_{\ve{R}+\ve{e}_-}
                          + \hat{b}_{\ve{R}+\ve{e}_--\ve{e}_+}s
                             + \hat{b}_{\ve{R}-\ve{e}_+}
                               \Big) + h.c.
 \\
  &\mathcal{H}_\text{Yuk}^{C_{2v}} = -\xi \sum_{\ve{R}}
                 s_{\ve{R}} \hat{a}^\dag_{\ve{R}}
                 \Big( - \hat{b}_{\ve{R}}
                      - \hat{b}_{\ve{R}+\ve{e}_-  }
                       - \hat{b}_{\ve{R}+\ve{e}_--\ve{e}_+}
                      + \hat{b}_{\ve{R}-\ve{e}_+}
                               \Big) + h.c.
 \\
  &\mathcal{H}_\mu^{C_{2v}} = \mu \sum_{\ve{R}} 
                 \Big( \hat{a}^\dag_{\ve{R}} \hat{a}_{\ve{R}} 
                     + \hat{b}^\dag_{\ve{R}} \hat{b}_{\ve{R}}
                               \Big)
\end{align}
\end{subequations}
In the above,  we have considered an  arbitrary set of Ising spins $s_{\ve{R}} = \pm 1 $.

Since equation \eqref{eq:ham_c2v_real} is real, the corresponding fermion determinant for $N_\sigma = 1$ is also real and therefore positive for even $N_\sigma$. 

For the sign to remain positive with odd $N_\sigma$, we have to dismiss $H_\mu$ and introduce Majorana fermions:
\begin{align}
	\hat{a}^{\dagger}_{\ve{R}}     &=  \tfrac{1}{2} \left( \hat{\gamma}_{\ve{R},1}  -  i \hat{\ve{\gamma}}_{\ve{R},2}  \right) 
	&
	\hat{b}^{\dagger}_{\ve{R}}     =  \tfrac{1}{2} \left( i \hat{\eta}_{\ve{R},1}  + \hat{\eta}_{\ve{R},2}  \right). 
\end{align}
In the Majorana basis, the Fermionic part of the Hamiltonian reads: 
\begin{subequations}
\begin{align}
\begin{split}
  &\mathcal{H}_0^{C_{2v}} = \frac{it}{2}\sum_{\ve{R}}
                        \hat{\ve{\gamma}}^{\rm T}_{\ve{R}}
                        \Big(  - \hat{\ve{\eta}}_{\ve{R}}
                             + \hat{\ve{\eta}}_{\ve{R}+\ve{e}_-}       + \hat{\ve{\eta}}_{\ve{R}+\ve{e}_--\ve{e}_+}
                             + \hat{\ve{\eta}}_{\ve{R}-\ve{e}_+}
                               \Big)
\end{split}
 \\
\begin{split}
  &\mathcal{H}_\text{Yuk}^{C_{2v}} = \frac{i \xi}{2} \sum_{\ve{R}}
                 s_{\ve{R}} \hat{\ve{\gamma}}^{\rm T}_{\ve{R}}
                 \Big( - \hat{\ve{\eta}}_{\ve{R},}
                      - \hat{\ve{\eta}}_{\ve{R}+\ve{e}_-}
                              - \hat{\ve{\eta}}_{\ve{R}+\ve{e}_--\ve{e}_+}
                      + \hat{\ve{\eta}}_{\ve{R}-\ve{e}_+}
                               \Big)
\end{split}
\end{align}
\end{subequations}
In the above,  $ \ve{\hat{\gamma}}^{T}_{\ve{R}}   = (\hat{\gamma}_{\ve{R},1} , \hat{\gamma}_{\ve{R},2} ) $  and a similar form holds for $\ve{\hat{\eta}}_{\ve{R}}$.  The fact that the Hamiltonian is diagonal in the Majorana  index   shows that it has a higher  $O(2N_{\sigma})$     as opposed to the apparent   $SU(N_{\sigma}) $ one  in  the fermion   representation.   It also has for consequence that  for the  $N_\sigma = 1$ case, the fermion determinant is nothing but the square of a Pfaffian  that takes  real values. Hence the negative sign problem is absent \cite{Li15,Huffman14}.      We close this subsection  by making contact with the work of  Ref.~\cite{Li16}. 
Let  $\ve{\mu}$  be a vector  of  Pauli matrices   acting on the Majorana index.  Adopting the notation of  Ref.~\cite{Li16},  we can define: 
\begin{eqnarray}
	& & \hat{T}^{-}_{1} \alpha  \ve{\hat{\gamma}}_{\ve{R}}  \left( \hat{T}^{-}_{1}  \right)^{-1}   = \overline{\alpha} i  \ve{\mu}_y  \ve{\hat{\gamma}}_{\ve{R}}  \nonumber   \\
	&& \hat{T}^{-}_{1} \alpha  \ve{\hat{\eta}}_{\ve{R}}  \left( \hat{T}^{-}_{1}  \right)^{-1}   = - \overline{\alpha} i \ve{\mu}_y  \ve{\hat{\eta}}_{\ve{R}} 
\end{eqnarray} 
and 
\begin{eqnarray}
	& & \hat{T}^{+}_{2} \alpha  \ve{\hat{\gamma}}_{\ve{R}}  \left( \hat{T}^{+}_{2}  \right)^{-1}   = \overline{\alpha}   \ve{\mu}_x  \ve{\hat{\gamma}}_{\ve{R}}  \nonumber   \\
	&& \hat{T}^{+}_{2} \alpha  \ve{\hat{\eta}}_{\ve{R}}  \left( \hat{T}^{+}_{2}  \right)^{-1}   = - \overline{\alpha}  \ve{\mu}_x  \ve{\hat{\eta}}_{\ve{R}}. 
\end{eqnarray}
Since  $\left[ \hat{T}^{+}_{2} , \mathcal{H}^{C_{2v}}  \right] = \left[ \hat{T}^{-}_{1} , \mathcal{H}^{C_{2v}}  \right]  =0 $    and $\hat{T}^{-}_{1} $ and $\hat{T}^{+}_{2} $  anti-commute,  our Hamiltonian belongs to the  so-called Majorana class, and is hence free of the negative sign problem.

%%%%%%%%%%%%%%%%%%%%%%%%%%%%%%%%%%%%%%%%%%%%%%%%%%%%%%%
\subsection{The \texorpdfstring{$C_{4v}$}{C4v} model} 

Consider  the spinor $ \hat{\ve{c}}^{\dagger}_{\ve{R}} = \left( \hat{a}^{\dagger}_{\ve{R}}, \hat{b}^{\dagger}_{\ve{R}} \right)$. With this notation, the fermionic part of the $C_{4v}$ model takes the form: 
\begin{subequations}
\begin{eqnarray}
   \mathcal{H}_0^{C_{4v}} & = & -t \sum_{\ve{R} \in A, \delta = \pm} \left(
                        \hat{\ve{c}}_{\ve{R}}^\dag   \,   \ve{e}_{+} \cdot {\ve{\tau}} \,  \hat{\ve{c}}_{\ve{R} + \delta \ve{e}_x}   
                        +  \hat{\ve{c}}_{\ve{R}}^\dag   \,   \ve{e}_{-} \cdot {\ve{\tau}} \,  \hat{\ve{c}}_{\ve{R} + \delta \ve{e}_y}   + h.c. \right)
 \\
  \mathcal{H}_\text{Yuk}^{C_{4v}}  & = & \xi \sum_{\ve{R}} 
                  s_{\ve{R}} \hat{\ve{c}}^\dag_{\ve{R}}   \ve{e}_y \cdot \ve{\tau}  \hat{\ve{c}}_{\ve{R}} 
\\
  \mathcal{H}_\mu^{C_{4v}} & = & \mu \sum_{\ve{R}} 
                 \hat{\ve{c}}^\dag_{\ve{R}} \hat{\ve{c}}_{\ve{R}}
\end{eqnarray}
\end{subequations}
In the above, $\ve{\tau}$  denotes a vector of Pauli  matrices that act  on the   \textit{orbital}    space,   $ \ve{e}_{\pm} = \frac{1}{\sqrt{2}} \left(  \ve{e}_x \pm \ve{e}_y\right) $,  and   $\ve{R} \in A $ denotes  the sum of the A sub-lattice, $(-1)^{R_x+R_y }  =1$.   We  have also  considered an arbitrary  set of Ising spins  $ s_{\ve{R}}  = \pm 1 $. 
Consider the relation, 
\begin{equation} 
	U^{\dagger} (\ve{e},\theta) \ve{\tau} U (\ve{e},\theta)  =  R(\ve{e},\theta) \ve{\tau}
\end{equation}
with $ U (\ve{e},\theta) \ve{\tau} =  e^{- i \theta \ve{e} \cdot \ve{\tau}/2}  $ an  SU(2)   rotation  of angle $\theta $  around axis $\ve{e}$ ($ |\ve{e}| = 1$ ) and $R(\ve{e},\theta)  $ an SO(3)  with same angle and  axis.   We can hence carry out a  canonical transformation, 
\begin{equation}
   \hat{d}_{\ve{R}}   =   U  \hat{c}_{\ve{R}}, 
\end{equation}
that  rotates $  \ve{e}_{+}    \rightarrow \ve{e}_z $,  $  \ve{e}_{-}    \rightarrow \ve{e}_x $,   and  $  \ve{e}_{y}    \rightarrow  - \frac{1}{\sqrt{2}} \left( \ve{e}_x - \ve{e}_z  \right) $  by combing a $\pi/4$  rotation around the z-axis  and   subsequently  a $\pi/2$  rotation around the x-axis.      After  this canonical transformation, the Hamiltonian is real, and takes the form: 
\begin{subequations}
\begin{eqnarray}
   \mathcal{H}_0^{C_{4v}} & = & -t \sum_{\ve{R} \in A, \delta = \pm} \left(
                        \hat{\ve{d}}_{\ve{R}}^\dag   \,    {\tau}_z \,  \hat{\ve{d}}_{\ve{R} + \delta \ve{e}_x}   +
                         \hat{\ve{d}}_{\ve{R}}^\dag   \,   {\tau}_x \,  \hat{\ve{d}}_{\ve{R} + \delta \ve{e}_y}   + h.c. \right)
%\nonumber
\\
  \mathcal{H}_\text{Yuk}^{C_{4v}}  & = & - \frac{\xi}{\sqrt{2}} \sum_{\ve{R}} 
                  s_{\ve{R}} \hat{\ve{d}}^\dag_{\ve{R}}   \left( \tau_x - \tau_z \right)   \hat{\ve{d}}_{\ve{R}} 
\\
%\nonumber
  \mathcal{H}_\mu^{C_{4v}} & = & \mu \sum_{\ve{R}} 
                 \hat{\ve{d}}^\dag_{\ve{R}} \hat{\ve{d}}_{\ve{R}}
\end{eqnarray}
\end{subequations}

We can now express the model in terms of  Majorana fermions  and choose the following representation for 
$(-1)^{R_x+R_y } =1$, 
\begin{eqnarray}
	\hat{\ve{d}}^{\dagger}_{\ve{R}}     =  \frac{1}{2} \left( \hat{\ve{\gamma}}_{\ve{R},1}  -  i \hat{\ve{\gamma}}_{\ve{R},2}  \right) 
\end{eqnarray}
and for  $(-1)^{R_x+R_y } =-1$,
\begin{eqnarray}
	\hat{\ve{d}}^{\dagger}_{\ve{R}}     =  \frac{1}{2} \left( i \hat{\ve{\gamma}}_{\ve{R},1}  + \hat{\ve{\gamma}}_{\ve{R},2}  \right).
\end{eqnarray}
   Let $ \ve{\mu}  $ be a vector  of Pauli spin matrices that acts on the Majorana index.      With this choice, the Hamiltonian then takes the form:
\begin{eqnarray}
   \mathcal{H}_0^{C_{4v}} & = & \frac{i t}{2} \sum_{\ve{R} \in A, \delta = \pm} \left(
                        \hat{\ve{\gamma}}_{\ve{R}}^{\rm T}  \,    {\ve{\tau}_z} \,  \hat{\ve{\gamma}}_{\ve{R} + \delta \ve{e}_x}   +
                         \hat{\ve{\gamma}}_{\ve{R}}^{T}   \,   {\ve{\tau}}_x \,  \hat{\ve{\gamma}}_{\ve{R} + \delta \ve{e}_y}   \right)
 \nonumber \\
  \mathcal{H}_\text{Yuk}^{C_{4v}}  & = & \frac{\xi}{4\sqrt{2}} \sum_{\ve{R}} 
                  s_{\ve{R}} \hat{\ve{\gamma}}^{\rm T}_{\ve{R}}   \left(  \ve{\tau}_x  \ve{\mu}_{y}- \ve{\tau}_z  \ve{\mu}_y  \right)   \hat{\ve{\gamma}}_{\ve{R}} 
\\
  \mathcal{H}_\mu^{C_{4v}} & = & \mu \sum_{\ve{R}} 
                 \left( 2 - \hat{\ve{\gamma}}^{\rm T}_{\ve{R}} \mu_y \hat{\ve{\gamma}}_{\ve{R}}
                               \right)
\end{eqnarray}
 
Using the notation of Ref.~\cite{Li16}    we define: 
\begin{equation}
	\hat{T}^{-}_{1} \alpha  \hat{\ve{\gamma}}_{\ve{R}}  \left( \hat{T}^{-}_{1}  \right)^{-1}   = \overline{\alpha} i \ve{\tau}_y  \ve{\mu}_x  \hat{\ve{\gamma}}_{\ve{R}} 
\end{equation}
and 
\begin{equation}
	\hat{T}^{-}_{2} \alpha  \hat{\ve{\gamma}}_{\ve{R}}  \left( \hat{T}^{-}_{2}  \right)^{-1}   = \overline{\alpha} i \ve{\tau}_y  \ve{\mu}_z  \hat{\ve{\gamma}}_{\ve{R}}
\end{equation}
that satisfy 
\begin{equation}
	\left[ \mathcal{H}^{C_{4v}}, \hat{T}^{-}_{1} \right]  = \left[ \mathcal{H}^{C_{4v}}, \hat{T}^{-}_{2}  \right] = 0.
\end{equation}
Both above  symmetries square to (-1)  and anti-commute   with each other.  This hence places us in the Kramers class, see Ref.~\cite{Li16,Wu04},   and no  negative sign problem occurs.

%%%%%%%%%%%%%%%%%%%%%%%%%%%%%%%%%%%%%%%%%%%%%%%%%%%%%%%
\section{Fourier transformed models}
%%%%%%%%%%%%%%%%%%%%%%%%%%%%%%%%%%%%%%%%%%%%%%%%%%%%%%%

We define the Fourier transformation as:
\begin{subequations}
\label{eq:fouriertraf}
\begin{align}
\begin{pmatrix}
\hat{a}_{\ve{R},\sigma}^\dag
\\
\hat{b}_{\ve{R},\sigma}^\dag
\end{pmatrix}
&=
\frac{1}{\sqrt{N}} \sum_{\ve{k}} \e^{-i\ve{k}\ve{R}}
\begin{pmatrix}
\hat{a}_{\ve{k},\sigma}^\dag
\\
\hat{b}_{\ve{k},\sigma}^\dag
\end{pmatrix}
  \\
 s^z_{\ve{R}} &= \frac{1}{N}\sum_{\ve{q}} s^z_{\ve{q}} \e^{i\ve{q}\ve{R}}
\end{align}
\end{subequations}
with this definition, both models take the form
\begin{subequations}
\label{eq:ham-fourier}
\begin{align}
  \mc{H} &= -\sum_{\sigma=1}^{N_\sigma} \sum_{\ve{k}} \hat{a}_{\ve{k},\sigma}^\dag \Bigg( 
         \hat{b}_{\ve{k},\sigma} Z_0(\ve{k}) 
      + \frac{\xi}{N}\sum_{\ve{q}}  \hat{b}_{\ve{k}-\ve{q},\sigma} s^z_{\ve{q}} Z_\text{Yuk}(\ve{k}) 
      \Bigg) + h.c.
     + H_\text{Ising}
\end{align}
with
\begin{align}
  Z_0(\ve{k}) &= \left\{
  \begin{array}{ll}
  2 t \left( \e^{i\frac{\pi}{4}} \cos k_x  + \e^{-i\frac{\pi}{4}} \cos k_y \right) \e^{-i\ve{k}_y}
    & C_{2v} \text{ model}
  \\
  2 t \left( \e^{i\frac{\pi}{4}} \cos k_x  + \e^{-i\frac{\pi}{4}} \cos k_y \right)
    & C_{4v} \text{ model}
  \end{array}
  \right.
  \\
  Z_\text{Yuk}(\ve{k}) &= \left\{
  \begin{array}{ll}
  i 2 \left( -\e^{i\frac{\pi}{4}} \sin k_x  + \e^{-i\frac{\pi}{4}} \sin k_y \right) \e^{-i\ve{k}_y} 
    & C_{2v} \text{ model}
  \\
  i & C_{4v} \text{ model.}
  \end{array}
  \right. 
\end{align}
\end{subequations}

%%%%%%%%%%%%%%%%%%%%%%%%%%%%%%%%%%%%%%%%%%%%%%%%%%%%%%%
\section{Symmetries}
%%%%%%%%%%%%%%%%%%%%%%%%%%%%%%%%%%%%%%%%%%%%%%%%%%%%%%%

\subsection{The \texorpdfstring{$C_{2v}$}{C2v} model}
The First model has a $C_{2v}$ symmetry, consisting of two reflections: $T_+$ and $T_-$ on $\ve{e}_\pm = \ve{e}_x \pm \ve{e}_y$, the $\pi$ rotation needed by the point group can be obtained as $T_\pi = T_+ \cdot T_-$. $T_-$ invariance hinges on the Z$_2$ Ising symmetry, $s^z \rightarrow -s^z$, and is therefore broken in the ordered phase.

The $C_{2v}$ symmetry pins the Dirac points (up to a gauge choice) to $(\pi/2, \pm \pi/2)$, while in the ordered phase, meandering parallel to $\ve{e}_+$ is possible.

To  show  this  symmetry, we   expand  the  momentum- and real-space vectors as:
$\ve{k} = k_+\ve{e}_+ + k_-\ve{e}_-$ and $\ve{R} = R_+\ve{e}_+ + R_-\ve{e}_-$.

The first reflection $T_+$   reads:

\begin{align}
\hat{T}_+^{-1} 
\begin{pmatrix}
 \hat{a}^\dag_{\ve{k},\sigma} \\
 \hat{b}^\dag_{\ve{k},\sigma} 
\end{pmatrix} \hat{T}_+ &= 
\begin{pmatrix}
 \hat{b}^\dag_{(k_+,-k_-),\sigma} \phantom{\e^{-i k_+}} \\
 \hat{a}^\dag_{(k_+,-k_-),\sigma}  \e^{-i k_+}
\end{pmatrix}
\label{eq:T1}
\\
\hat{T}_+^{-1} s^z_{\ve{q}} \hat{T}_+ &= s^z_{(q_+,-q_-)}
\end{align}

Inserting the above  in  Eq.~\eqref{eq:ham-fourier}, we obtain:
\begin{align*}
&\hat{T}_+^{-1} \mc{H}^{C_{2v}} \hat{T}_+ = \\
&=
 -\sum_{\ve{k}} \e^{i k_+} \hat{b}_{(k_+,-k_-),\sigma}^\dag \Bigg( 
         \hat{a}_{(k_+,-k_-),\sigma} Z_0^{C_{2v}}(\ve{k})
       + \xi\sum_{\ve{q}}  \hat{a}_{(k_+-q_+,-k_-+q_-),\sigma} \hat{s}^z_{(q_+,-q_-)} Z_\text{Yuk}^{C_{2v}}(\ve{k}) 
      \Bigg) + h.c.
     + H_\text{Ising}
\\
&= -\sum_{\ve{k}} \e^{-i k_+} \hat{a}_{(k_+,k_-),\sigma}^\dag \Bigg( 
         \hat{b}_{(k_+,k_-),\sigma} \bar{Z}_0^{C_{2v}}(k_+,-k_-)
       + \xi\sum_{\ve{q}}  \hat{b}_{(k_+-q_+,k_--q_-),\sigma} \hat{s}^z_{(q_+,q_-)} \bar{Z}_\text{Yuk}^{C_{2v}}(k_+,-k_-) 
      \Bigg) + h.c.
     + H_\text{Ising}
\\
& \qquad \left[
\begin{array}{rl}
\bar{Z}_0^{C_{2v}}(k_+,-k_-)        &= Z_0^{C_{2v}}(\ve{k}) \e^{i k_+} \\
\bar{Z}_\text{Yuk}^{C_{2v}}(k_+,-k_-) &= Z_\text{Yuk}^{C_{2v}}(\ve{k}) \e^{i k_+}
\end{array}
\right]
\\
&= -\sum_{\ve{k}} \hat{a}_{(k_+,k_-),\sigma}^\dag \Bigg( 
         \hat{b}_{(k_+,k_-),\sigma} Z_0^{C_{2v}}(\ve{k})
       + \xi\sum_{\ve{q}} \hat{b}_{\ve{k}-\ve{q},\sigma} \hat{s}^z_{\ve{q}} Z_\text{Yuk}^{C_{2v}}(\ve{k}) 
      \Bigg) + h.c.
     + H_\text{Ising}
\\
&=\mc{H}^{C_{2v}}
\end{align*}

In real space, $\hat{T}_+$ translates to:
\begin{align}
\hat{T}_+^{-1} 
\begin{pmatrix}
\hat{a}^\dag_{\ve{R},\sigma}
 \\ 
\hat{b}^\dag_{\ve{R},\sigma}
\end{pmatrix} 
\hat{T}_+ 
&= 
\begin{pmatrix}
 \hat{b}^\dag_{(R_+  ,-R_-),\sigma}
 \\
 \hat{a}^\dag_{(R_++1,-R_-),\sigma}
\end{pmatrix}
\\
\hat{T}_+^{-1} s^z_{\ve{R}} \hat{T}_+ &= s^z_{(R_+,-R_-)}
\end{align}

The second reflection $T_-$ can be expressed as:

\begin{align}
\hat{T}_-^{-1} 
\begin{pmatrix}
 \hat{a}^\dag_{\ve{k},\sigma} \\
 \hat{b}^\dag_{\ve{k},\sigma} 
\end{pmatrix} \hat{T}_- &= 
\begin{pmatrix}
 \hat{b}^\dag_{(-k_+,k_-),\sigma} \phantom{\e^{i k_-}}\\
 \hat{a}^\dag_{(-k_+,k_-),\sigma} \e^{i k_-}
\end{pmatrix}
\label{eq:T2}
\\
\hat{T}_-^{-1} s^z_{\ve{q}} \hat{T}_- &= -s^z_{(-q_+,q_-)}
\end{align}

Inserting  into Eq.~\eqref{eq:ham-fourier}, we obtain:
\begin{align*}
&\hat{T}_-^{-1} \mc{H}^{C_{2v}} \hat{T}_- =
\\
&= -\sum_{\ve{k}} \e^{-i k_-} \hat{b}_{(-k_+,k_-),\sigma}^\dag \Bigg( 
         \hat{a}_{(-k_+,k_-),\sigma} Z_0(\ve{k})
       + \xi\sum_{\ve{q}}  \hat{a}_{(-k_++q_+,k_--q_-),\sigma} (-\hat{s}^z_{(-q_+,q_-)}) Z_\text{I}(\ve{k}) 
      \Bigg) + h.c.
     + H_\text{Ising}
\\
&= -\sum_{\ve{k}} \e^{i k_-} \hat{a}_{(k_+,k_-),\sigma}^\dag \Bigg( 
         \hat{b}_{(k_+,k_-),\sigma} \bar{Z}_0(-k_+,k_-)
       + \xi\sum_{\ve{q}}  \hat{b}_{(k_+-q_+,k_--q_-),\sigma} (-\hat{s}^z_{(q_+,q_-)}) \bar{Z}_\text{I}(-k_+,k_-) 
      \Bigg) + h.c.
     + H_\text{Ising}
\\
& \qquad \left[
\begin{array}{rl}
\bar{Z}_0(-k_+,k_-)        &= Z_0^{C_{2v}}(\ve{k}) \e^{-i k_-} \\
\bar{Z}_\text{I}(-k_+,k_-) &= -Z_\text{Yuk}^{C_{2v}}(\ve{k}) \e^{-i k_-}
\end{array}
\right]
\\
&= -\sum_{\ve{k}} \hat{a}_{\ve{k},\sigma}^\dag \Bigg( 
         \hat{b}_{\ve{k},\sigma} Z_0^{C_{2v}}(\ve{k})
       + \xi\sum_{\ve{q}}  \hat{b}_{\ve{k}-\ve{q}} \hat{s}^z_{\ve{q}} Z_\text{Yuk}^{C_{2v}}(\ve{k})
      \Bigg) + h.c.
     + H_\text{Ising}
\\
&=\mc{H}^{C_{2v}}
\end{align*}

In real space, $\hat{T}_-$ translates to:
\begin{align}
\hat{T}_-^{-1} 
\begin{pmatrix}
\hat{a}^\dag_{\ve{R},\sigma}
 \\ 
\hat{b}^\dag_{\ve{R},\sigma}
\end{pmatrix} 
\hat{T}_- 
&= 
\begin{pmatrix}
 \hat{b}^\dag_{(-R_+, R_-),\sigma}
 \\
 \hat{a}^\dag_{(-R_+, R_--1),\sigma}
\end{pmatrix}
\\
\hat{T}_-^{-1} s^z_{\ve{R}} \hat{T}_- &= s^z_{(-R_+,R_-)}
\end{align}

%%%%%%%%%%%%%%%%%%%%%%%%%%%%%%%%%%%%%%%%%%%%%%%%%%%%%%%
\subsection{The \texorpdfstring{$C_{4v}$}{C4v} model}
The Second model has a $C_{4V}$ symmetry, consisting of a rotation by $\tfrac{\pi}{2}$ and reflections on the  x and y axis.

The corresponding operators are in momentum space:
\begin{align}
\hat{T}_{\pi/2}^{-1} 
\begin{pmatrix}
  \hat{a}_{\ve{k},\sigma}^{\dag}
  \\
  \hat{b}_{\ve{k},\sigma}^{\dag}  
\end{pmatrix}
\hat{T}_{\pi/2}
&=  
\begin{pmatrix}
  \hat{b}_{(-k_y,k_x),\sigma}^{\dag}
  \\
  \hat{a}_{(-k_y,k_x),\sigma}^{\dag}  
\end{pmatrix}
&
\hat{T}_x^{-1} 
\begin{pmatrix}
  \hat{a}_{\ve{k},\sigma}^{\dag}
  \\
  \hat{b}_{\ve{k},\sigma}^{\dag}  
\end{pmatrix}
\hat{T}_x
&=  
\begin{pmatrix}
  \hat{a}_{(k_x,-k_y),\sigma}^{\dag}
  \\
  \hat{b}_{(k_x,-k_y),\sigma}^{\dag}  
\end{pmatrix}
&
\hat{T}_{y}^{-1} 
\begin{pmatrix}
  \hat{a}_{\ve{k},\sigma}^{\dag}
  \\
  \hat{b}_{\ve{k},\sigma}^{\dag}  
\end{pmatrix}
\hat{T}_{y}
&=  
\begin{pmatrix}
  \hat{a}_{(-k_x,k_y),\sigma}^{\dag}
  \\
  \hat{b}_{(-k_x,k_y),\sigma}^{\dag}  
\end{pmatrix}
\\
\hat{T}_{\pi/2}^{-1} \hat{s}^{z}_{\ve{q}} \hat{T}_{\pi/2}
&=
-\hat{s}^{z}_{ (-q_y,q_x)}
&
\hat{T}_{x}^{-1} \hat{s}^{z}_{\ve{q}} \hat{T}_{x}
&=
\hat{s}^{z}_{ (q_x,-q_y)}
&
\hat{T}_{y}^{-1} \hat{s}^{z}_{\ve{q}} \hat{T}_{y}
&=
\hat{s}^{z}_{ (-q_x,q_y)}
\end{align}

And in real space:
\begin{align}
\hat{T}_{\pi/2}^{-1} 
\begin{pmatrix}
  \hat{a}_{\ve{R},\sigma}^{\dag}
  \\
  \hat{b}_{\ve{R},\sigma}^{\dag}  
\end{pmatrix}
\hat{T}_{\pi/2}
&=  
\begin{pmatrix}
  \hat{b}_{(-R_y,R_x),\sigma}^{\dag}
  \\
  \hat{a}_{(-R_y,R_x),\sigma}^{\dag}  
\end{pmatrix}
&
\hat{T}_x^{-1} 
\begin{pmatrix}
  \hat{a}_{\ve{R},\sigma}^{\dag}
  \\
  \hat{b}_{\ve{R},\sigma}^{\dag}  
\end{pmatrix}
\hat{T}_x
&=  
\begin{pmatrix}
  \hat{a}_{(R_x,-R_y),\sigma}^{\dag}
  \\
  \hat{b}_{(R_x,-R_y),\sigma}^{\dag}  
\end{pmatrix}
&
\hat{T}_{y}^{-1} 
\begin{pmatrix}
  \hat{a}_{\ve{R},\sigma}^{\dag}
  \\
  \hat{b}_{\ve{R},\sigma}^{\dag}  
\end{pmatrix}
\hat{T}_{y}
&=  
\begin{pmatrix}
  \hat{a}_{(-R_x,R_y),\sigma}^{\dag}
  \\
  \hat{b}_{(-R_x,R_y),\sigma}^{\dag}  
\end{pmatrix}
\\
\hat{T}_{\pi/2}^{-1} \hat{s}^{z}_{\ve{R}} \hat{T}_{\pi/2}
&=
-\hat{s}^{z}_{ (-R_y,R_x)}
&
\hat{T}_{x}^{-1} \hat{s}^{z}_{\ve{R}} \hat{T}_{x}
&=
\hat{s}^{z}_{ (R_x,-R_y)}
&
\hat{T}_{y}^{-1} \hat{s}^{z}_{\ve{R}} \hat{T}_{y}
&=
\hat{s}^{z}_{ (-R_x,R_y)}
\end{align}
In the Ising ordered phase, the Ising symmetry $\hat{s}^{z}_{\ve{R}} \rightarrow -\hat{s}^{z}_{\ve{R}}$ is broken, which reduces $\hat{T}_{\pi/2}$ to $\hat{T}_{\pi}$, such that the $C_{4v}$ symmetry is reduced to $C_{2v}$. This reduced symmetry allows the cones to meander.

\bigskip
\noindent\textit{Particle-hole symmetry:}

 $\hat{T}_{\text{ph}}$ This particle-hole symmetry implies that energy eigenstates satisfy $E(\ve{k}) = - E(-\ve{k} + \ve{Q})$, $\ve{Q} = (\pi, \pi)$. 

\begin{align}
\hat{T}_\text{ph}^{-1}
\alpha
\begin{pmatrix}
  \hat{a}_{\ve{R},\sigma}^\dag
  \\
  \hat{b}_{\ve{R},\sigma}^\dag 
\end{pmatrix}
\hat{T}_\text{ph}
&=
\overline{\alpha} (-1)^{R_x + R_y}
\begin{pmatrix}
  \hat{a}_{\ve{R},\sigma}
  \\
  \hat{b}_{\ve{R},\sigma}
\end{pmatrix}
\label{eq:T-phR}
\\
\hat{T}_\text{ph}^{-1}
\alpha s^z_{\ve{R}}
\hat{T}_\text{ph}
&=
- \overline{\alpha} s^z_{\ve{R}}
\end{align}

\begin{align}
\hat{T}_\text{ph}^{-1} 
\alpha
\begin{pmatrix}
  \hat{a}^\dag_{\ve{k},\sigma}
  \\
  \hat{b}^\dag_{\ve{k},\sigma} 
\end{pmatrix}
\hat{T}_\text{ph}
&=
\overline{\alpha}
\begin{pmatrix}
  \hat{a}_{\ve{Q}-\ve{k},\sigma}
  \\
  \hat{b}_{\ve{Q}-\ve{k},\sigma}
\end{pmatrix}
\label{eq:T-phK}
\quad
\ve{Q} =
\begin{pmatrix}
\pi \\ \pi
\end{pmatrix}
\\
\hat{T}_\text{ph}^{-1}
\alpha s^z_{\ve{q}}
\hat{T}_\text{ph}
&=
-\overline{\alpha} s^z_{-\ve{q}}
\end{align}

Inserting in  Eq.~\eqref{eq:ham-fourier}, we obtain:
\begin{align*}
\hat{T}_\text{ph}^{-1} \mc{H}^{C_{4v}} \hat{T}_\text{ph} 
&=
-\sum_{\ve{k}} \hat{a}_{\ve{Q}-\ve{k}} \Bigg( 
         \hat{b}^\dag_{\ve{Q}-\ve{k}} \bar{Z}_0^{C_{4v}}(\ve{k})
      +\frac{\xi}{N} \sum_{\ve{q}} \hat{b}^\dag_{\ve{Q}-\ve{k}+\ve{q}} 
      \left( -s^z_{-\ve{q}} \right) \bar{Z}_\text{Yuk}^{C_{4v}}(\ve{k}) 
      \Bigg) + h.c.
     + H_\text{Ising}
\\
&= \phantom{-} \sum_{\ve{k}} \hat{a}^\dag_{\ve{Q}-\ve{k}} \Bigg( 
         \hat{b}_{\ve{Q}-\ve{k}} Z_0^{C_{4v}}(\ve{k})
      +\frac{\xi}{N} \sum_{\ve{q}} \hat{b}_{\ve{Q}-\ve{k}+\ve{q}} 
      \left( -s^z_{-\ve{q}} \right) Z_\text{Yuk}^{C_{4v}}(\ve{k}) 
      \Bigg) + h.c.
     + H_\text{Ising}
\\
&= \phantom{-} \sum_{\ve{k}} \hat{a}^\dag_{\ve{k}} \Bigg( 
         \hat{b}_{\ve{k}} Z_0^{C_{4v}}(\ve{Q}-\ve{k})
      +\frac{\xi}{N} \sum_{\ve{q}} \hat{b}_{\ve{k}-\ve{q}} 
      \left( -s^z_{\ve{q}} \right) Z_\text{Yuk}^{C_{4v}}(\ve{Q}-\ve{k}) 
      \Bigg) + h.c.
     + H_\text{Ising}
\\
& \qquad \left[
\begin{array}{rl}
Z_0^{C_{4v}}(\ve{Q}-\ve{k})          &= -Z_0^{C_{4v}}(\ve{k}) \\
Z_\text{Yuk}^{C_{4v}}(\ve{Q}-\ve{k}) &= Z_\text{Yuk}^{C_{4v}}(\ve{k})
\end{array}
\right]
\\
&= -\sum_{\ve{k}} \hat{a}^\dag_{\ve{k}} \Bigg( 
         \hat{b}_{\ve{k}} Z_0^{C_{4v}}(\ve{k})
      +\frac{1}{N} \sum_{\ve{q}} \hat{b}_{\ve{k}-\ve{q}} 
      \left( s^z_{\ve{q}} \right) Z_\text{Yuk}^{C_{4v}}(\ve{k}) 
      \Bigg) + h.c.
     + H_\text{Ising}
\\
&= \mc{H}^{C_{4v}}
\end{align*}

As a  result of this symmetry,  the single particle spectral function satisfies
$A(\ve{k}, \omega) =  A(-\ve{k} + \ve{Q}, -\omega)$, with $\ve{Q} = (\pi, \pi )$.

\clearpage
%%%%%%%%%%%%%%%%%%%%%%%%%%%%%%%%%%%%%%%%%%%%%%%%%%%%%%%
\section{Mean-field approximation}
\label{sec:meanfield}
%%%%%%%%%%%%%%%%%%%%%%%%%%%%%%%%%%%%%%%%%%%%%%%%%%%%%%%
%
In the Mean-field approximation, we expand Eq.~\eqref{eq:ham-fourier} around $\langle \hat{s}^z_{\ve{R}} \rangle \equiv \phi$. The resulting Mean-field Hamiltonian reads:
\begin{align}
\mc{H}_\text{MF} &= 
  \sum_{\sigma=1}^{N_\sigma} \sum_{\ve{k}} \hat{K}_{\ve{k},\sigma}(\phi\xi)
  +
  \sum_{\ve{R}} \hat{I}_{\ve{R}}(\phi)
\end{align}
With:
\begin{align*}
\hat{K}_{\ve{k},\sigma}(\phi\xi) &= 
  -\hat{a}_{\ve{k},\sigma}^\dag \hat{b}_{\ve{k},\sigma}
  Z(\ve{k}, \phi\xi)
&
Z(\ve{k}, \phi\xi) &= Z_0(\ve{k}) + \phi \xi Z_\text{Yuk}(\ve{k})
&
\hat{I}_{\ve{R}}(\phi) &= 
  -4 J \left( \phi\hat{s}^z_{\ve{R}} - \tfrac{1}{2}\phi^2 \right) - h\hat{s}^x_{\ve{R}}
\end{align*}
The fermionic dispersion is 
\begin{equation}
\label{eq:mf-dispersion}
\pm \left| Z(\ve{k}, \phi\xi) \right|.
\end{equation}
\bigskip

\noindent
To determine the nature of the zero-temperature phase transition, we determine the order parameter $\phi$ for a given transverse field $h$ by minimizing the ground state energy $E_{0, \text{MF}} = \lim_{\beta \rightarrow \infty} E_\text{MF} = \lim_{\beta \rightarrow \infty} \Braket{\mc{H}_\text{MF}}_\text{MF}$.

\begin{align*}
E_\text{MF} &= \Braket{\mc{H}_\text{MF}}_\text{MF}
= \frac
  {{\rm Tr}\big(\exp(-\beta \mc{H}_\text{MF}) \mc{H}_\text{MF} \big)}
  {{\rm Tr} \big(\exp(-\beta \mc{H}_\text{MF})\big)}
\\
&= N_\sigma \sum_{\ve{k}} |Z(\ve{k}, \phi\xi)| \frac
        {1 - \exp\left(\beta |Z(\ve{k}, \phi\xi)| \right)}
        {1 + \exp\left(\beta |Z(\ve{k}, \phi\xi)| \right)}
+
L^2 \left( 2\phi^2 - \sqrt{h^2 + 16\phi^2} \tanh\left(\beta\sqrt{h^2 + 16\phi^2}\right) \right)
\end{align*}

\begin{align}
\label{eq:MF-Energy}
\lim_{\beta \rightarrow \infty} \frac{E_\text{MF}}{L^2}
&= \underbrace{- \frac{N_\sigma}{L^2} \sum_{\ve{k}} |Z(\ve{k}, \phi\xi)|}_{\epsilon_\text{F}(L, \phi \xi)}
+
\underbrace{\left( 2\phi^2 - \sqrt{h^2 + 16\phi^2} \right)}_{\epsilon_\text{I}(\phi, h)}
\\
&\partial_{\phi} \epsilon_\text{F}(L, \phi) = 
  -\frac{N_\sigma}{L^2} \sum_{\ve{k}} \frac{1}{|Z(\ve{k}, \phi)|}
  \Big(  
   \Re\left(Z_0(\ve{k}) \bar{Z}_\text{Yuk}(\ve{k})\right)
   + \phi |Z_\text{Yuk}(\ve{k})|^2
  \Big)
\\
&\partial^2_{\phi} \epsilon_\text{F}(L, \phi) = 
  -\frac{N_\sigma}{L^2} \sum_{\ve{k}} \left( 
  \frac{|Z_\text{Yuk}(\ve{k})|^2}{|Z(\ve{k}, \phi)|} - 
  \frac{\Re\left(Z_0(\ve{k}) \bar{Z}_\text{Yuk}(\ve{k})\right)}
       {|Z(\ve{k}, \phi)|^3}
  \Big(
    \Re\left(Z_0(\ve{k}) \bar{Z}_\text{Yuk}(\ve{k})\right) + \phi |Z_\text{Yuk}(\ve{k})|
  \Big)
  \right)
\end{align}

Equation \eqref{eq:MF-Energy} separates into a fermionic and an Ising part, $\epsilon_\text{F}$ and $\epsilon_\text{I}$. While $\epsilon_\text{I}$ has a well-behaved, closed form, $\epsilon_\text{F}$ has some non-analytic points on finite lattices (see Fig.~\ref{fig:meanfield_F}). Namely $\partial^2_\phi \epsilon_\text{F}$ diverges, if $Z(\ve{k}, \phi\xi)$  vanishes. 

\begin{figure}
\centering
\includegraphics[width=.3\linewidth]{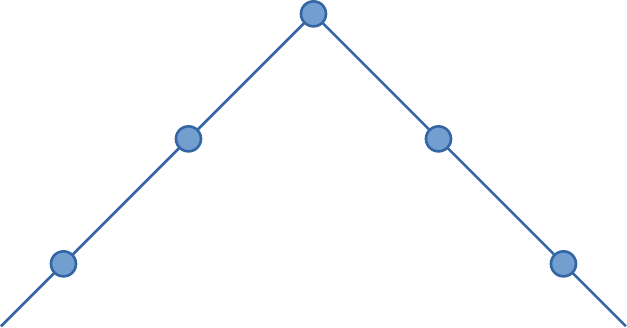}
\hspace{.05\linewidth}
\includegraphics[width=.3\linewidth]{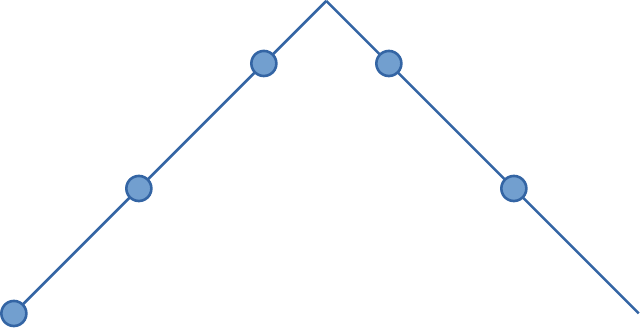}
\caption{
\label{fig:Finite_size_Dirac}
Sketch for understanding finite-size artifacts of Dirac systems as a  function of  $\ve{k}$-quantization. Shown is the valence band of a one-dimensional Dirac cone on a lattice of size $L=5$. We can see that the choice of  the momenta  quantization   (i.e.  boundary  conditions)  affects the energy.  Left: Dirac   cone  belongs to the set  of  finite size $\ve{k}$-points. Right: Dirac cone is between two finite-size $\ve{k}$-points. Left has a 
higher ground state energy,  and a level  crossing  occurs  when a $\ve{k}$-point crosses the Fermi surface.      In nematic  transitions   translation symmetry  is  not broken  such that momenta is well defined  and the $\ve{k}$  quantization  for a  given lattice  size remains unchanged.  However  the  position of the Dirac  cone  meanders.    The energy level  crossing that  
 originates  is  reminiscent  from that obtained  when  twisting  boundary  conditions  \cite{Scalapino93,Assaad94}. 
}
\end{figure}

\begin{figure}[h]
\includegraphics[width=\linewidth]{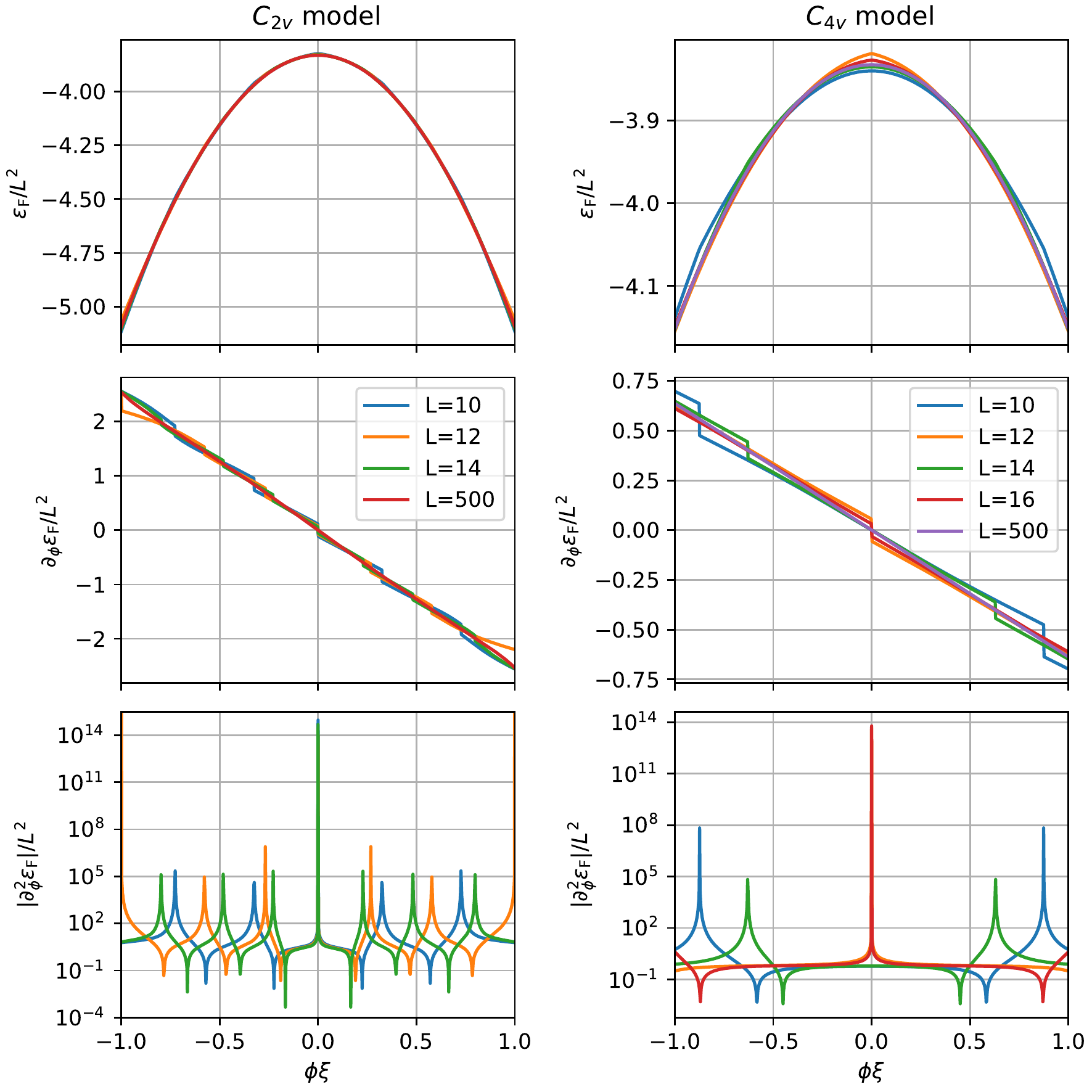}
\caption{
\label{fig:meanfield_F}
Fermionic part of Mean-field ground state energy.
}
\end{figure}

This corresponds to a finite size artifact which can be qualitatively understood with the help of Fig.~\ref{fig:Finite_size_Dirac}. 
Essentially,  $\phi$,   shifts  the    single   particle  energy   and   produces level  crossing    reminiscent of   those  produced   when 
twisting  boundary  conditions   \cite{Scalapino93,Assaad94}.  Fig. ~\ref{fig:Finite_size_Dirac}    shows  the valence band of a one-dimensional Dirac cone on a lattice of size 5 $\ve{k}$ points  at  two different  twists.      As a  function of the  twist  the   $\ve{k}$-point   will cross the  Fermi  surface    and  at this  crossing point   a  singularity in  the kinetic  energy  --  corresponding to a level crossing --   will appear.    This is  explicitly  shown  in  Fig.~\ref{fig:meanfield_c2v_xi75}.    This observation   means that   the  thermodynamic limit and   the  derivative $\partial_{\phi} $  do not commute:  one should   first   take  the thermodynamic limit   prior   to carrying out the derivative. 

To avoid this artifact, we consider two different approaches: 
\begin{enumerate}
\item Chose a weaker coupling $\xi$, such that the Dirac points do not    cross  the  Fermi   surface  in proximity to the critical point. However,  choosing a small $\xi$ may result in a slow flow from the 3d Ising fixed point of the unperturbed Ising model to   nematic  criticality. 
\item Chose antiperiodic boundary conditions in space for the fermions, so to shift the $\ve{k}$-points   away  from the Fermi surface:  Fig.~\ref{fig:meanfield_c2v_latt}.      However,  this  choice   results in   large    size effects   presumably  due to the   boundary-condition induced finite size  gap. 
\end{enumerate}
It turns out the first option is   the  best  choice  and  that $\xi$   can be chosen  large enough so as  to  minimize  the  aforementioned crossover  effects. 
\bigskip

The $C_{4v}$ model (Fig.~\ref{fig:meanfield_c4v}) shows different behaviors between systems with linear size $L = 2 + 4\mathbb{N}$ and $L = 4\mathbb{N}$.  At   $L = 4\mathbb{N}$ and  periodic boundary  conditions, the Dirac points in the disordered phase are located at  $\ve{k}$-points resolved by the finite lattice.  This is  not the case at  $L = 2 + 4\mathbb{N}$ (cf. Fig.~\ref{fig:meanfield_c4v_latt}). As a result, the $L = 4\mathbb{N}$  sizes have a smoothed out phase transition.
Nevertheless, both $L = 4\mathbb{N}$  and $L = 2 + 4\mathbb{N}$ converge to the same result in the thermodynamic limit. The Monte-Carlo simulations also have odd-even effects, as elaborated in Section \ref{sec:odd-even} of this supplemental. It turns out that even system sizes produce nicer numerical results for the phase transition.

\begin{figure}[h]
\includegraphics[width=\linewidth]{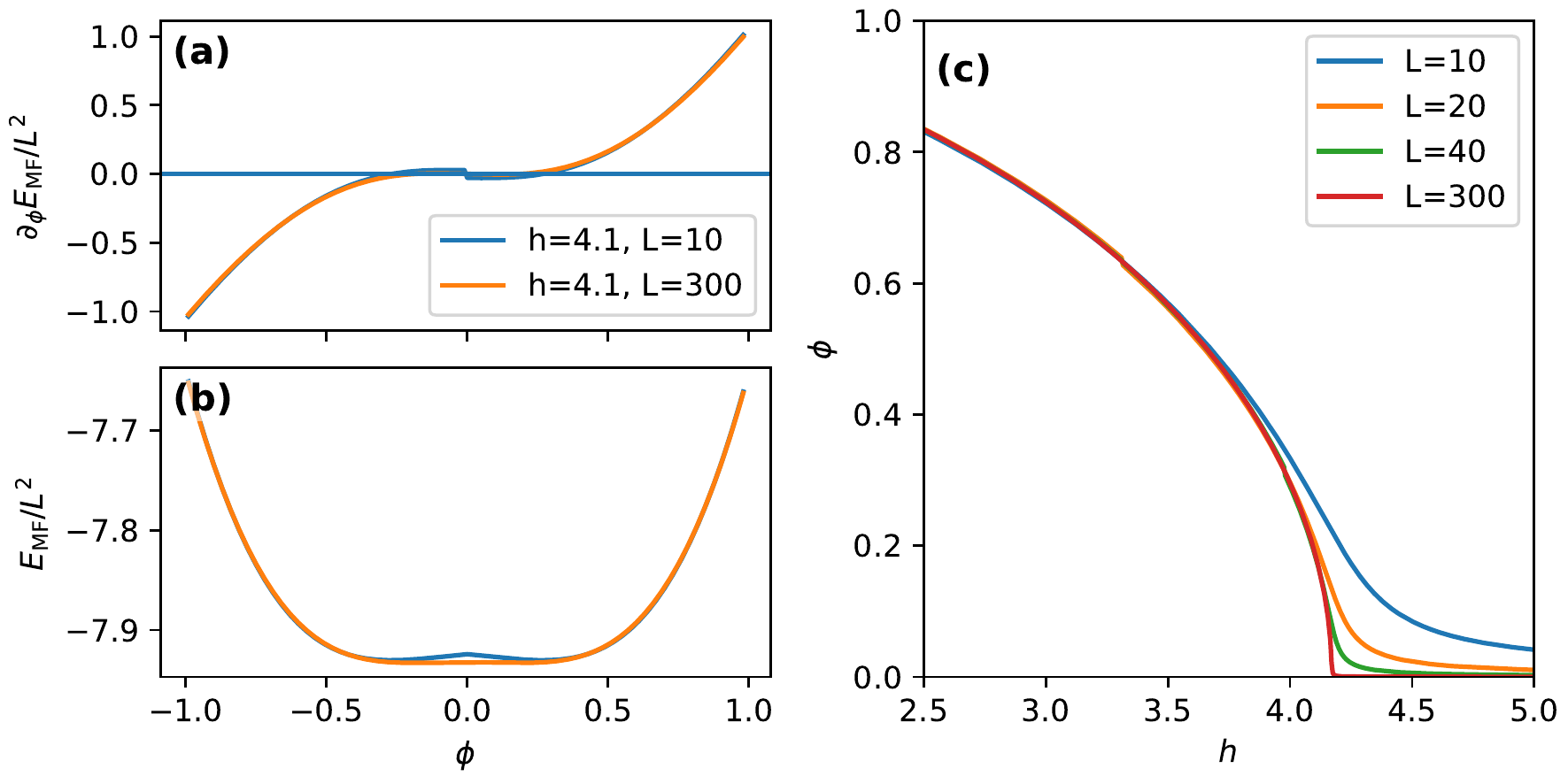}
\caption{
\label{fig:meanfield_c2v_xi25}
Mean-field results for the $C_{2v}$ model with $\xi = 0.25$.
}
\end{figure}

\begin{figure}
\includegraphics[width=\linewidth]{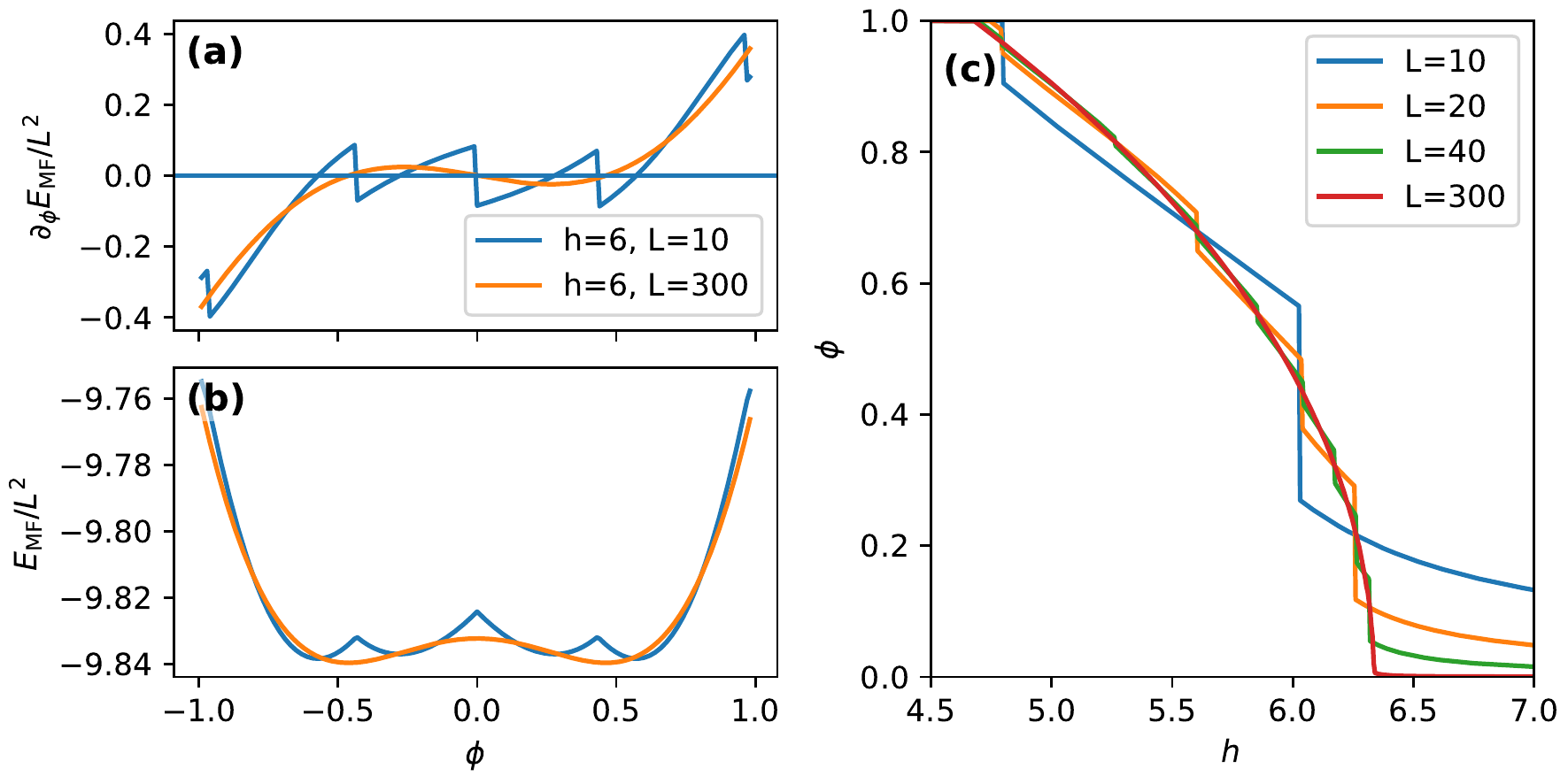}
\caption{
\label{fig:meanfield_c2v_xi75}
Mean-field results for the $C_{2v}$ model with coupling strength $\xi = 0.75$. 
}
\end{figure}

\begin{figure}[h]
\centering
\includegraphics[width=.45\linewidth]{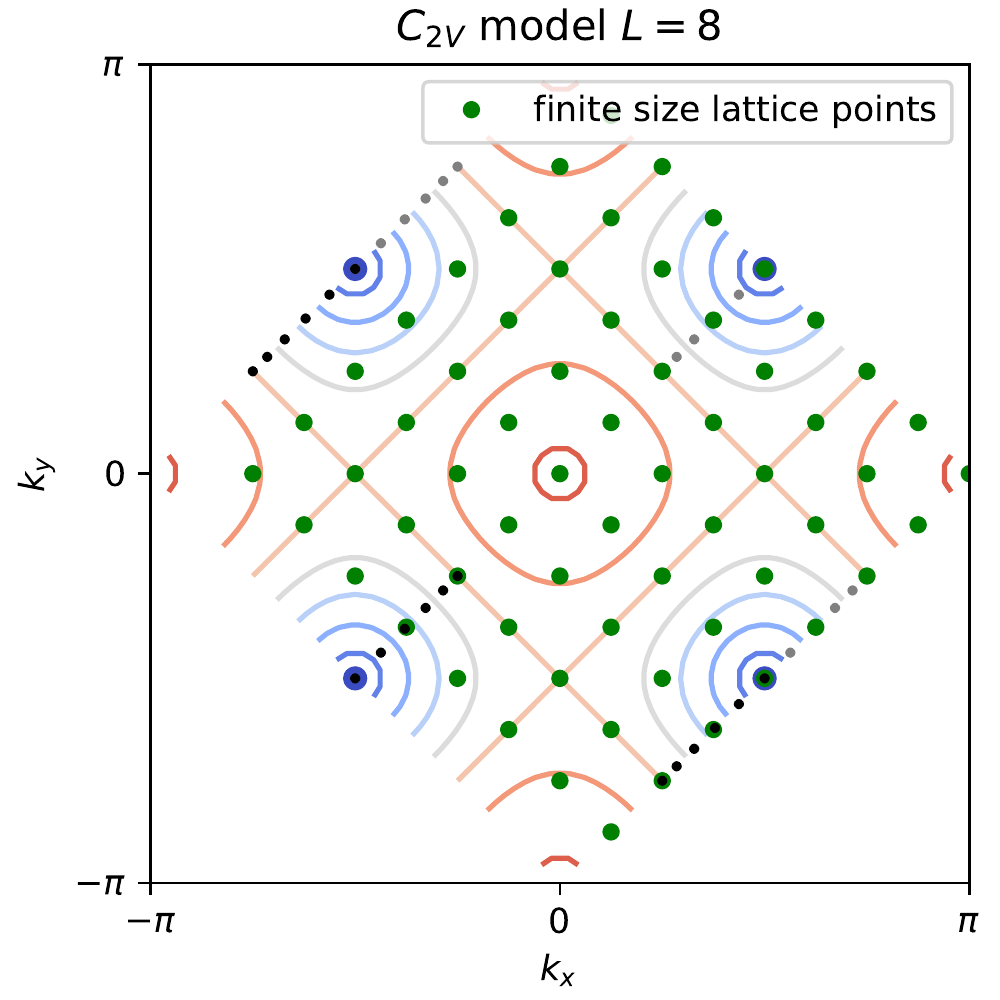}
\hspace{.05\linewidth}
\includegraphics[width=.45\linewidth]{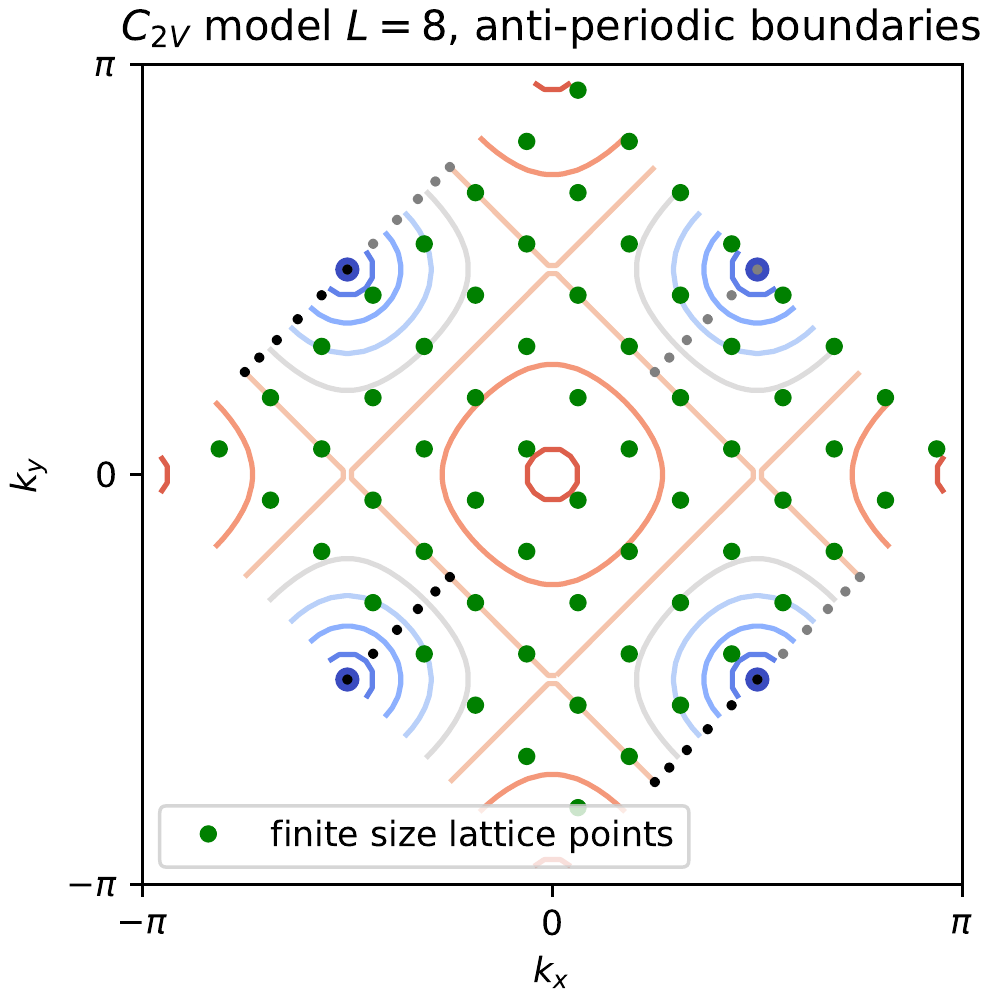}
\caption{
\label{fig:meanfield_c2v_latt}
Brillouin zone of $C_{2v}$ model with $\ve{k}$ points of a $8*8$ lattice. Left: With periodic boundary conditions. Right: With antiperiodic boundary conditions for movement parallel to $(1,-1)$. Also sketched: Dispersion in disordered phase and trajectory of Dirac cones.
}
\end{figure}

\begin{figure}[h]
\includegraphics[width=\linewidth]{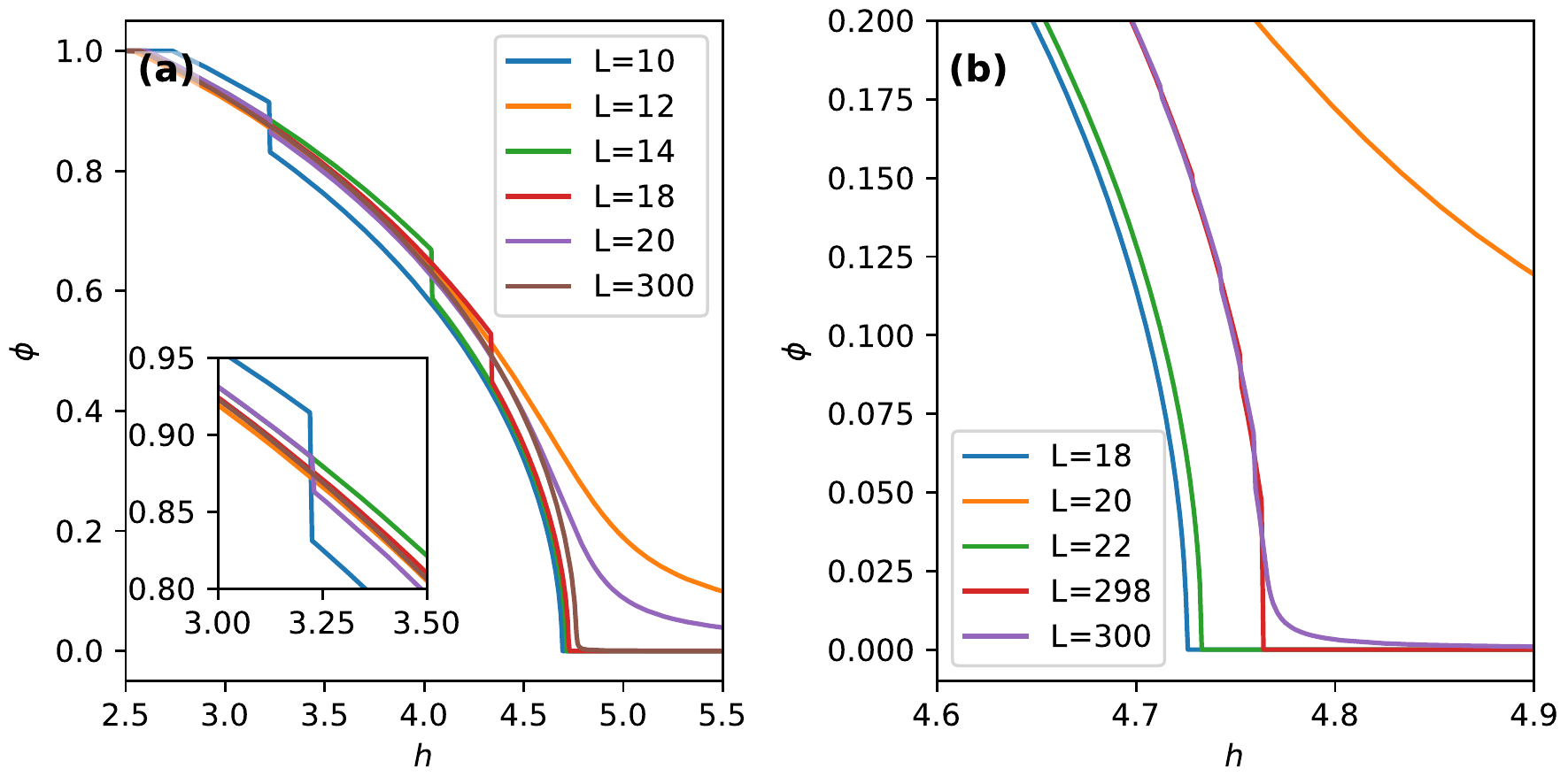}
\caption{
\label{fig:meanfield_c4v}
Mean-field results for the $C_{4v}$ model.
}
\end{figure}

\begin{figure}[h]
\centering
\includegraphics[width=.45\linewidth]{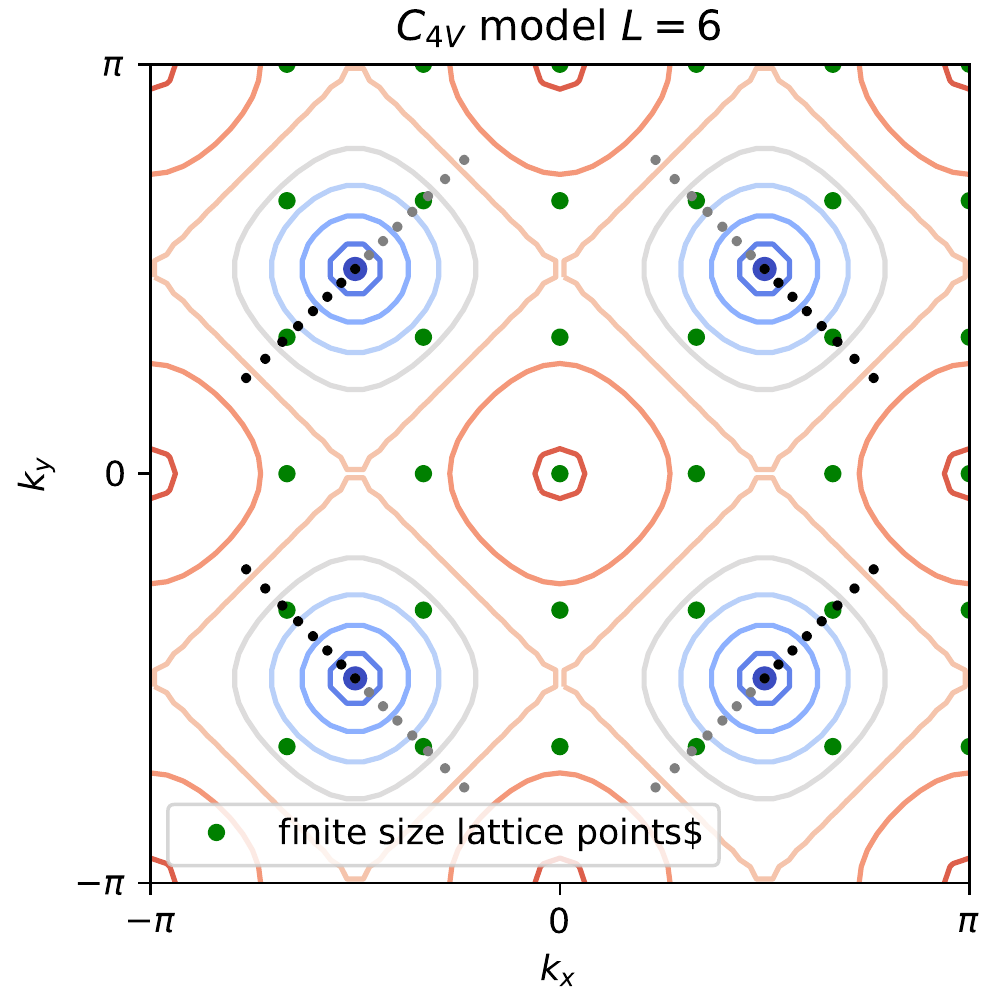}
\hspace{.05\linewidth}
\includegraphics[width=.45\linewidth]{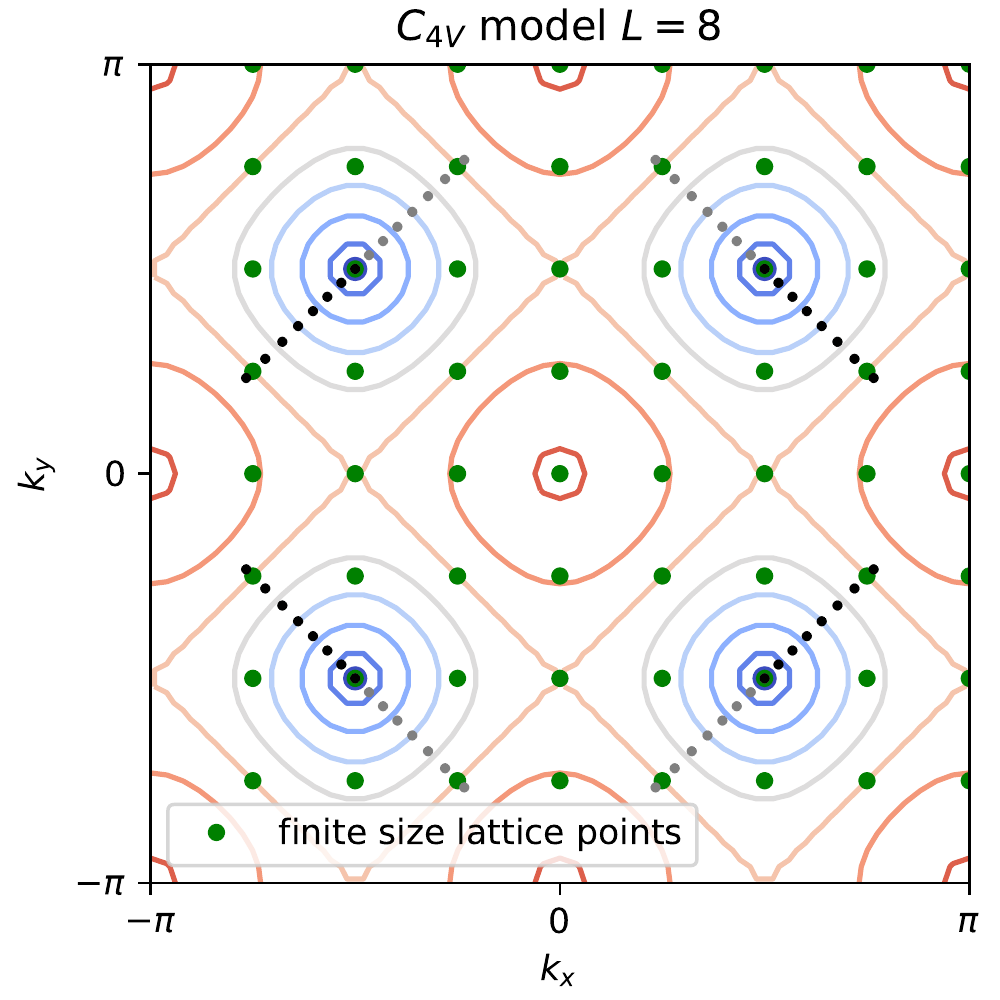}
\caption{
\label{fig:meanfield_c4v_latt}
Brillouin zone of $C_{4v}$ model with $\ve{k}$ points of $6*6$ and $8*8$ lattice. Also sketched: Dispersion in disordered phase and trajectory of Dirac cones. Left: $6*6$ lattice, the Dirac cones in the disordered phase are each centered between four $\ve{k}$ points. Right: $8*8$ lattice, the Dirac cones in the disordered phase are directly resolved by the $\ve{k}$ points.
}
\end{figure}

\ 
\clearpage
%%%%%%%%%%%%%%%%%%%%%%%%%%%%%%%%%%%%%%%%%%%%%%%%%%%%%%%
\section{Low energy models}
%%%%%%%%%%%%%%%%%%%%%%%%%%%%%%%%%%%%%%%%%%%%%%%%%%%%%%%

We derive the low energy model from Eq.~\eqref{eq:ham-fourier} by expansion in $\ve{k}$ around the nodal points $\ve{K}_i$ 
and  for a  scalar Ising field $\phi(\ve{q})$:
\begin{align}
\mc{H} &= 
-\sum_{\sigma=1}^{N_\sigma} \sum_{i} \int\!\text{d}\ve{\kappa}\; 
  \hat{a}_{\ve{K}_i+\ve{\kappa},\sigma}^\dag 
  \Bigg( 
         \hat{b}_{\ve{K}_i+\ve{\kappa},\sigma} Z_0(\ve{K}_i+\ve{\kappa}) 
      \nonumber\\
      & +
      \frac{1}{2\pi}\int\!\text{d}\ve{q}\; \hat{b}_{\ve{K}_i+\ve{\kappa}-\ve{q},\sigma} \phi({\ve{q}}) Z_\text{Yuk}(\ve{K}_i+\ve{\kappa}) 
      \Bigg) + h.c.
     + H_\text{Ising}\{\phi\}
\end{align}
In  leading  order in $ \ve{\kappa} $ we obtain:
\begin{align}
\mc{H} &= 
-\sum_{\sigma=1}^{N_\sigma} \sum_{i} \int\!\text{d}\ve{\kappa}\; 
  \hat{a}_{\ve{K}_i+\ve{\kappa},\sigma}^\dag 
  \Bigg( 
         \hat{b}_{\ve{K}_i+\ve{\kappa},\sigma} \ve{\kappa} \nabla Z_0(\ve{K}_i)  
      \nonumber\\
      & +
      \frac{1}{2\pi}\int\!\text{d}\ve{q}\; \hat{b}_{\ve{K}_i+\ve{\kappa}-\ve{q},\sigma} \phi({\ve{q}}) Z_\text{Yuk}(\ve{K}_i) 
      \Bigg) + h.c.
     + H_\text{Ising}\{\phi\}.
\end{align}

\noindent
Introducing the Fourier transformations:
\begin{align*}
\begin{pmatrix}
\hat{a}^i_{\sigma}(\ve{r}) \\
\hat{b}^i_{\sigma}(\ve{r})
\end{pmatrix}
&= 
\frac{1}{2\pi} \int\!\text{d}\ve{\kappa}\; 
\e^{i\ve{\kappa} \ve{r}}
\begin{pmatrix}
\hat{a}_{\ve{K}_i+\ve{\kappa},\sigma} \\
\hat{b}_{\ve{K}_i+\ve{\kappa},\sigma}
\end{pmatrix}
\\
\phi(\ve{r}) &= 
\frac{1}{2\pi} \int\!\text{d}\ve{\kappa}\; 
\e^{i\ve{q} \ve{r}} \phi(\ve{q}).
\end{align*}

\noindent
and defining:
\begin{align*}
  \ve{v}_i &\equiv \nabla Z_0(\ve{K}_i)
  &
  I_i &\equiv Z_\text{Yuk}(\ve{K}_i)
\end{align*}

\noindent
The Hamiltonian takes the form:
\begin{align}
\mc{H} &= 
-\sum_{\sigma=1}^{N_\sigma} \sum_{i} \int\!\text{d}\ve{r}\; 
  {\hat{a}^i_{\sigma}}{} (\ve{r})^\dag
  \Big( i\ve{v}_i \cdot \nabla_{\ve{r}} + \phi(\ve{r}) I_i \Big) 
  \hat{b}^i_{\sigma}(\ve{r}) 
   + h.c. + H_\text{Ising}\{\phi\}
\label{eq:low_energy}
\end{align}

%%%%%%%%%%%%%%%%%%%%%%%%%%%%%%%%%%%%%%%%%%%%%%%%%%%%%%%
\subsection{The \texorpdfstring{$C_{2v}$}{C2v} model}

\noindent
The $C_{2v}$ model has the nodal points $\ve{K}_\pm = \begin{pmatrix} \pi/2 \\ \pm\pi/2 \end{pmatrix}$. By defining the four-component Dirac spinor
\begin{align*}
  \Psi_\sigma(\ve{r}) &= \begin{pmatrix}
  \hat{a}^+_{\sigma}(\ve{r}) & 
  \hat{b}^+_{\sigma}(\ve{r}) & 
  \hat{a}^-_{\sigma}(\ve{r}) & 
  \hat{b}^-_{\sigma}(\ve{r}) 
  \end{pmatrix}^{\rm T}
\end{align*}
and
\begin{align*}
\tau_1 &= \begin{pmatrix}
0 & 1 \\ 1 & 0
\end{pmatrix}
&
\tau_2 &= \begin{pmatrix}
0 & -i \\ i & 0
\end{pmatrix}
&
\tau_3 &= \begin{pmatrix}
1 & 0 \\ 0 & -1
\end{pmatrix}.
\end{align*}

\noindent
Eq.~\eqref{eq:low_energy} can be written as
\begin{align*}
\mc{H}^{C_{2v}} &= 
\sum_{\sigma=1}^{N_\sigma} \int\!\text{d}\ve{r}\; 
  \Psi_\sigma^\dag(\ve{r}) \Bigg[
         2i t \begin{pmatrix}
         \tau_2 & 0 \\
         0 & -\tau_1
         \end{pmatrix} \partial_{r_+}
         +
         2i t \begin{pmatrix}
         \tau_1 & 0 \\
         0 & -\tau_2
         \end{pmatrix} \partial_{r_-}
         +
         2\sqrt{2}\xi \phi(\ve{r})
         \begin{pmatrix}
         -\tau_2 & 0 \\
         0 & -\tau_1
         \end{pmatrix}
  \Bigg]\Psi_\sigma(\ve{r})  + H_\text{Ising}\{\phi\}.
\end{align*}

\noindent
Introducing the gamma matrices
\begin{align*}
\gamma_0 &= \begin{pmatrix}
  -\tau_3 & 0 \\ 0 & -\tau_3
\end{pmatrix}
&
\gamma_1 &= \begin{pmatrix}
  \tau_1 & 0 \\ 0 & \tau_2
\end{pmatrix}
&
\gamma_2 &= \begin{pmatrix}
  -\tau_2 & 0 \\ 0 & -\tau_1
\end{pmatrix}
&
\{\gamma_\alpha, \gamma_\beta\} &= 2 \delta_{\alpha\beta},
\end{align*}

\noindent
we can write the action in the  form
\begin{align}
  S^{C_{2v}} = &\int\!\text{d}^D x\, \sum_{\sigma=1}^{N_\sigma} \Big[
   \Psi_{\sigma}^\dag(x) \big[
                 \openone \partial_\tau
                 +v \gamma_0\gamma_1 \partial_+
                 +v \gamma_0\gamma_2 \partial_-
                 +g \phi(x) \gamma_2
               \big] \Psi_\sigma(x) 
               \Big]+ S_\text{Ising}(\{ \phi \})
\end{align}

%%%%%%%%%%%%%%%%%%%%%%%%%%%%%%%%%%%%%%%%%%%%%%%%%%%%%%%
\subsection{The \texorpdfstring{$C_{4v}$}{C4v} model}
The $C_{4v}$ model has the nodal points $\ve{K}_{+\pm} = \begin{pmatrix} \pi/2 \\ \pm\pi/2 \end{pmatrix}$ and $\ve{K}_{-\pm} = -\ve{K}_\pm$. By defining the eight-component Dirac spinor
\begin{align*}
  \Psi_\sigma(\ve{r}) &= \begin{pmatrix}
  \hat{a}^{++}_{\sigma}(\ve{r}) & 
  \hat{b}^{++}_{\sigma}(\ve{r}) & 
  \hat{a}^{-+}_{\sigma}(\ve{r}) & 
  \hat{b}^{-+}_{\sigma}(\ve{r}) &
  \hat{a}^{+-}_{\sigma}(\ve{r}) & 
  \hat{b}^{+-}_{\sigma}(\ve{r}) & 
  \hat{a}^{--}_{\sigma}(\ve{r}) & 
  \hat{b}^{--}_{\sigma}(\ve{r}) 
  \end{pmatrix}^{\rm T}
\end{align*}

Eq.~\eqref{eq:low_energy} can be written as
\begin{align*}
\mc{H}^{C_{4v}} &= 
\sum_{\sigma=1}^{N_\sigma} \int\!\text{d}\ve{r}\; 
  \Psi_{\sigma}^\dag(\ve{r}) \Bigg[ \\
  &
      2i t
      \Bigg(
         \begin{pmatrix}
         \tau_1 & 0 \\
         0 & -\tau_1
         \end{pmatrix} 
         \oplus
         \begin{pmatrix}
         -\tau_2 & 0 \\
         0 & \tau_2
         \end{pmatrix}
      \Bigg)
         \partial_{r_+}
         +
         2i t
      \Bigg(
         \begin{pmatrix}
         -\tau_2 & 0 \\
         0 & \tau_2
         \end{pmatrix}
         \oplus
         \begin{pmatrix}
         \tau_1 & 0 \\
         0 & -\tau_1
         \end{pmatrix}
      \Bigg)
         \partial_{r_-}
         +
         2\sqrt{2}\xi \phi(\ve{r})
      \Bigg(
         \begin{pmatrix}
         \tau_2 & 0 \\
         0 & \tau_2
         \end{pmatrix}
         \oplus
         \begin{pmatrix}
         \tau_2 & 0 \\
         0 & \tau_2
         \end{pmatrix}
      \Bigg)
  \\
  &\qquad\Bigg] \Psi_{\sigma}(\ve{r}) + H_\text{Ising}\{\phi\}
\end{align*}

\noindent
Introducing the gamma matrices
\begin{align*}
\tilde{\gamma}_0 &= \begin{pmatrix}
  \tau_3 & 0 \\ 0 & -\tau_3
\end{pmatrix}
&
\tilde{\gamma}_1 &= \begin{pmatrix}
  \tau_1 & 0 \\ 0 & \tau_1
\end{pmatrix}
&
\tilde{\gamma}_2 &= \begin{pmatrix}
  \tau_2 & 0 \\ 0 & \tau_2
\end{pmatrix}
&
\{\tilde{\gamma}_\alpha, \tilde{\gamma}_\beta\} &= 2 \delta_{\alpha\beta}
\end{align*}

\noindent
We can write the action in the compact form
\begin{align}
  S^{C_{4v}} = &\int\!\text{d}^D x\, \sum_{\sigma=1}^{N_\sigma} \Psi_{\sigma}^\dag(x) \Big[
     \openone \partial_\tau
     +v \left( \tilde{\gamma}_0\tilde{\gamma}_1 \oplus \tilde{\gamma}_0\tilde{\gamma}_2 \right) \partial_+
     +v \left( \tilde{\gamma}_0\tilde{\gamma}_2 \oplus \tilde{\gamma}_0\tilde{\gamma}_1 \right) \partial_-
     +g \phi(x) \left( \tilde{\gamma}_2 \oplus \tilde{\gamma}_2\right)
               \Big] \Psi_{\sigma}(x) + S_\text{Ising}(\{ \phi \})
\end{align}

\clearpage
%%%%%%%%%%%%%%%%%%%%%%%%%%%%%%%%%%%%%%%%%%%%%%%%%%%%%%%
\section{Renormalization group flow}
\begin{figure}[t]
\includegraphics[scale=1]{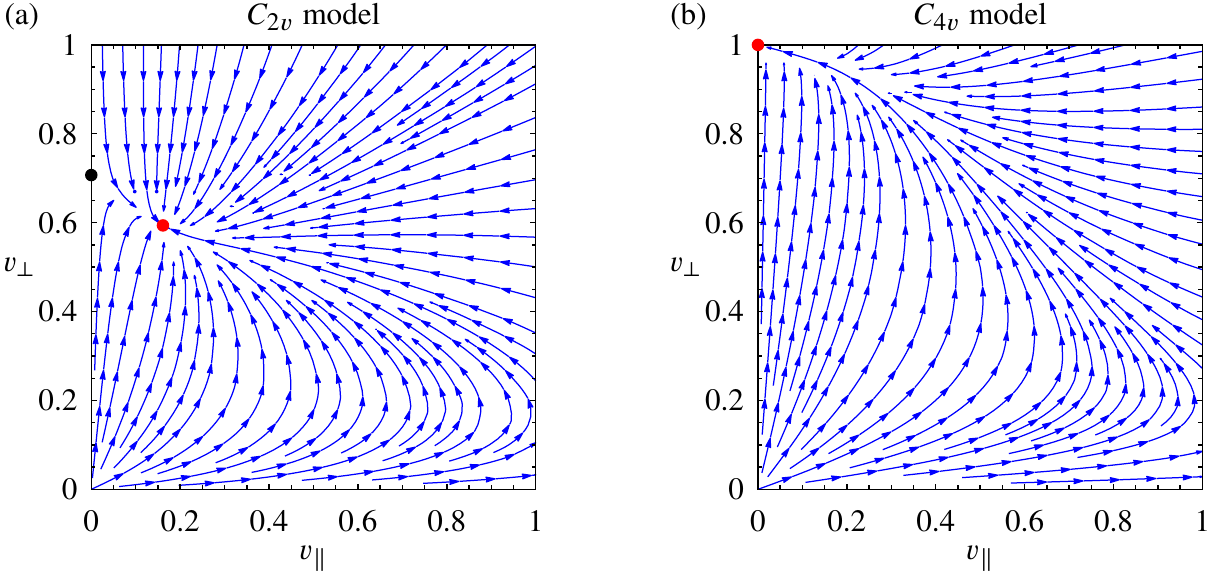}
\caption{Renormalization group flow in the $v_\parallel$-$v_\perp$ plane for (a) the $C_{2v}$ model and (b) the $C_{4v}$ model. Arrows denote flow towards infrared. The fixed point at $(v_\parallel^*, v_\perp^*) = (0,1/\sqrt{a_-})$ and $(g^2/v_\parallel)_* = 12\sqrt{a_-}\epsilon/N'$ is unstable in the $C_{2v}$ model [black dot in (a)], but stable in the $C_{4v}$ model [red dot in (b)]. In the $C_{2v}$ model, there is a nontrivial stable fixed point at $(v_\parallel^*, v_\perp^*) = (0.1611, 0.5942)$, with $g^2_* = 0.2123\epsilon$ for $N' = 4$ [red dot in (a)]. For visualization purposes, we have fixed the ratio $g^2/(v_\perp v_\parallel)$ to its value at the respective stable fixed point (red dots) in these plots.}
\label{fig:rg-flow}
\end{figure}

%%%%%%%%%%%%%%%%%%%%%%%%%%%%%%%%%%%%%%%%%%%%%%%%%%%%%%%
In this section, we present details of the renormalization group (RG) analysis of the continuum field theories.
Due to the lack of Lorentz and continuous spatial rotational symmetries in the low-energy models, the Fermi and bosonic velocities, as well as their anisotropies, will in general receive different loop corrections.
In order to appropriately take this multiple dynamics~\cite{Meng12,Janssen15} into account, it is useful to employ a regularization in the frequency only, which preserves the property that the different momentum components can be rescaled independently. This allows us to keep the boson velocities $c \equiv c_+ = c_-$ fixed, i.e., we measure the Fermi velocities in units of $c=1$. Integrating over the ``frequency shell'' $\Lambda/b \leq |\omega| \leq \Lambda$ with $b>1$ and \emph{all} momenta causes the velocities and couplings to flow at criticality $r=0$ as
\begin{align}
\label{eq:flow-1}
	\frac{\du v_\parallel}{\du \ln b} & =
	\frac12 (\eta_\phi - \eta_+ - 2 \eta_\psi) v_\parallel - F(v_\parallel,v_\perp) g^2,
%	-\frac{5g^2}{3} v_\parallel - \frac{2g^2}{15} (5 N_\sigma-7 v_\parallel) (v_\perp-1)
%	\frac{g^2}{15} \left[v_\parallel \left(14 v_\perp - 39 \right)-10 N_\sigma \left(v_\perp-1\right)\right]
	%
	\\
	\frac{\du v_\perp}{\du \ln b} & =
	\frac12 (\eta_\phi - \eta_- - 2 \eta_\psi) v_\perp + F(v_\perp,v_\parallel) g^2,
%	- \frac{g^2}{15 v_\parallel} (10 N_\sigma+7 v_\parallel) (v_\perp-1)-\frac{g^2}{3} 
%	-\frac{g^2}{15 v_\parallel} \left[10 N_\sigma (v_\perp-1)+v_\parallel (7 v_\perp-2)\right]
	%
	\\
	\frac{\du g^2}{\du \ln b} & = 
	\left(\epsilon - \frac{\eta_++\eta_-}{2} - 2 \eta_\psi\right) g^2 - 2 G(v_\parallel, v_\perp) g^4,
	\\
\label{eq:flow-4}
	\frac{\du \lambda}{\du \ln b} & = 
	\left(\epsilon - \frac{\eta_++\eta_-}{2} - \eta_\phi\right) \lambda - 18 \lambda^2 + \frac{N' g^4}{16 v_\parallel v_\perp},
\end{align}
with the anomalous dimensions $\eta_\psi = g^2 H(v_\parallel,v_\perp)$, $\eta_\phi = N' g^2/(12v_\parallel v_\perp)$, and $\eta_\pm = a_\pm N' g^2 v_\perp/(12v_\parallel)$, to the one-loop order. Here, the angular integrals are performed in $d=2$, while the dimensions of the couplings are counted in general $d$~\cite{Vojta00a, Janssen14}. 
At the present order, the flows of the two models differ only in the definition of the coefficients $a_\pm$, with $a_+ = 0$, $a_- = 2$ ($a_+=a_-=1$) in the $C_{2v}$ ($C_{4v}$) model, and the number of spinor components $N' = 4N_\sigma$ ($N' = 8N_\sigma$).
Our regularization scheme allows the evaluation of the one-loop integrals in closed form, leading to the functions
\begin{align}
	F(v_1, v_2) & = \frac{1}{\pi}  \int_{-\infty}^\infty \int_{-\infty}^\infty \frac{v_1 q_1^2 \du q_1 \du q_2}{\left(1+q_1^2+q_2^2\right)^2 \left(1+v_1^2 q_1^2+v_2^2 q_2^2\right)}
	\nonumber \\
	& = \frac{v_1 \left[v_1 \left(v_2^2-1\right) \sqrt{\frac{1-v_1^2}{v_2^2-1}}+(v_1+v_2) \sin ^{-1}\left({v_1}{\sqrt{\frac{v_2^2-1}{v_2^2-v_1^2}}}\right)-(v_1+v_2) \csc ^{-1}\left(\sqrt{\frac{v_2^2-v_1^2}{v_2^2-1}}\right)\right]}{\left(v_1^2-1\right) \left(v_2^2-1\right) (v_1+v_2) \sqrt{\frac{1-v_1^2}{v_2^2-1}}}\,,
	\\
	G(v_1, v_2) & = \frac{1}{2\pi}  \int_{-\infty}^\infty \int_{-\infty}^\infty \frac{1-v_1^2 q_1^2+v_2^2 q_2^2}{\left(1+q_1^2+q_2^2\right) \left(1+v_1^2 q_1^2+v_2^2 q_2^2\right)^2}\du q_1 \du q_2
	\nonumber \\
	& = \frac{\left(v_2^2-1\right) (v_1 v_2-1) \sqrt{\frac{1-v_1^2}{v_2^2-1}}+\left(v_1^2+v_2^2-2\right) \csc ^{-1}\left(\frac{1}{v_1}\sqrt{\frac{v_2^2-v_1^2}{v_2^2-1}}\right)-\left(v_1^2+v_2^2-2\right) \csc ^{-1}\left(\sqrt{\frac{v_2^2-v_1^2}{v_2^2-1}}\right)}{2 \left(v_1^2-1\right) \left(v_2^2-1\right)^2 \sqrt{\frac{1-v_1^2}{v_2^2-1}}}
	\nonumber \\ & \quad
	+\frac{v_1 \left[\left(v_2^2-1\right) \sqrt{\frac{1-v_1^2}{v_2^2-1}}-v_1 (v_1+v_2) \sin ^{-1}\left({\sqrt{\frac{v_2^2-1}{v_2^2-v_1^2}}}\right)+v_1 (v_1+v_2) \csc ^{-1}\left(\frac{1}{v_1}\sqrt{\frac{v_2^2-v_1^2}{v_2^2-1}}\right)\right]}{2 \left(v_1^2-1\right) \left(v_2^2-1\right) (v_1+v_2) \sqrt{\frac{1-v_1^2}{v_2^2-1}}}
	\nonumber \\ & \quad
	+\frac{v_2^4 (v_1+v_2) \sin ^{-1}\left({\sqrt{\frac{v_2^2-1}{v_2^2-v_1^2}}}\right)}{2 v_2^2 \left(v_2^2-1\right)^2 (v_1+v_2) \sqrt{\frac{1-v_1^2}{v_2^2-1}}}
	%
%	\nonumber \\ & \quad
	%
	+\frac{\left(v_1^3+v_1^2 v_2+v_1 v_2^2+v_2^3\right) \sin ^{-1}\left({v_1}{\sqrt{\frac{v_2^2-1}{v_2^2-v_1^2}}}\right)- 2 \left(v_1^2-1\right) v_2^3 \sqrt{\frac{1-v_1^2}{v_2^2-1}}}{4 \left(v_1^2-1\right) v_2^2 \left(v_2^2-1\right) (v_1+v_2) \sqrt{\frac{1-v_1^2}{v_2^2-1}}}
	\nonumber \\ & \quad
	- \frac{\left(v_2^2+1\right) \left[v_1^3 \left(2 v_2^2-1\right)+v_1^2 v_2 \left(2 v_2^2-1\right)-v_1 v_2^2-v_2^3\right] \csc ^{-1}\left(\frac{1}{v_1}\sqrt{\frac{v_2^2-v_1^2}{v_2^2-1}}\right)}{4 \left(v_1^2-1\right) v_2^2 \left(v_2^2-1\right)^2 (v_1+v_2) \sqrt{\frac{1-v_1^2}{v_2^2-1}}}\,,
	\nonumber \\
	H(v_1, v_2)  & = \frac{1}{\pi}  \int_{-\infty}^\infty \int_{-\infty}^\infty \frac{\du q_1 \du q_2}{\left(1+q_1^2+q_2^2\right)^2 \left(1+v_1^2 q_1^2+v_2^2 q_2^2\right)}
	\nonumber \\	
	& = \frac{\left(v_1^2+v_2^2-2 v_1^2 v_2^2\right) \left[\csc ^{-1}\left(\sqrt{\frac{v_2^2-v_1^2}{v_2^2-1}}\right)
	- \csc ^{-1}\left(\frac{1}{v_1}\sqrt{\frac{v_2^2-v_1^2}{v_2^2-1}}\right)\right]}{\left(v_1^2-1\right) \left(v_2^2-1\right)^2 \sqrt{\frac{1-v_1^2}{v_2^2-1}}}
	%
%	\nonumber \\ & \quad
	%		
	-\frac{(v_1 v_2-1) \sqrt{\frac{1-v_1^2}{v_2^2-1}}}{\left(v_1^2-1\right) \left(v_2^2-1\right) \sqrt{\frac{1-v_1^2}{v_2^2-1}}}\,.
\end{align}
The above one-loop flow equations admit a nontrivial fixed point that is characterized by anisotropic Fermi velocities $v_\parallel^* = 0$ and $v_\perp^* = 1/\sqrt{a_-}>0$, and vanishing $g^2_*$ and $\lambda^*$, but finite ratio $(g^2/v_\parallel)_* = 12\sqrt{a_-}\epsilon/N' + \mathcal O(\epsilon^2)$.
Perturbations of the couplings $g^2$ and $\lambda$ and the Fermi velocity $v_\perp$ away from this fixed point turn out be irrelevant; however, the flow of $v_\parallel$ near the fixed point is
\begin{align}
	\left.\frac{\du v_\parallel}{\du \ln b}\right|_{g^2_*,v_\perp^*} & = 
	\frac{\epsilon}{2} (1-a_+) v_\parallel - \frac{20\epsilon}{N'} v_\parallel^2 + \mathcal O(v_\parallel^3).
\end{align}
Hence, in the $C_{4v}$ model with $a_+ = a_- = 1$, $v_\parallel$ is marginally irrelevant, rendering the fixed point stable.
The fixed point represents a quantum critical point with maximally anisotropic Fermi velocities $(v_\parallel^*, v_\perp^*) = (0,1)$ and boson anomalous dimensions, describing the temporal and spatial decays of the order-parameter correlations, as $\eta_\phi = \epsilon$ and $\eta_+ = \eta_- = \epsilon$, respectively.
The fermion anomalous dimension becomes $\eta_\psi = 0$.
In the vicinity of this fixed point, the flow of $v_\parallel$ can be integrated out analytically, reading
\begin{align}
	v_\parallel(b) \simeq \frac{N'}{20\epsilon \ln b},
\end{align}
where we have assumed $b \gg 1$ for simplicity. This demonstrates that the Fermi velocity flow in the vicinity of the $C_{4v}$ fixed point is logarithmically slow, reflecting the fact that $v_\parallel$ is marginally irrelevant at this fixed point. This indicates that exponentially large lattice sizes are needed to ultimate reach the fixed point.
By contrast, in the $C_{2v}$ model with $a_+ = 0$ and $a_- = 2$, $v_\parallel$ is a relevant parameter near the maximal-anisotropy fixed point and flows to larger values. By numerically integrating out the flow, we find that the parameters $v_\perp$, $v_\parallel$, and $g^2$ flow to a new nontrivial stable fixed point at which the boson anomalous dimensions satisfy a sum rule, $\eta_++\eta_- + 2\eta_\phi  = 2\epsilon$ with $0 = \eta_+ < \eta_-, \eta_\phi < \epsilon$. The fixed point is located at
$(v_\parallel^*, v_\perp^*)  = (0.1611, 0.5942)$ and
$(g^{2}_{*},\lambda^*) = (0.2123,0.1755)\epsilon + \mathcal O(\epsilon^2)$
for $N' = 4N_\sigma = 4$. We find the corresponding anomalous dimensions as
$(\eta_\phi,\eta_+,\eta_-,\eta_\psi) = (0.7391, 0, 0.5219, 0.1643)\epsilon + \mathcal O(\epsilon^2)$, reflecting again the fact that the character of the stable fixed point in the $C_{2v}$ model is different from the one of the $C_{4v}$ model.
The different behaviors of the Fermi velocities in the two models is illustrated in Fig.~\ref{fig:rg-flow}, which shows the renormalization group flow in the $v_\parallel$-$v_\perp$ plane. For visualization purposes, we have fixed the ratios $g^2/(v_\perp v_\parallel)$ to their values at the respective stable fixed points in these plots. We have explicitly verified that $g^2/(v_\perp v_\parallel)$ corresponds to an irrelevant parameter near these fixed points (marked as red dots in Fig.~\ref{fig:rg-flow}).

\begin{figure}[t]
\includegraphics[width=\linewidth]{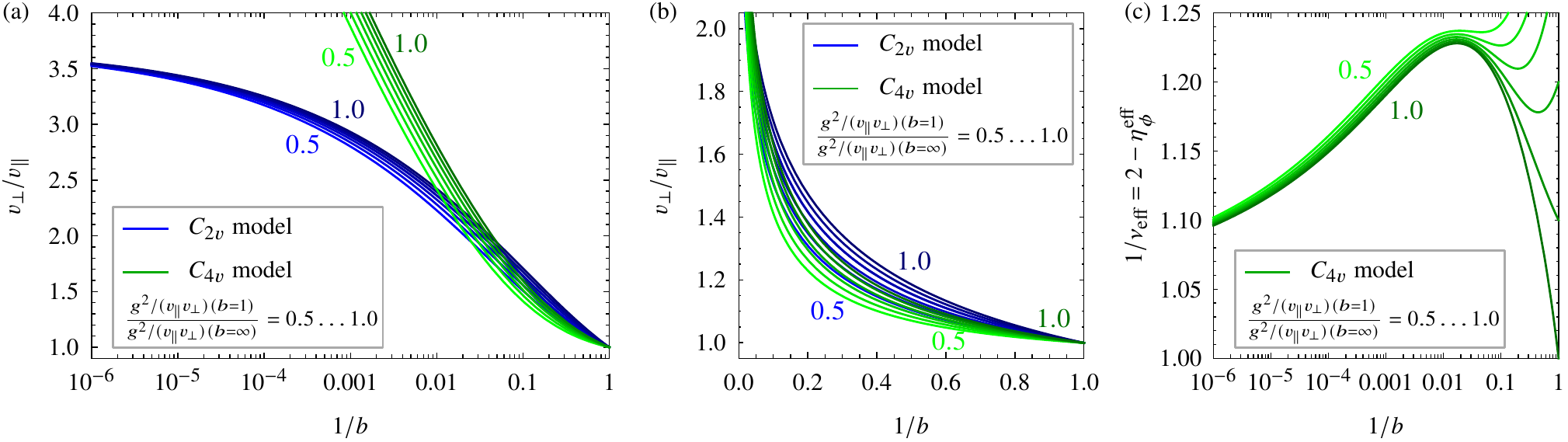}
\caption{(a,b) Ratio of Fermi velocities $v_\perp/v_\parallel$ as function of RG scale $1/b$ in the $C_{2v}$ model (blue) and $C_{4v}$ model (green) for different starting values of the interaction parameter $g^2/(v_\parallel v_\perp)$ at the ultraviolet scale $b=1$. 
Here, we have numerically integrated out the full RG flow in the $(v_\parallel, v_\perp, g^2)$ parameter space, assuming initial velocities
%ultraviolet starting values of 
$v_\parallel(b=1) = v_\perp(b=1) = 0.25$, and $g^2/(v_\parallel v_\perp)(b=1)$ between 50\% and 100\% of the value at the respective stable fixed point. (a) Semilogarithmic plot, demonstrating the finite infrared anisotropy in the $C_{2v}$ model and the logarithmic divergence in the $C_{4v}$ model. (b) Same data as in (a), but using a linear plot, illustrating the similarity of the anisotropy flows in the two models on the high-energy scale, to be compared with the QMC data shown in Fig.~3(e) of the main text.
(c) Effective correlation-length exponent $1/\nu_\text{eff}$ as function of RG scale $1/b$ in the $C_{4v}$ model in semilogarithmic plot, defined according to Eq.~\eqref{eq:nu-eff}, illustrating the drifting of the exponents and the quasiuniversal behavior for $1/b \lesssim 0.05$. We have used the same ultraviolet starting values as in (a,b).}
\label{fig:anisotropy}
\end{figure}

To make further contact with the QMC data displayed in Fig.~3(e) of the main text, we show in Fig.~\ref{fig:anisotropy}(a,b) the Fermi velocity ratio $v_\perp/v_\parallel$ as function of RG scale $1/b$ in the two models, assuming an isotropic ratio $v_\perp/v_\parallel = 1$ at the ultraviolet scale $b=1$, for different initial values of the interaction parameter $g^2/(v_\parallel v_\perp)$. We emphasize that a sizable deviation between the two models is observable only at very low energies $1/b \lesssim 0.01$, while the RG flows in the high-energy regime are very similar for the employed starting values. 
Identifying the RG energy scale $1/b$ roughly with the inverse lattice size $1/L$, this result explains why the lattice sizes available in our simulations are too small to detect a substantial difference in the finite-size scaling of $v_\perp/v_\parallel$.
This also implies that the estimates for the critical exponent obtained from the finite-size analysis of the QMC data describes only an intermediate regime, in which the RG flow is not yet fully integrated out.
Let us illustrate this point further for the case of the $C_{4v}$ model. In this case, we can define a scale-dependent \emph{effective} correlation-length exponent by using the scaling relation
\begin{align}
\label{eq:nu-eff}
	1/\nu_\text{eff}(b) = 2 - \eta_\phi^\text{eff}(b),
\end{align}
where $\eta_\phi^\text{eff}(b) = N' g^2(b) /[12 v_\perp(b) v_\perp(b)]$ is the effective boson anomalous dimension. This relation becomes exact in the vicinity of the $C_{4v}$ fixed point, for which $\lambda^\ast = 0$.
The effective correlation-length exponent $1/\nu_\text{eff}$ is plotted as function of the RG scale $1/b$ in Fig.~\ref{fig:anisotropy}(c) for different values of the initial interaction parameter $g^2/(v_\parallel v_\perp)$. We note that the approach to $\nu_\text{eff} \to 1$ in the deep infrared is extremely slow, with sizable deviations from the fixed-point value at intermediate scales.
Interestingly, while the behavior in the high-energy regime $1/b \gtrsim 0.05$ is nonuniversal and strongly depends on the particular starting values of the RG flow, a quasiuniversal regime emerges at intermediate energy $1/b \lesssim 0.05$, in which the exponents still drift, but have only a very weak dependence on the initial interaction parameters. This quasiuniversal behavior is a characteristic feature of systems with marginal or close-to-marginal operators~\cite{Nahum15, Nahum19}. Here, it arises from the slow flow of the velocity anisotropy ratio $v_\perp/v_\parallel$, which implies that the effective exponents will become functions of $v_\perp/v_\parallel$ only, but not of the ultraviolet starting values of the interaction parameters.
The quasiuniversality reflects the fact that there is only \emph{one} slowly decaying perturbation to the fixed point (i.e., the leading irrelevant operator), whereas all other perturbations decay quickly, and hence have died out once $1/b\lesssim 0.05$.
Importantly, the largest lattice sizes available in the QMC simulations appear to be just large enough to approach the quasiuniversal regime, if we again identify $1/b$ roughly with $1/L$. Reassuringly, for $L=20$, we therewith obtain the RG estimate $1/\nu_\text{eff} \simeq 1.20 \dots 1.25$, which is in the same ballpark as the estimate from the finite-size scaling analysis of the QMC data discussed in the main text.

%%%%%%%%%%%%%%%%%%%%%%%%%%%%%%%%%%%%%%%%%%%%%%%%%%%%%%%
\section{Observables}
%%%%%%%%%%%%%%%%%%%%%%%%%%%%%%%%%%%%%%%%%%%%%%%%%%%%%%%
 
In this section, we define the  observables used throughout this work  to  study  the   quantum phase  transition.   
We  have considered  quantities  based on bosonic and fermionic degrees  of freedom. 

\subsection{ Bosonic  degrees of freedom}

The structure factor $S(\ve{k})$ and susceptibility $\chi(\ve{k})$  are defined as
\begin{align}
S(\ve{k}) &= \frac{1}{N^2}\sum_{\langle \ve{R}, \ve{R}'\rangle} 
                             \left(
                               \langle s^z_{\ve{R}} s^z_{\ve{R}'}\rangle  
                               - \langle s^z_{\ve{R}}\rangle \langle s^z_{\ve{R}'}\rangle
                             \right) e^{i\ve{k} (\ve{R} - \ve{R}')}, 
\end{align}
and
\begin{align}
  \chi(\ve{k}) &= \int\!\text{d}\tau\, \frac{1}{N^2}\sum_{\langle \ve{R},\ve{R}'\rangle} 
                             \left(
                               \Braket{ s^z_{\ve{R}}(0) s^z_{\ve{R}'}(\tau) }
                               - \Braket{ s^z_{\ve{R}}(0) } \Braket{ s^z_{\ve{R}'}(\tau) }
                             \right) e^{i\ve{k} (\ve{R} - \ve{R}')}.
\end{align}
Both $S(\ve{k}=0)$ and $\chi(\ve{k}=0)$ are suitable order parameters for the paramagnetic-ferromagnetic phase transision. Note  that in  the main text, these observables are defined without subtraction of the background $\langle s^z_{\ve{R}}\rangle \langle s^z_{\ve{R}'}\rangle$, or $\Braket{ s^z_{\ve{R}}(0) } \Braket{ s^z_{\ve{R}'}(\tau) }$. Generically, this is generally equivalent, since in a fully ergodic simulation   the  background   vanishes by symmetry.    In fact,   the  global move    in the Monte Carlo sampling that flips all the  spins   has an acceptance of unity such that   the background is 
identical to zero. 
In some  case,  it is convenient to omit  this  global move. In fact to image   the meandering of the cones,  Sec.~\ref{sec:gap},   we   omitted  the global  move  so as to achieve $\braket{s^z} > 0$ and observe the displaced Dirac cones.

By definition, renormalization group   invariant  quantities  have   vanishing   scaling dimension.  They  can be derived 
from  the  correlation function and  susceptibilities  in terms  of the correlation ratios   $R_S$ and $R_\chi$
\begin{align}
R_O = 1 - \frac{O(\ve{k}_{\text{min}})}{O(\ve{k}=\ve{0})}
\quad \text{with} \quad O = S, \chi,
\end{align}
where $\ve{k}_{\text{min}}$ corresponds to the longest wave length on the   considered  lattice.  Generally, one can chose instead of $\ve{k}_{\text{min}}$ any wave-vector that approaches $\ve{k}=0$ as $1/L$ to achieve the same asymptotic behavior.   However, we have found that 
using  $\ve{k}_{\text{min}}$   works best for us.
Another   RG-invariant  quantity  is the Binder ratio,  $B$,  defined as
\begin{align}
  B &= \frac{1}{2}\left( 3 - \frac{ \Braket{(s^z)^4} }{ \Braket{(s^z)^2}^2 } \right).
\end{align}

To  provide  further information on the nature of the transition, we have considered  the derivative of the free energy, 
\begin{align}
  \frac{1}{N}\frac{\partial F}{\partial h} = \Braket{ \frac{1}{N} \sum_{\ve{R}} s^x_{\ve{R}} }  \equiv X 
\end{align}

%%%%%%%%%%%%%%%%%%%%%%%%%%%%%%%%%%%%%%%%%%%%%%%%%%%%%%%
\subsection{Fermionic degrees of freedom}
The fermionic observables consist of the momentum-resolved single-particle gap $\Delta_\text{sp}(\ve{k})$ which we use to image the  meandering of Dirac points.  We furthermore use  this  quantity to   determine  the velocity anisotropy.  
    
\subsubsection{Fermionic single-particle gap} 

\label{sec:gap}  
To properly define $\Delta_\text{sp}(\ve{k})$, we first introduce an energy   basis:  
\begin{align}
  \mc{H} \Ket{\Psi^N_n(\ve{k})} = E^N_n(\ve{k}) \Ket{\Psi^N_n(\ve{k})}, 
\end{align}
where $\Ket{\Psi^N_n(\ve{k})}$ are also eigenstates of particle number $\hat{N}$ and momentum $\hat{\ve{k}}$ operators:
\begin{align}
  \hat{N} \Ket{\Psi^N_n(\ve{k})} &= N \Ket{\Psi^N_n(\ve{k})}
  &
  \hat{\ve{k}} \Ket{\Psi^N_n(\ve{k})} &= \ve{k} \Ket{\Psi^N_n(\ve{k})}
\end{align}
In this basis, the gap is:
\begin{align}
  \Delta_\text{sp}(\ve{k}) = E_0^{N_0+1}(\ve{k}) - E_0^{N_0},
\end{align}
where $N_0$ is the particle number of the half-filled system.

Now consider the time-displaced Green function
\begin{align}
  G(\ve{k},\tau) = \Braket{\hat{c}_{\ve{k}}(\tau) \hat{c}_{\ve{k}}^\dag}
  \quad \text{with} \quad
  \hat{c}_{\ve{k}}(\tau) = e^{\tau \mc{H}} \hat{c}_{\ve{k}} e^{-\tau \mc{H}}.
\end{align}
Assuming a unique ground state,   the $T=0$ Green   function reads: 
\begin{align}
  \lim_{\beta \rightarrow \infty} G(\ve{k},\tau)
  = \Braket{\Psi^{N_0}_0 | \hat{c}_{\ve{k}}(\tau) \hat{c}_{\ve{k}}^\dag | \Psi^{N_0}_0} 
  = \sum_n e^{-\tau \left(E_n^{N_0+1}(\ve{k}) - E_0^{N_0}\right)} 
       \left|\Braket{\Psi^{N_0+1}_n(\ve{k}) | \hat{c}_{\ve{k}}^\dag | \Psi^{N_0}_0} \right|^2.
\end{align}
Provided that the wave function renormalization, $\left|\Braket{\Psi^{N_0+1}_n(\ve{k}) | \hat{c}_{\ve{k}}^\dag | \Psi^{N_0}_0} \right|^2$ is finite and that $\ket{\Psi^{N_0+1}_0(\ve{k})}$ is non-degenerate, then
\begin{align}
  \label{eq:green_gap}
  \lim_{\tau \rightarrow \infty} \lim_{\beta \rightarrow \infty} G(\ve{k},\tau)
   = e^{-\tau \left(E_0^{N_0+1}(\ve{k}) - E_0^{N_0}\right)} 
       \left|\Braket{\Psi^{N_0+1}_0(\ve{k}) | \hat{c}_{\ve{k}}^\dag | \Psi^{N_0}_0} \right|^2
 \end{align}
and we can extract $\Delta_\text{sp}(\ve{k}) = E_0^{N_0+1}(\ve{k}) - E_0^{N_0}  $ by   fitting   the tail of $G(\ve{k},\tau)$   to an exponential form. 

\begin{figure}
\includegraphics[width=\linewidth]{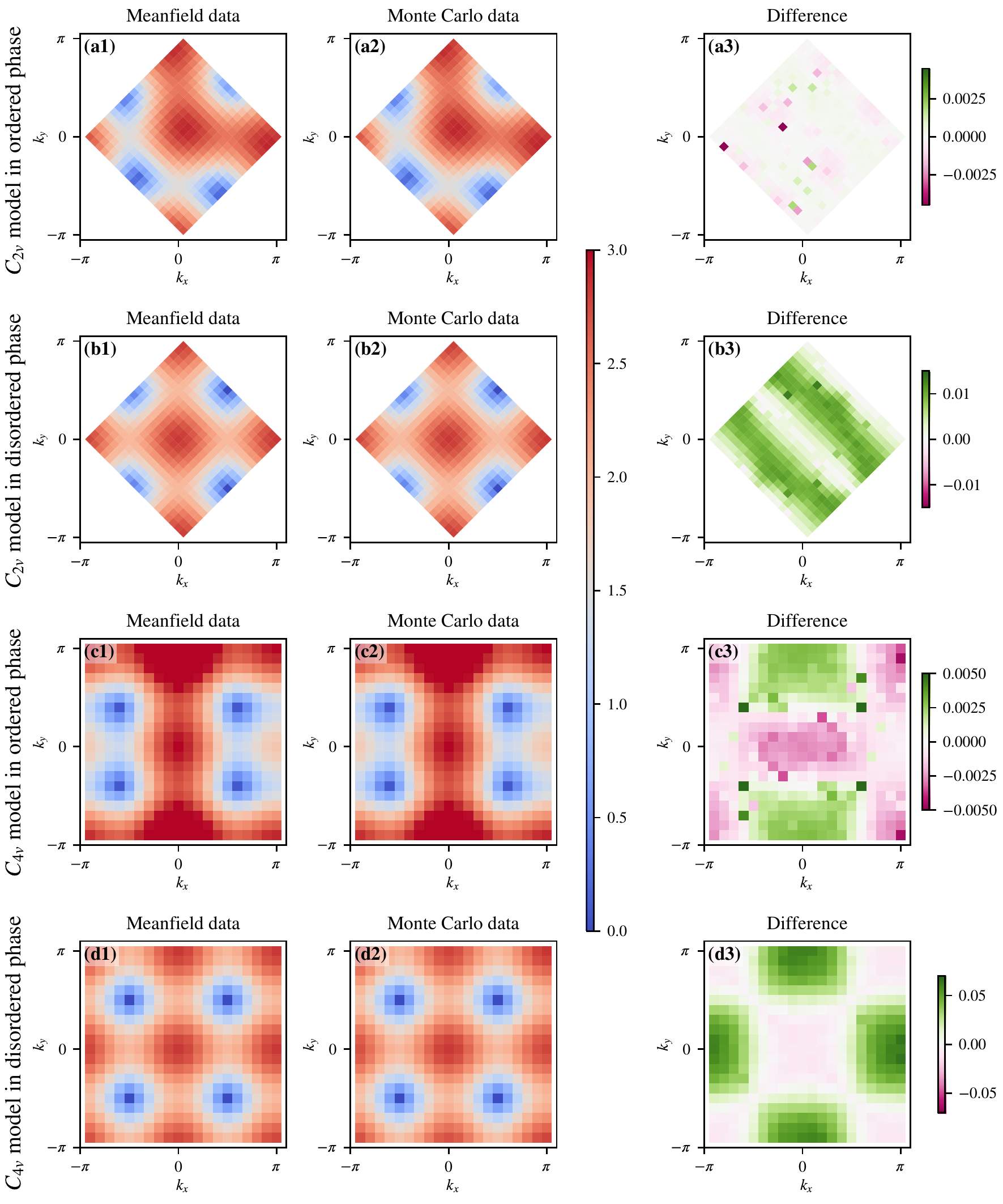}
\caption{
\label{fig:dispersion_mc_vs_mf}
Testing the dispersion of the fermionic single particle gap obtained from Monte Carlo data against mean field results. The first column shows the mean field results according to Eg.~\eqref{eq:mf-dispersion}, while the second column shows numerical results obtained by employing Eq.~\eqref{eq:green_gap} and the last column shows mean field minus Monte Carlo results.
\textbf{(a)}~$C_{2v}$ model in ordered phase, at $h=1.0$. The value of the mean field parameter $\phi$ is set to $\langle s^z \rangle$ from the Monte Carlo simulation. 
\textbf{(b)}~$C_{2v}$ model in disordered phase, at $h=5.0$. The value of the mean field parameter is set to $\phi=0$.
\textbf{(c)}~Same as (a), but for the $C_{4v}$ model.
\textbf{(d)}~Same as (b), but for the $C_{4v}$ model.
}
\end{figure}

In Fig.~\ref{fig:dispersion_mc_vs_mf} we show that this approach works, by comparing the dispersions deep in the disordered and ordered phases to mean field results. Note that  in a fully ergodic Monte Carlo  simulation  we would  sample both options for breaking the Ising $\mathbb{Z}_2$ symmetry.   As  mentioned  previously    and  to produce the  results of  Figs.~\ref{fig:dispersion_mc_vs_mf}(a2, c2),  we  have  omitted  the  global move that  flips all  the spins and comes with a unit acceptance. 

We observe a slight systematic derivation between mean field results and Monte Carlo data in the disordered phase. This  stems  from 
fluctuations of the order parameter  in the  vicinity  of the  critical point.  

\subsubsection{Fermi velocity anisotropy \texorpdfstring{$v_\perp / v_\parallel$}{vperp/vparallel}}
 With  $\Delta_\text{sp}(\ve{k})$ we  can extract the anisotropy at the nodal points $\ve{K}$,  via, 
 \begin{align}
\label{eq:v_anisotropy}
\frac{v_\perp}{v_\parallel} &= \lim_{\delta \rightarrow 0}
\frac{\Delta_\text{sp}(\ve{K}+\delta \ve{e}_\perp) - \Delta_\text{sp}(\ve{K})}
     {\Delta_\text{sp}(\ve{K}+\delta \ve{e}_\parallel) - \Delta_\text{sp}(\ve{K})}.
\end{align}
Where $\ve{e}_\perp$ and $\ve{e}_\parallel$ are unit vectors perpendicular and parallel to the meandering direction of the Dirac cones. We have considered three different strategies for approaching this limit on the finite size lattices, that are all equivalent in the thermodynamic limit:

\subparagraph{1. The direct approach:}
The most straightforward implementation of Eq.~\eqref{eq:v_anisotropy} on a finite lattice would be
\begin{align}
\label{eq:v_anisotropy1}
\frac{v_\perp}{v_\parallel} &=
\frac{\Delta_\text{sp}(\ve{K}+\ve{\delta}_\perp) - \Delta_\text{sp}(\ve{K})}
     {\Delta_\text{sp}(\ve{K}+\ve{\delta}_\parallel) - \Delta_\text{sp}(\ve{K})},
\end{align}
where $\ve{\delta}_\perp$, $\ve{\delta}_\parallel$ are the shortest distances  from the nodal point  on the finite-size $\ve{k}$-Lattice.

\subparagraph{2. Manually setting the finite size gap $\Delta_\text{sp}(\ve{K}) = 0$:} 
This approach makes sense, since we know that in the themordynamic limit the gap vanishes.
With this strategy, Eq.~\eqref{eq:v_anisotropy} takes the form
\begin{align}
\label{eq:v_anisotropy2}
\frac{v_\perp}{v_\parallel} &=
\frac{\Delta_\text{sp}(\ve{K}+\ve{\delta}_\perp)}
     {\Delta_\text{sp}(\ve{K}+\ve{\delta}_\parallel)}.
\end{align}

\subparagraph{3. Avoid the nodal points:}
Another approach for avoiding the finite size gap is to measure one step away from it:
\begin{align}
\label{eq:v_anisotropy3}
\frac{v_\perp}{v_\parallel} &=
\frac{\Delta_\text{sp}(\ve{K}+2\ve{\delta}_\perp) - \Delta_\text{sp}(\ve{K}+\ve{\delta}_\perp)}
     {\Delta_\text{sp}(\ve{K}+2\ve{\delta}_\parallel) - \Delta_\text{sp}(\ve{K}+\ve{\delta}_\parallel)},
\end{align}

\begin{figure}
\centering
\includegraphics[width=\linewidth]{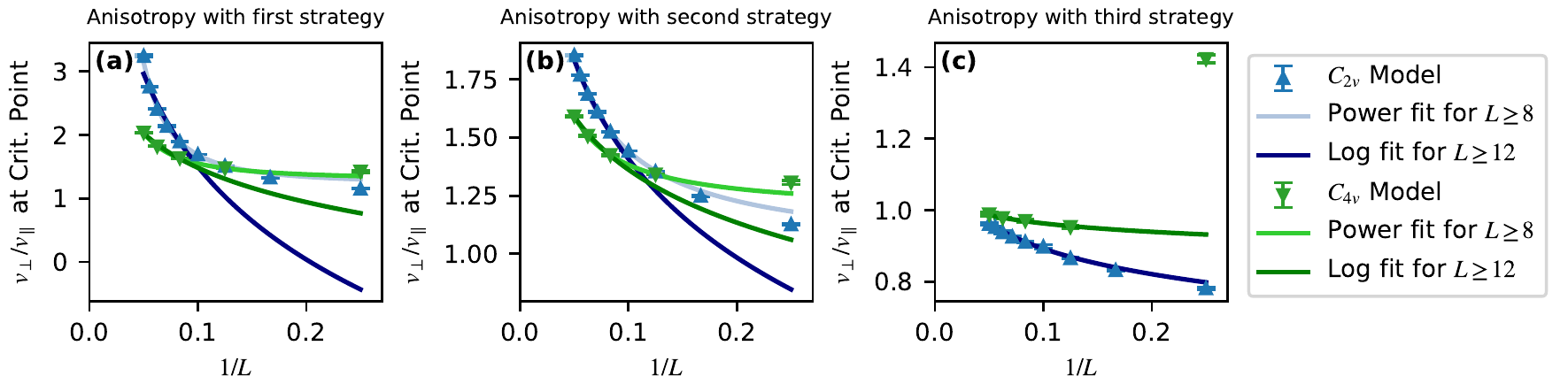}
\caption{
\label{fig:v_anisotropy_test}
Fermi velocity anisotropy at the  critical point as function of $1/L$, determined   with  (a):~Eq.~\eqref{eq:v_anisotropy1},  (b):~Eq.~\eqref{eq:v_anisotropy2},  (c):~Eq.~\eqref{eq:v_anisotropy3}.  Power law and logarithmic fits are shown, except for (c), where only a logarithmic fit is performed. }
\end{figure}

The results for these different approaches are shown in Fig.~\ref{fig:v_anisotropy_test}.
The third strategy results in velocity anisotropies $<1$, while Fig.~2(c) in the main text clearly shows that $v_\perp / v_\parallel > 1$ at the critical point. This implies that the approach strongly underestimates the anisotropy due to the fact that the considered lattices sizes are too small for not measuring directly at the nodal point.

The other two approaches, while not giving quantitatively the same results, are qualitatively equivalent. We have opted to use the second strategy, corresponding to Eq.~\eqref{eq:v_anisotropy2}.

%%%%%%%%%%%%%%%%%%%%%%%%%%%%%%%%%%%%%%%%%%%%%%%%%%%%%%%
\section{Critical exponents}
%%%%%%%%%%%%%%%%%%%%%%%%%%%%%%%%%%%%%%%%%%%%%%%%%%%%%%%

\subsection{Correlation length exponent \texorpdfstring{$\nu$}{nu} from  RG invariant quantities.}

A  renormalization  group   quantity,  $R$,   has  by  definition a vanishing scaling  dimension.      Consider a system  at  temperature $\beta$,    of  size 
$L_{+}  \times L_{-} $    with  a single  relevant  coupling $h$. Under a  renormalization   group transformation that   rescales  $L_{+}    \rightarrow L_{+}/b$  with  $b>1$,  we   expect \cite{Goldenfeld}:
\begin{equation}
	   R( (h - h_c) ,   \beta,  L_{+}, L_{-}) =  R( (h-h_c)',   \beta/b^{z},  L_{+}/b, L_{-}/b^{1+ \Delta  z }). 
\end{equation}
In the above  $\Delta  z \neq 0 $  encodes  the  difference in scaling between the   the $L_{+} $ and $L_{-}$  directions. 
Linearization of the RG  transformation,   $(h-h_c)'   =  b^{1/\nu}  (h-h_c)$  and  setting   the scale  $b= L$   as  well as   $L_{-} = L_{+} = L  $, in accordance to our  simulations,   yields:  
\begin{equation}
  \label{eq:rg-invariant-scaling}
  R \left( (h - h_c) ,   \beta, L  \right)     =   f \left( L^{1/\nu} (h -h_c),  L^z/\beta ,  L^{ - \Delta z } , L^{-\omega} \right). 
\end{equation}
In the above we  have  accounted for   possible  corrections to scaling $ L^{-\omega} $.   In the presence  of a single  length scale  
$  \Delta  z = 0  $,  such that  the  generic  finite size scaling form  is recovered.    

\begin{figure}
\centering
\includegraphics[scale=1]{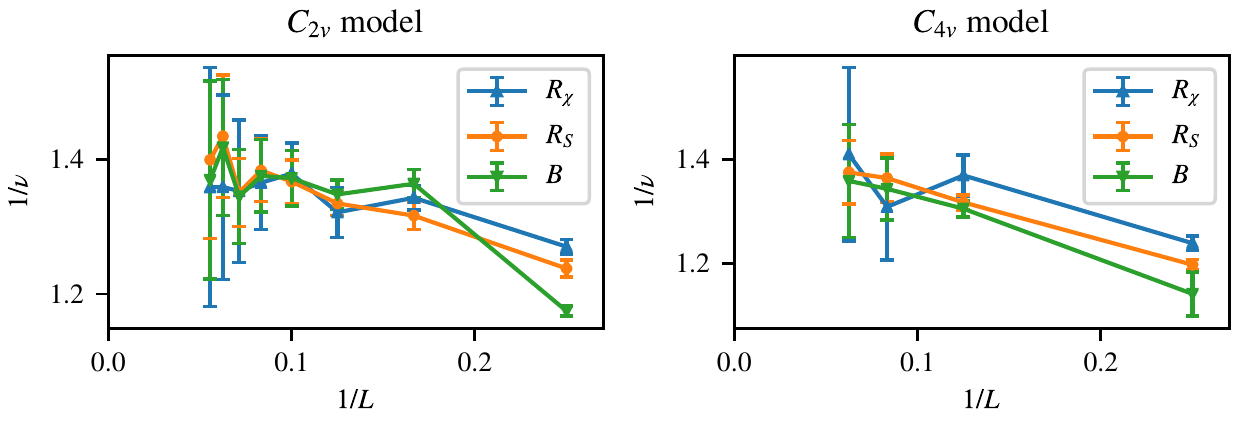}
\caption{
\label{fig:nu_extrapolate}
Critical exponent $1/\nu$ of $C_{2v}$ ($C_{4v}$) model from pairwise data collapse of RG-invariant quantities, using linear system sizes $L$ and $L+2$ ($L+4$). }
\end{figure}

Since in  our  simulations the temperature is representative of the ground state, we can neglect the dependence on $L^{z}/\beta$. Up to corrections to scaling, $\omega$, and  the possibility of  $ \Delta z \neq 0 $,  which   would result in another  correction to  scaling term,  the  data for different lattice sizes  cross at the critical field $h_\mathrm{c}$ and should collapse when plotted as function of $(h-h_c)L^{1/\nu}$. The results for such data collapses are shown in Tables~\ref{tab:nu_c2v}, \ref{tab:nu_c4v}. Furthermore, Fig.~\ref{fig:nu_extrapolate} shows $1/\nu$ for the  $C_{2v}$ and $C_{4v}$  models  from pairwise data collapse of RG-invariant quantities, using system sizes $L$ and $L+2$ ($L+4$). Both suggest a relatively well converged result for $L \geq 12$.    Although seemingly  converged, our system sizes are  too small to   detect a logarithmic  drift in the  exponents.

\begin{table}
\centering
\caption{
\label{tab:nu_c2v}
Data collapse results for RG-invariant quantities of $C_{2v}$ model.}
\begin{tabular}{crccr}
\toprule
Observables & Used system sizes       &        $h_c$             &         $1/\nu$          & $\chi^2$ \\
\midrule
$R_S$    &  8, 10, 12, 14, 16, 18, 20 &  $3.272715 \pm 0.000074$ &  $1.358934 \pm 0.001824$ &   2.4 \\
$R_S$    &     10, 12, 14, 16, 18, 20 &  $3.272222 \pm 0.000065$ &  $1.373499 \pm 0.002887$ &   1.8 \\
$R_S$    &         12, 14, 16, 18, 20 &  $3.272304 \pm 0.000099$ &  $1.376110 \pm 0.006104$ &   1.9 \\
$R_S$    &             14, 16, 18, 20 &  $3.272617 \pm 0.000138$ &  $1.375085 \pm 0.007038$ &   1.8 \\
$R_S$    &                 16, 18, 20 &  $3.272521 \pm 0.000274$ &  $1.368184 \pm 0.014756$ &   1.9 \\
$R_S$    &                     18, 20 &  $3.273322 \pm 0.000449$ &  $1.373454 \pm 0.022000$ &   1.9 \\
$B$      &  8, 10, 12, 14, 16, 18, 20 &  $3.270955 \pm 0.000089$ &  $1.317011 \pm 0.002273$ &  13.4 \\
$B$      &     10, 12, 14, 16, 18, 20 &  $3.271532 \pm 0.000141$ &  $1.344485 \pm 0.004371$ &   4.6 \\
$B$      &         12, 14, 16, 18, 20 &  $3.272535 \pm 0.000175$ &  $1.352409 \pm 0.006242$ &   3.0 \\
$B$      &             14, 16, 18, 20 &  $3.273181 \pm 0.000152$ &  $1.361056 \pm 0.007298$ &   2.6 \\
$B$      &                 16, 18, 20 &  $3.273783 \pm 0.000210$ &  $1.370788 \pm 0.013022$ &   2.2 \\
$B$      &                     18, 20 &  $3.274325 \pm 0.000464$ &  $1.340845 \pm 0.028414$ &   1.8 \\
$R_\chi$ &  8, 10, 12, 14, 16, 18, 20 &  $3.281138 \pm 0.000030$ &  $1.421387 \pm 0.000002$ &  22.0 \\
$R_\chi$ &     10, 12, 14, 16, 18, 20 &  $3.279752 \pm 0.000073$ &  $1.381046 \pm 0.000002$ &   7.3 \\
$R_\chi$ &         12, 14, 16, 18, 20 &  $3.275155 \pm 0.000500$ &  $1.369774 \pm 0.001609$ &   5.8 \\
$R_\chi$ &             14, 16, 18, 20 &  $3.277021 \pm 0.000233$ &  $1.338662 \pm 0.010808$ &   2.2 \\
$R_\chi$ &                 16, 18, 20 &  $3.276434 \pm 0.000176$ &  $1.342788 \pm 0.004183$ &   2.2 \\
$R_\chi$ &                     18, 20 &  $3.275856 \pm 0.000693$ &  $1.369095 \pm 0.034679$ &   2.4 \\
\bottomrule
\end{tabular}
\end{table}  

\begin{table}
\centering
\caption{
\label{tab:nu_c4v}
Data collapse results for RG-invarian quantities of $C_{4v}$ model.}
\begin{tabular}{crccr}
\toprule
Observables & Used system sizes &        $h_c$     &         $1/\nu$   & $\chi^2$ \\
\midrule
$R_S$    &  8, 12, 16, 20 &  $3.64606 \pm 0.00007$ & $1.328 \pm 0.006$ & 18.7 \\
$R_S$    &     12, 16, 20 &  $3.64886 \pm 0.00011$ & $1.381 \pm 0.011$ &  3.2\\
$R_S$    &         16, 20 &  $3.65108 \pm 0.00022$ & $1.402 \pm 0.023$ &  1.7\\
$B$      &  8, 12, 16, 20 &  $3.64108 \pm 0.00009$ & $1.254 \pm 0.007$ & 73.2\\
$B$      &     12, 16, 20 &  $3.64818 \pm 0.00012$ & $1.340 \pm 0.014$ &  5.1\\
$B$      &         16, 20 &  $3.65138 \pm 0.00025$ & $1.362 \pm 0.026$ &  1.8\\
$R_\chi$ &  8, 12, 16, 20 &  $3.66319 \pm 0.00027$ & $1.309 \pm 0.018$ & 16.5\\
$R_\chi$ &     12, 16, 20 &  $3.65708 \pm 0.00031$ & $1.368 \pm 0.028$ &  3.1\\
$R_\chi$ &         16, 20 &  $3.65537 \pm 0.00070$ & $1.428 \pm 0.092$ &  2.4\\
\bottomrule
\end{tabular}
\end{table}

\subsection{Scaling dimensions and scaling anisotropy}

Next, we examine the scaling dimension of the   bosonic  field  from the Ising spin correlations: 
\begin{align}
S(\ve{x}) = 
\Braket{\hat{s}^z_{\bf R}(\tau) \hat{s}^z_{{\bf 0}}(0)}
\end{align}
where $\ve{x} = (\ve{R}, \tau)$ is a space-time coordinate. 
The models considered in this research    are not  Lorentz invariant   such that the  scaling dimension  acquires a  direction dependence.  
Following  Eq.~\eqref{eq:rg-invariant-scaling} we   expect: 
\begin{align}
S\left( r \hat{\ve{d}}_*, h \right) \propto 
  \frac{1}{ | r \hat{\ve{d}}_*|^{2 \Delta_{s,*}}} f\left( L^z/\beta, (h-h_c)L^{1/\nu}, L^{-\Delta z} , L^{-\omega} \right)
\end{align}
where  $\hat{\ve{d}}_*$    defines  the  direction.

\begin{figure}
\includegraphics[width=\linewidth]{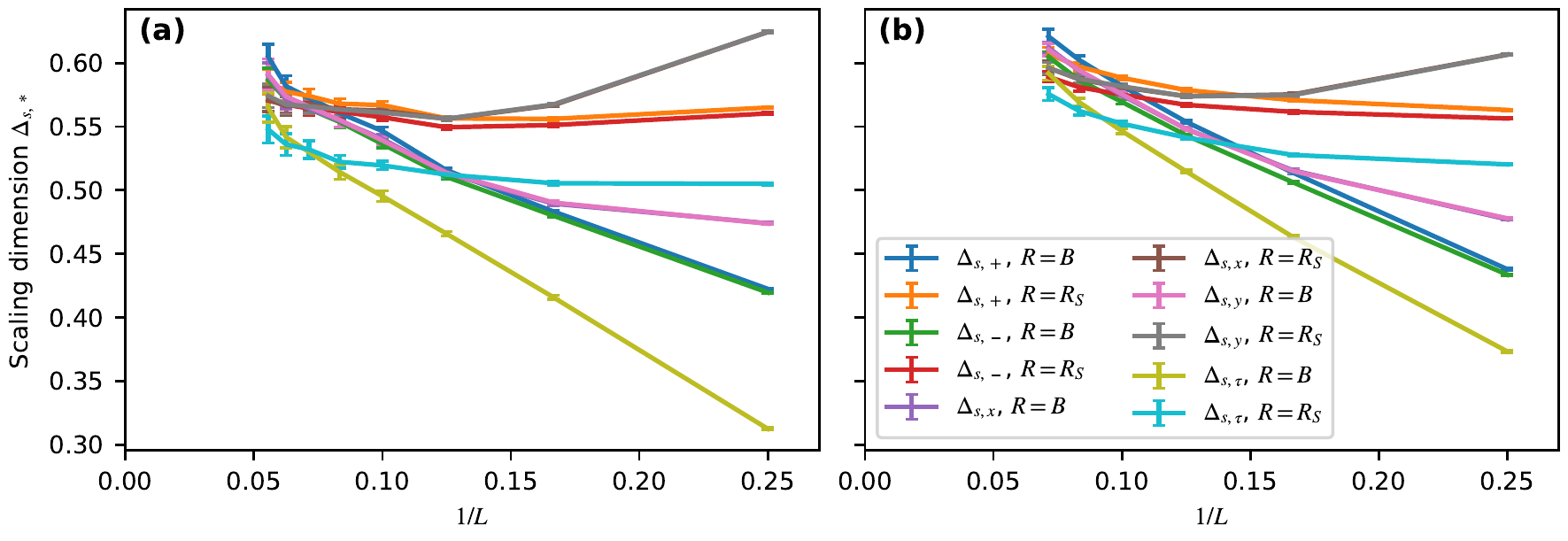}
\caption{
\label{fig:scaling_anisotropy_c2v}
Scaling dimension of Ising field of $C_{2v}$ model. For \textbf{(a)} $\xi=0.25$, \textbf{(b)} $\xi=0.4$. Note: $\Delta_{s,y}$, $R=B$ is indistinguishable from  $\Delta_{s,x}$, $R=B$ and $\Delta_{s,y}$, $R=R_S$  is identical to  $\Delta_{s,x}$, $R=R_S$.
}
\end{figure}

\begin{figure}
  \begin{minipage}[b]{0.5\textwidth}
    \includegraphics[width=\linewidth]{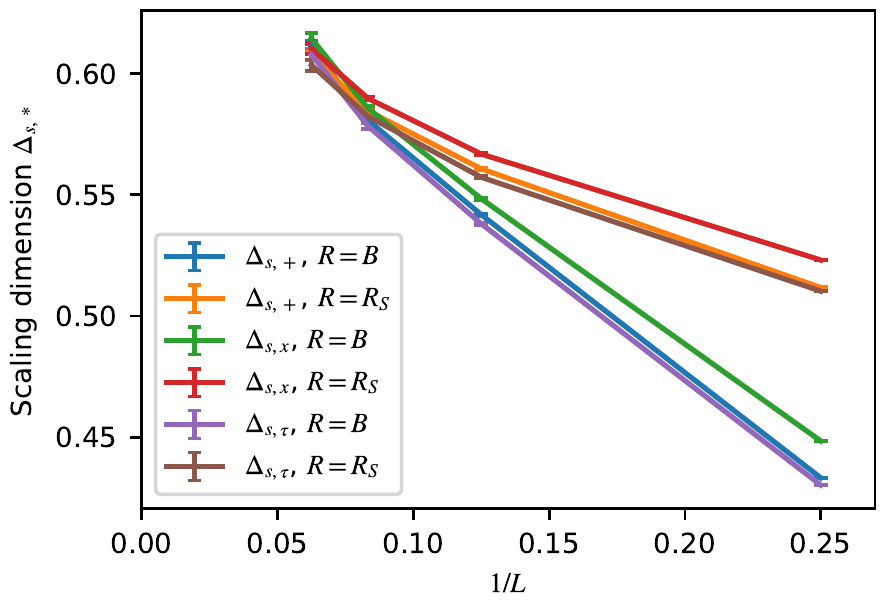}
    \caption{
      \label{fig:scaling_anisotropy_c4v}
      Scaling dimension of Ising field of $C_{4v}$ model.
    }
  \end{minipage}
  \begin{minipage}[b]{0.09\textwidth}
    $\phantom{.}$
  \end{minipage}
  \begin{minipage}[b]{0.39\textwidth}
    \includegraphics[scale=1]{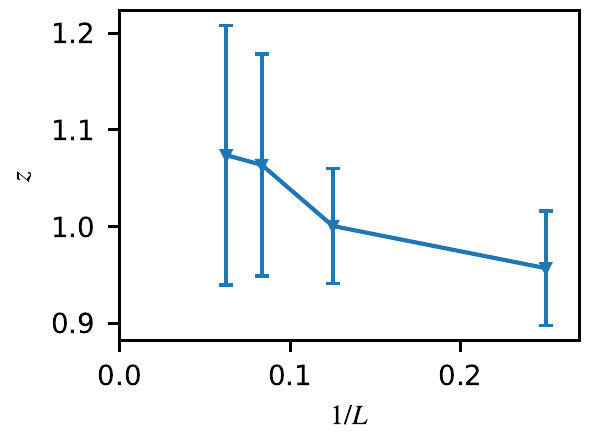}
    \caption{
      \label{fig:z}
      Dynamical exponent $z$ of $C_{4v}$ model.
    }
  \end{minipage}
\end{figure}

To determine the scaling dimensions, we consider $S(L \hat{\ve{d}}_*, h)$, for different system sizes $L$ and use an RG-invariant quantity $R$ to replace in leading order $f( L^z/\beta, (h-h_c)L^{1/\nu},  L^{- \Delta z }, L^{-\omega} ) = \tilde{f}(R)$. Using this form, we perform data collapses using system sizes $L$ and $L+2$ ($L$ and $L+4$), where the only free parameter is $\Delta_{s,*}$. The considered directions are defined in Table~\ref{tab:scaling_directions}, the $C_{4v}$ symmetry of the second model enforces $\Delta_{s,x}=\Delta_{s,y}$ and $\Delta_{s,+}=\Delta_{s,-}$. As the results in Figs.~\ref{fig:scaling_anisotropy_c2v} and \ref{fig:scaling_anisotropy_c4v} show, we cannot resolve a scaling anisotropy between the chosen directions.      We conjecture  that  anisotropies in the exponents will emerge in the   infrared limit. Given the very  slow  flow  we  believe that  our  numerical simulations are  not in a position to probe  these energy scales. 

\begin{table}
\caption{
\label{tab:scaling_directions}
Considered directions for the scaling dimension.
}
\begin{tabular}{cc}
\toprule
$*$ & $\hat{\ve{d}}_*$ \\
\midrule
$x$    & $(\hat{\ve{e}}_x, 0)$ \\
$y$    & $(\hat{\ve{e}}_y, 0)$ \\
$+$    & $\frac{1}{2}(\hat{\ve{e}}_x+\hat{\ve{e}}_y, 0)$ \\
$-$    & $\frac{1}{2}(\hat{\ve{e}}_x-\hat{\ve{e}}_y, 0)$ \\
$\tau$ & $(\ve{0}, 0.3)$ \\
\bottomrule
\end{tabular}
\end{table}

\subsection{Dynamical exponent \texorpdfstring{$z$}{z}}
To determine the dynamical exponent of the $C_{4v}$ model, assume isotropic scaling in space, as suggested by the RG analysis. Then the RG-invariant quantities follow a the critical point the form
\begin{align}
  \label{eq:rg-invariant_iso}
  R = f( L^z/\beta, (h-h_c)L^{1/\nu}, L^{-\omega} ).
\end{align}
At the crossing points $h_*(L)$, with $R(h_*(L), L) = R(h_*(L), L+\Delta_L)$ and $\Delta_L = 4$, we measure $R(\beta)$. Omitting corrections to scaling  leads to  $R(L, \beta) = f( L^z/\beta)$. From this we derive
\begin{align}
  z = \frac
    {\log\left( \frac{\partial_\beta R(L)}{\partial_\beta R(L+\Delta_L)} \right)}
    {\log\left(\frac{L+\Delta_L}{L} \right))}.
\end{align}

The results  are shown in Fig.~\ref{fig:z},  and  are  consistent  with  $z=1$ as  suggested in the  RG analysis. 

\newpage
%%%%%%%%%%%%%%%%%%%%%%%%%%%%%%%%%%%%%%%%%%%%%%%%%%%%%%%
\section{Odd-even effects} \label{sec:odd-even}
\begin{figure}[H]
\includegraphics[width=\textwidth]{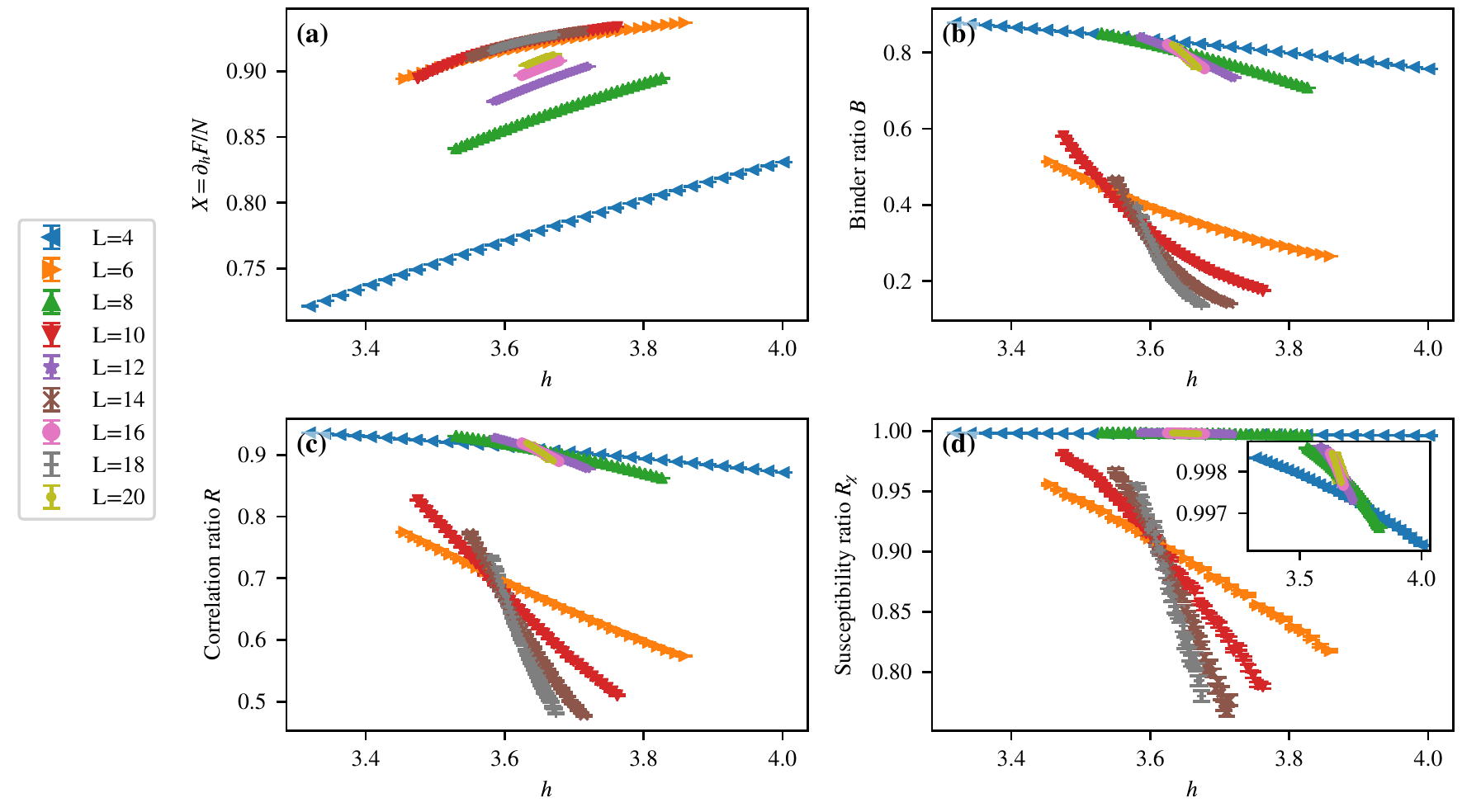}
\caption{
\label{fig:X_RG-inv}
Derivative of free energy and three RG-invariant quantities, showing a continuous transition around $h\approx 3.65$. 
Notable is an odd-even effect between linear system sizes $L \in 4\mathbb{N}$ (=even) and $L \in 4\mathbb{N}+2$ (=odd). 
}
\end{figure}

\begin{figure}[H]
\includegraphics[width=\linewidth]{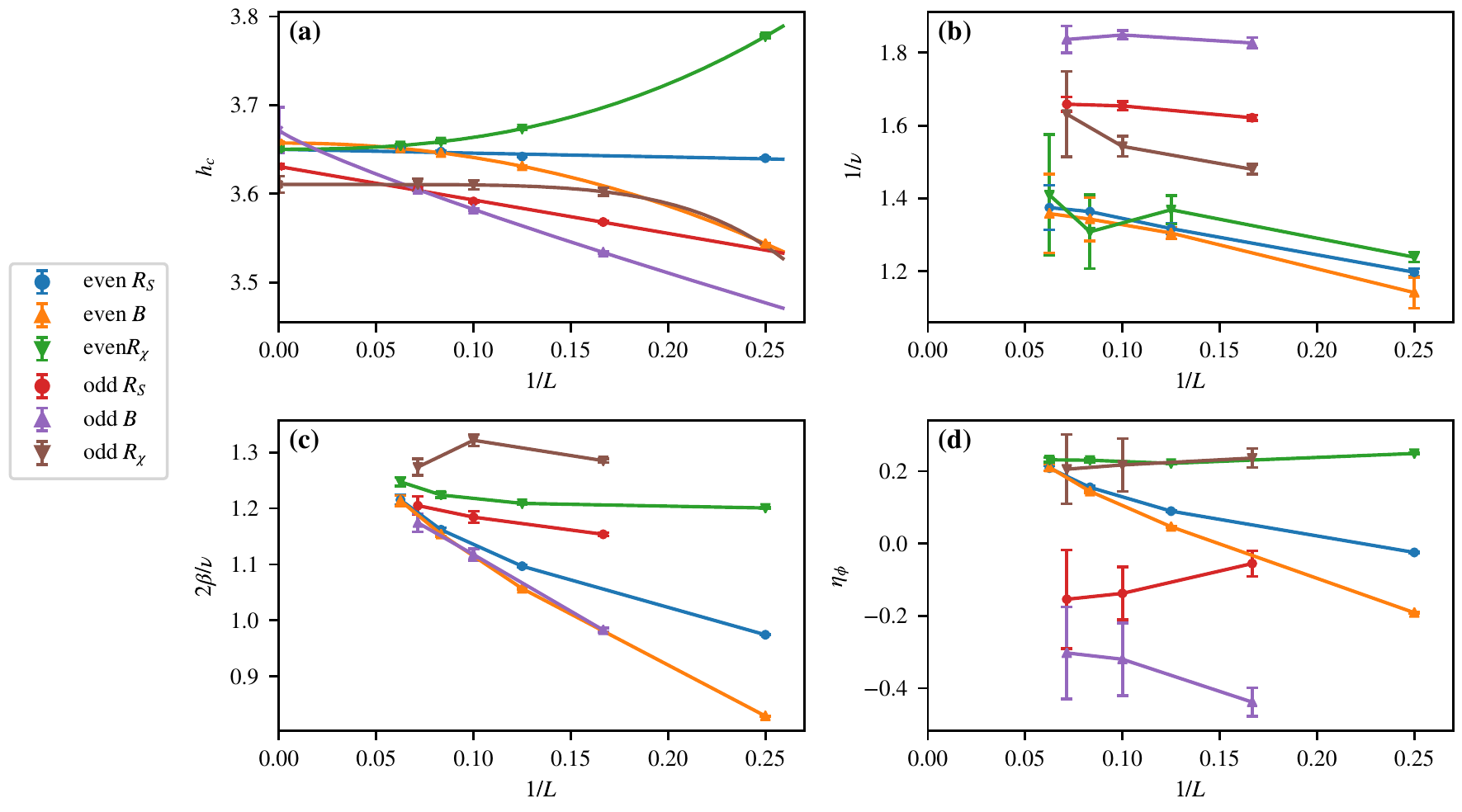}
\caption{
\label{fig:exponents_appendix}
Demonstration of  odd-even effects for  the $C_{4v}$ model.
{\textbf a}: Critical field $h_c$, extracted from the three RG-invariant quantities as determined by the crossing points  between
 linear system sizes $L$ and $L+4$. Odd and even system sizes show different behavior. Even  system size shows better convergence.
{\textbf b}: Critical exponent $1/\nu$ as determined by data collapse of the three RG-invariant quantities, correlation ratio $R$, Binder ratio $B$ and susceptibility ratio $R_\chi$, for linear system sizes $L$ and $L+4$.  Odd and even system sizes show different behavior. Even system size shows better convergence.
{\textbf c}: Critical exponent $2\beta/\nu$ as  determined  with Eq. \ref{eq:exponent_b}.
{\textbf d}: Critical exponent $\eta_\phi$ as  determined with  Eq. \ref{eq:eta_phi}.
}
\end{figure}

%%%%%%%%%%%%%%%%%%%%%%%%%%%%%%%%%%%%%%%%%%%%%%%%%%%%%%%
The $C_{4v}$ model,  has  strong  odd-even effects.  For   linear system sizes $L \in 4\mathbb{N}$ ($\equiv$even)  and  periodic  boundary conditions,  the Dirac points are  included in  the discrete set of $\ve{k}$   vectors. 
This is not the case for  odd lattices,  $L \in 4\mathbb{N}+2$.   Interestingly, the value of the  Binder and  correlation  ratios depend on this choice of the boundary, see Fig.~\ref{fig:X_RG-inv}(b),(c),(d)).    We believe that this stems form the fact that  both quantities do not have a well defined thermodynamic limit at $h = h_c$. i.e.  $ \lim_{L\rightarrow  \infty }  R_O( h = h_c) $   is  mathematically not defined.   However, the free energy, see Fig.~\ref{fig:X_RG-inv}(a), 
the critical field, see Fig.~\ref{fig:exponents_appendix}(a)
the exponents, see Figs.~\ref{fig:exponents_appendix}(b-d), should ultimately converge to the same value.  For odd lattices  corrections to scaling are  larger. 

The critical exponents $2\beta/\nu$ and $\eta$ in Figs.~\ref{fig:exponents_appendix}(c,d),  stem from the scaling assumptions
\begin{align}
  S(\ve{k}=0, h=h_c, L) &\propto L^{2\beta/\nu},
  & 
  \chi(\ve{k}=0, h=h_c, L) &\propto L^{2-\eta_\phi}
\end{align}
where we omitted, as before, the dependence on  the inverse temperature $\beta$ and  on corrections to scaling. Replacing $h_c$ by the crossing point $h_*(L)$ of an RG-invariant quantity $R$, meaning $R(h_*(L), L) = R(h_*(L), L+4)$ with $R \in \{R_S, R_\chi, B\}$, we obtain:
\begin{align}
  \label{eq:exponent_b}
  2\beta/\nu &= \log\left( \frac{S(\ve{k}=0, L+4, h=h_*(L)}
                             {S(\ve{k}=0, L, h=h_*(L)} \right) 
             / \log\left( \frac{L+4}{L} \right), 
  \\
  \label{eq:eta_phi}
  \eta_\phi &= 2 - \log\left( \frac{\chi(\ve{k}=0, L+4, h=h_*(L)}
                             {\chi(\ve{k}=0, L, h=h_*(L)} \right) 
             / \log\left( \frac{L+4}{L} \right).
\end{align}

\begin{figure}[H]
\centering
  \begin{minipage}[c]{0.4\textwidth}
    \includegraphics[width=\linewidth]{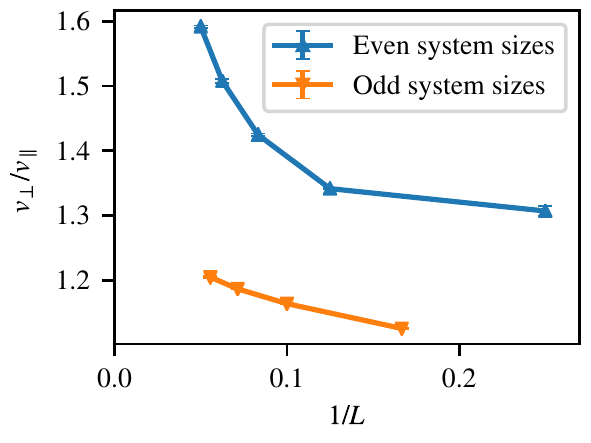}
  \end{minipage}
  \begin{minipage}[c]{0.3\textwidth}
    \caption{
\label{fig:v_anisotropy_appendix}
Odd-even effects for the $C_{4v}$ model on the anisotropy velocity of Dirac cones at the critical point.
    }
  \end{minipage}
\end{figure}

As apparent in Fig.~\ref{fig:exponents_appendix}, 
$h_c$ has the smallest corrections to scaling when determined  from $R_S$. However  the  smallest corrections to scaling are  when determining the critical exponents $2\beta/\nu$, $\eta_\phi$ and $z$, are  obtained   by using  $h_c$  as  determined from $R_\chi$.  Finally, the velocity anisotropy at the critical point grows in both cases, but is much smaller for odd system sizes, Fig.~\ref{fig:v_anisotropy_appendix}.

\newpage
%%%%%%%%%%%%%%%%%%%%%%%%%%%%%%%%%%%%%%%%%%%%%%%%%%%%%%%
\section{Other values for \texorpdfstring{$N_\sigma$}{Nsigma} and \texorpdfstring{$\xi$}{xi}}
%%%%%%%%%%%%%%%%%%%%%%%%%%%%%%%%%%%%%%%%%%%%%%%%%%%%%%%

In this section, we   report  the  result of additional simulations for  different  values of   $N_\sigma$ and $\xi$. For the $C_{2v}$ model, we show how at higher couplings, $\xi$,  discontinuities occur due to level crossings, as already described in  Sec.~\ref{sec:meanfield}. For the $C_{4v}$ model, we show that the transition stays continuous for  all  considered parameters.

\subsection{The \texorpdfstring{$C_{2v}$}{C2v} model}
\begin{figure}[hb]
\includegraphics[width=\textwidth]{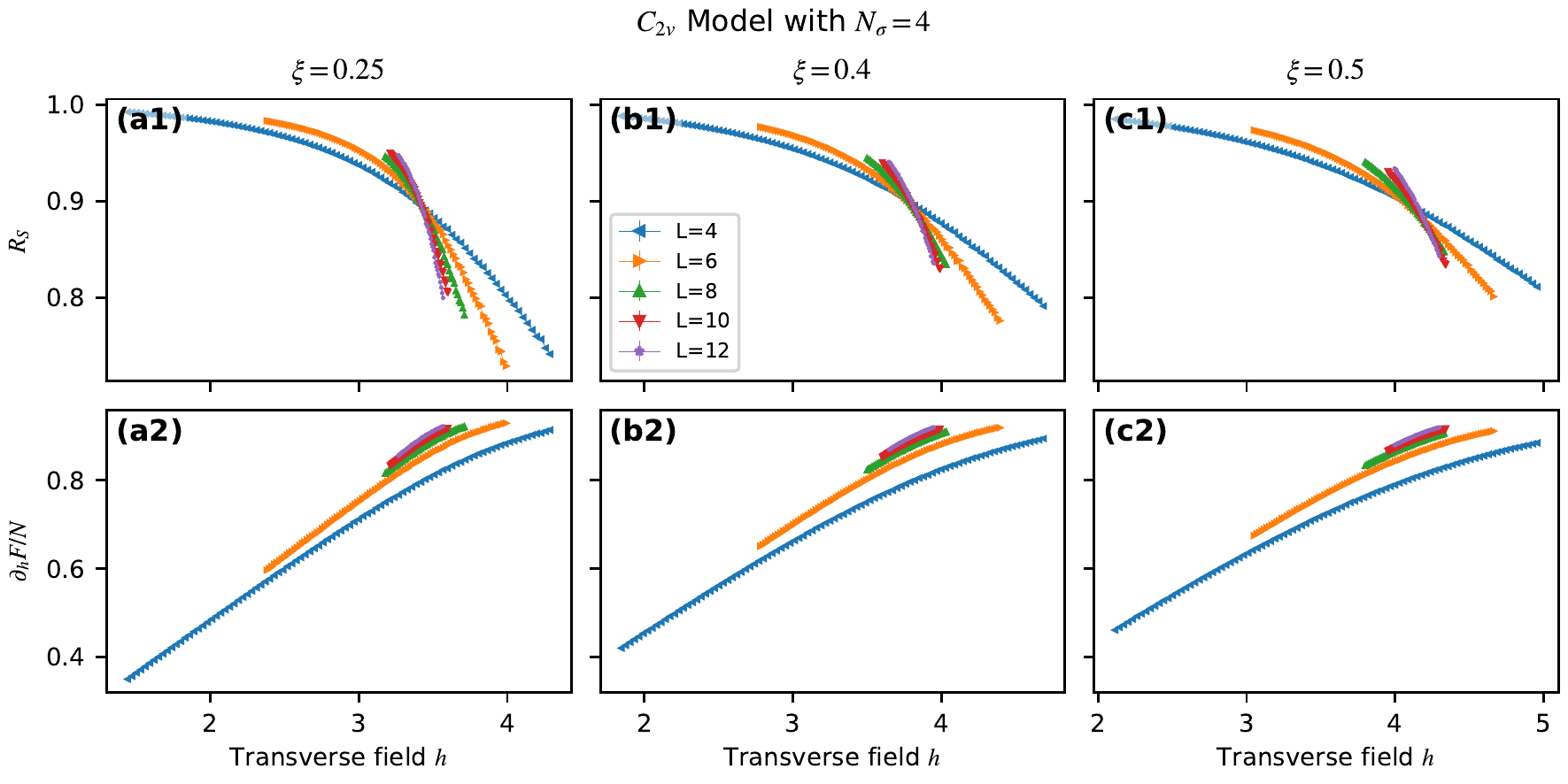}
\caption{
\label{fig:c2v_SU4}
Structure factor correlation ratio and derivative of free energy for the  $C_{2v}$ model at $N_\sigma=4$ and $\xi \in \{0.25, 0.4, 0.5\}$. The data is consistent with continuous transitions.
}
\end{figure}

\begin{figure}[H]
\includegraphics[width=\textwidth]{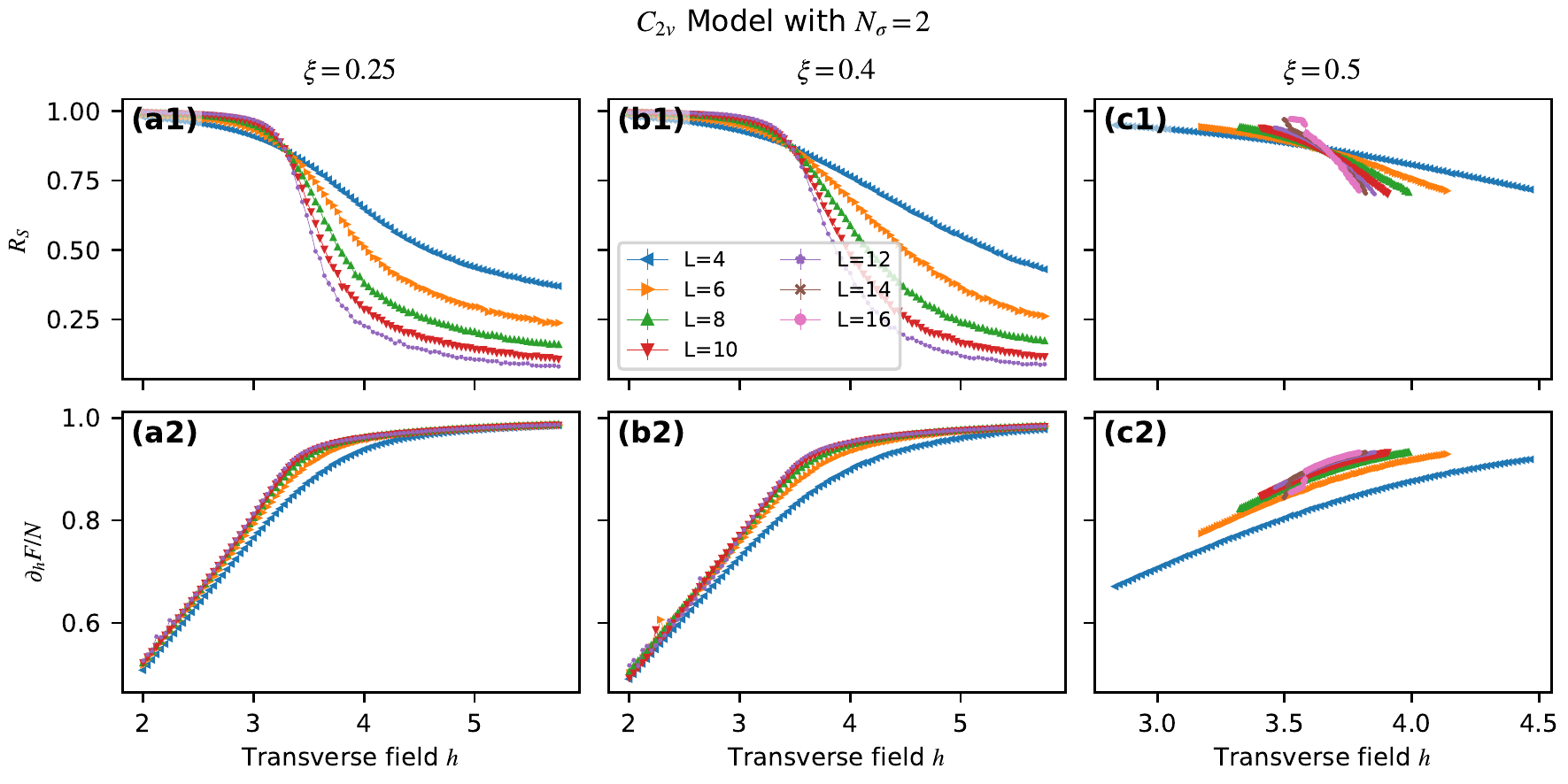}
\includegraphics[width=\textwidth]{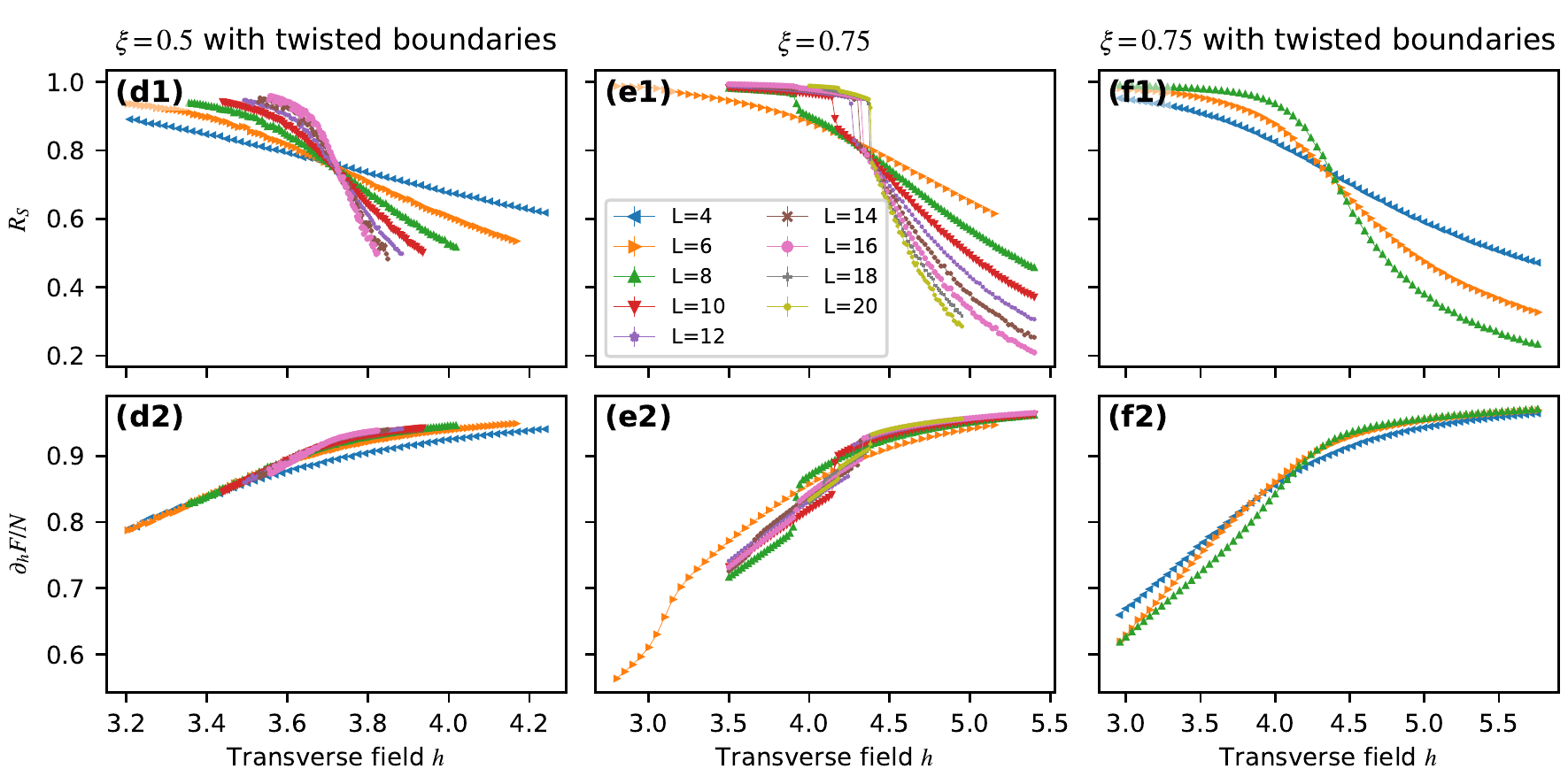}
\caption{
\label{fig:c2v_SU2}
Structure factor correlation ratio and derivative of free energy for the  $C_{2v}$ model at $N_\sigma=2$ and $\xi \in \{0.25, 0.4, 0.5, 0.75\}$. At $\xi=0.5$ and $\xi=0.75$ discontinuities due to level crossing emerge.  They  can be avoided by twisting the boundary conditions of the fermionic degrees of freedom. 
}
\end{figure}

Fig.~\ref{fig:c2v_SU4} shows the structure factor correlation ratio and derivative of free energy for the  $C_{2v}$ model at $N_\sigma=4$ and $\xi \in \{0.25, 0.4, 0.5\}$.  For these parameters we  observe a continuous phase transition. Fig.~\ref{fig:c2v_SU2} plots the same observables for $N_\sigma=2$. For lower values of the coupling $\xi$ the curves are also smooth, but at $\xi =0.5$ discontinuities appear, which get more pronounced at $\xi=0.75$.  At $\xi=0.75$, one can observe multiple discontinuities for a single system size, e.g. at $h\approx 4.2$ and $h\approx 4.4$ for $L=20$. These discontinuities occur due to level crossings, as already described in the mean field part in Sec.~\ref{sec:meanfield}. As shown in Fig.~\ref{fig:c2v_SU2}(d,f) and elaborated in the mean field section, they can be avoided by twisting the boundary conditions of the fermionic degrees of freedom.

\subsection{The \texorpdfstring{$C_{4v}$}{C4v} model}

\begin{figure}
\includegraphics[width=\textwidth]{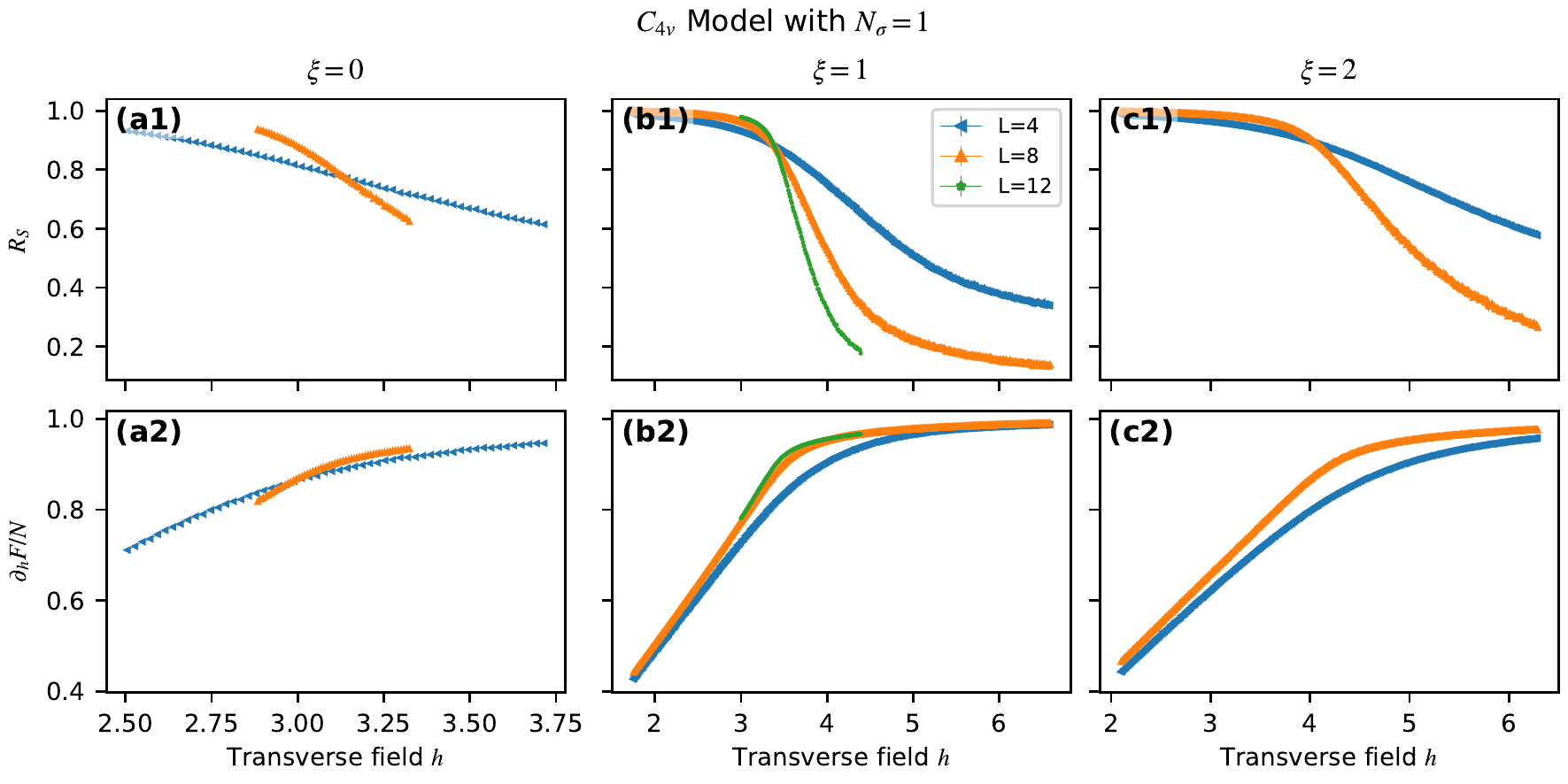}
\caption{
\label{fig:c4v_SU1}
Structure factor correlation ratio and derivative of free energy for the  $C_{4v}$ model at $N_\sigma=1$ and $\xi \in \{0, 1, 2\}$. The data shows continuous transitions.
}
\end{figure}

\begin{figure}
\includegraphics[width=\textwidth]{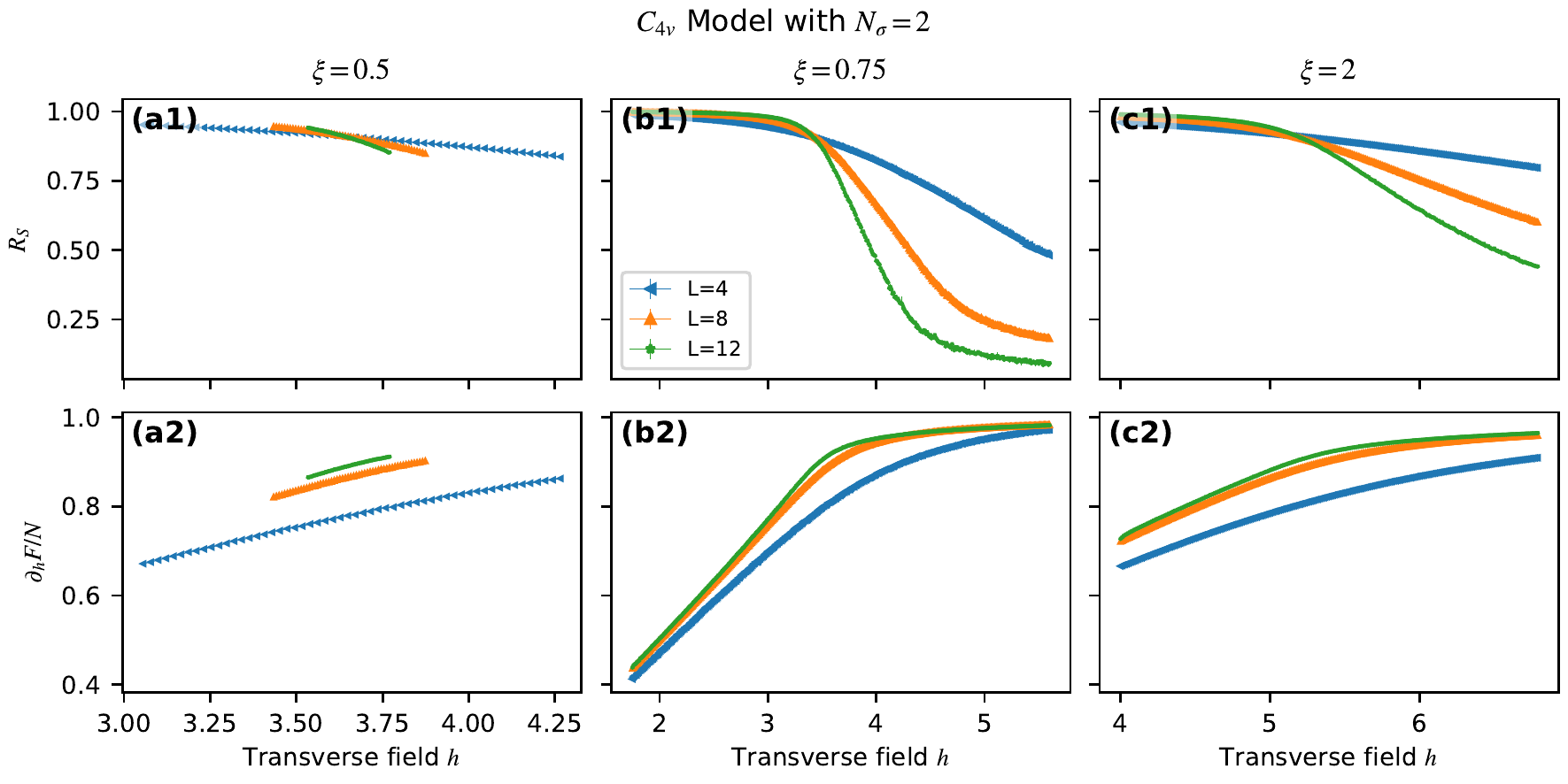}
\caption{
\label{fig:c4v_SU2}
Structure factor correlation ratio and derivative of free energy of the  $C_{4v}$ model at $N_\sigma=2$ and $\xi \in \{0.5, 0.75, 2\}$. The data shows continuous transitions.
}
\end{figure}

Figs.~\ref{fig:c4v_SU1},\ref{fig:c4v_SU2} have the same layout as the previous figures and show only continuous transitions for various combinations of $N_\sigma \in \{1, 2\}$, $\xi \in \{0.5, 0.75, 1, 2\}$.   We  also show data at  $\xi = 0$, which corresponds to the  transverse-field Ising model.

\end{document}